\documentclass[a4paper]{article}
\usepackage[a4paper,left=1.2in, right=1.2in,top=1.1in, bottom=1.1in]{geometry}

\begin{document}

\title{Three jet production and gluon saturation effects in p-p and p-Pb
collisions within high-energy factorization}

\author{A. van Hameren, P. Kotko, K. Kutak \vspace{10pt} \\ \textit{\small The H.\ Niewodnicza\'nski Institute of Nuclear Physics} \\ \textit{\small Polish Academy of Sciences}\\ \textit{\small Radzikowskiego 152, 31-342 Cracow, Poland}}
\maketitle
\begin{abstract}
We analyze three-jet production in the central-forward and forward
rapidity regions in proton-proton and proton-lead collisions at LHC
energies. Our calculation relies on high-energy factorization with
a single off-shell gluon obeying small $x$ evolution equation which includes saturation. The calculations are
made using two independent Monte Carlo codes implementing tree-level
gauge invariant off-shell matrix elements. We calculate differential
cross sections for azimuthal decorrelations and unbalanced jet transverse
momenta and discuss them in the context of differences in the evolution
of the unintegrated gluon densities.
\end{abstract}

\section{Introduction}

Jet production processes are excellent testing ground for perturbative
QCD, notably because their analysis does not require a knowledge of
fragmentation functions which are subject to large errors. The only
non-perturbative input that enters theoretical calculations are thus
parton distribution functions (PDFs). They are defined by a particular
factorization scheme; for instance for the collinear factorization
(see \cite{Collins:2011zzd} for a review) the PDFs undergo linear
Dokshitzer-Gribov-Lipatov-Altarelli-Parisi
evolution equations. It is however known, that at high energies reachable
nowadays at LHC, certain types of logarithms occurring in the perturbative
calculations can spoil the procedure. The general method is to resume
those logarithms giving rise to new types of evolution equations,
for instance the linear Balitski-Fadin-Kuraev-Lipatov (BFKL) equation \cite{Kuraev:1977fs,Balitsky:1978ic},
Catani-Ciafaloni-Fiorani-Marchesini (CCFM) \cite{Ciafaloni:1987ur,Catani:1989sg,Catani:1989yc}, the nonlinear
Balitski and Kovchegov (BK) equation \cite{Balitsky:1995ub,Kovchegov:1999yj} or nonlinear
extension of CCFM -- the Kutak-Golec-Biernat-Jadach-Skrzypek (KGBJS) equation \cite{Kutak:2011fu,Kutak:2012qk}.

An example of a situation that requires the resummation of high-energy
logarithms is the production of forward jets \cite{Mueller:1986ey,DelDuca:1993zw,Stirling:1994he,Marquet:2005vp}.
Large energies and rapidities of forward jets allows to
probe the small $x$ regime and thus it is an excellent testing ground for
various resummation schemes and gluon saturation phenomenon which
should occur at high-energy densities \cite{Gribov:1984tu,Mueller:1985wy}.
Forward jets are even more attractive nowadays as it is possible to study
them experimentally at LHC. Thanks to dedicated forward calorimeters,
the ATLAS and CMS detectors allow to reconstruct large-transverse-momentum
jets up to about 5 units of rapidity. This gives an opportunity to
study small $x$ effects experimentally and possibly access the kinematic region where
gluon saturation may enter the game. There are indeed hints that saturation
actually happens \cite{Stasto:2000er,Albacete:2010pg,Dumitru:2010iy,Kutak:2012rf}.
Various studies of forward jets were done in Refs. \cite{Deak:2009xt,Deak:2011ga,Ivanov:2012ms,Chachamis:2012mw,Hentschinski:2013bp}
. For the recent experimental studies see \cite{Chatrchyan:2011wm,Khachatryan:2011zj,Chatrchyan:2012gwa}.

In the present paper we study three-jet production at the LHC within
the high-energy factorization framework, which shall be reviewed in
Section \ref{sec:HighEnFact}. Multijet processes are interesting
particularly due to a bigger phase space -- by applying various cuts
different properties of gluon densities can be studied. For instance
by restricting two of the jets to balance each other on the transverse
plane, the third jet can access the gluon transverse space directly.
The detailed kinematics of the processes we consider is described
in Section \ref{sec:Kinem}. We present numerical results and discuss
their possible interpretation in Section \ref{sec:Results}. Finally,
we give overall summary in Section \ref{sec:Summary}.

\section{Factorization at high energies}

\label{sec:HighEnFact}

Let us now briefly recall some of the existing formalisms that may
be attempted to describe the observables at high energies. Before
doing so, let us however make some important remarks. First of all,
the full control over the calculation, in particular over its limitations,
can be achieved only when working within well established factorization
theorems of QCD. Besides the well-known collinear factorization there
are so-called transverse momentum dependent (TMD) factorization theorems.
They do work in certain processes (see \cite{Collins:2011zzd}) but
fail in some other. In general they are expected to fail in hadron-hadron
collisions (\cite{Collins:2007nk,Rogers:2010dm}, see also the short
summary in \cite{Mulders:2011zt}). The TMD factorizations involve
transverse momentum dependent gluon distribution functions, similar
to those that are often used in high-energy phenomenology. The problem
is, however, that the former are not universal (this is the reason
the factorization is violated) whereas the latter are often conjectured
to be universal. Let us thus comment on the factorization at the kinematic
limit we consider in the paper, namely the ``small $x$'' regime. It
deals with dense hadronic matter (especially for collisions with heavy
ions) for which the formalism of Color Glass Condensate (CGC) \cite{Gelis:2010nm}
proves to be very successful \cite{Albacete:2010pg}. Basically the
TMD factorizations do not deal with the small x limit, although it was
shown in \cite{Xiao:2010sp} that the universality of TMD PDFs is also violated
in that limit. On the other hand, in Ref. \cite{Dominguez:2011wm}
it is argued that an ``effective'' factorization within CGC (for
dilute-dense collisions) can be seen as an instance of TMD factorization
in case of dijets production in the limit of small unbalanced transverse
momentum. The factorization formula is then stated as a convolution
of a few universal transverse momentum dependent gluon densities and
matrix elements which are on-shell, but use off-shell kinematics.
In large $N_{c}$ limit those different gluon PDFs can be expressed in terms
of two fundamental quantities: the Weizs\"acker-Williams gluon density
\cite{McLerran:1993ni,McLerran:1993ka} and the ``dipole'' gluon
density (see eg. \cite{Kovchegov:2001sc,Kharzeev:2003wz} and references
therein; for the possible theoretical issues of those gluon densities
as seen from the TMD factorization point of view see \cite{Avsar:2011tz,Avsar:2012hj}).
Both densities are related to certain two-point Green functions. In
the more general case of multi-particle production higher correlators are
needed (within CGC they are expressed by means of certain averages
of the Wilson lines). However, as shown in Ref. \cite{Dominguez:2012ad}
in case of dilute-dense collisions and large $N_{c}$ limit only two-point
and four-point Green functions are needed. For recent applications
to multi-particle production see also \cite{Iancu:2013uva}.

We see, that the theoretical picture of the processes at very large
densities (very small x) is complicated. Unless one is outside the
saturation regime (i.e. in the BFKL or CCFM domain), only simple cases
like inclusive gluon production can be described in terms of single
transverse momentum dependent gluon density \cite{Blaizot:2004wu,Fujii:2006ab}. 

In the present paper we mainly concentrate on the proton-proton collisions
within the kinematic region accessible by contemporary experiments.
The nonlinear effects -- although present -- are actually rather weak. Therefore,
we shall use simple $k_{T}$-factorization with a single type of unintegrated
gluon density, incorporating, however, the nonlinear evolution (when
necessary we shall compare the results to the BFKL evolution with sub-leading corrections included \cite{Kwiecinski:1997ee}). Although
simplified, such an approach was proved to be very interesting phenomenologically
\cite{Kutak:2012rf}.

As a reference for the $k_{T}$-factorization we take the works of
Catani, Ciafaloni and Hautmann (CCH) \cite{Catani:1990eg,Catani:1990ka,Catani:1990xk,Catani:1993rn,Catani:1993ww,Catani:1994sq}
(the following type of factorization formula appeared also much earlier in \cite{Gribov:1984tu}).
Originally it was stated for heavy quark pair production at tree-level;
however we shall assume that one can extend it for more complicated
final states including gluons, with the complications explained below.
The factorization is expressed by the following formula (see Fig.
\ref{fig:factoriz}A); for some partonic final state $X$ and two
initial state hadrons $A$, $B$ we have\begin{multline}
d\sigma_{AB\rightarrow X}=\int \frac{d^{2}k_{T\, A}}{\pi}\int\frac{dx_{A}}{x_{A}}\,\int \frac{d^{2}k_{T\, B}}{\pi} \int\frac{dx_{B}}{x_{B}}\,\\
\mathcal{F}_{g^{*}/A}\left(x_{A},k_{T\, A}\right)\,\mathcal{F}_{g^{*}/B}\left(x_{B},k_{T\, B}\right)\, d\hat{\sigma}_{g^{*}g^{*}\rightarrow X}\left(x_{A},x_{B},k_{T\, A},k_{T\, B}\right),\label{eq:HEN_fact_1}\end{multline}
where $\mathcal{F}_{g^{*}/H}$ are transverse-momentum-dependent densities
 of the off-shell gluons $g^{*}$ inside $H$ (to be discussed below)
and $\hat{\sigma}_{g^{*}g^{*}\rightarrow X}$ is the high-energy 
 hard cross section for the process $g^{*}\left(k_{A}\right)g^{*}\left(k_{B}\right)\rightarrow X$.
The momenta of the off-shell gluons that enter the hard cross section
are defined to be\begin{equation}
k_{A}^{\mu}=x_{A}p_{A}^{\mu}+k_{T\, A}^{\mu},\,\,\,\,\, k_{B}^{\mu}=x_{B}p_{B}^{\mu}+k_{T\, B}^{\mu},\label{eq:kAkB}\end{equation}
where $p_A\cdot k_T = p_B\cdot k_T = 0$. 
The hard amplitude with the external off-shell gluons is defined
by means of certain high-energy (or eikonal) projectors, i.e. the
off-shell leg (including the propagator) with momentum $k_{A}$, $k_{B}$
is contracted with $\left|\vec{k}_{T\, A}\right|p_{A}^{\mu}$, 
$\left|\vec{k}_{T\, B}\right|p_{B}^{\mu}$
respectively (see Fig. \ref{fig:factoriz}B). As already mentioned,
in the original CCH works the production of a heavy
quark pair was considered. In that case the corresponding off-shell amplitude is
gauge invariant, fundamentally due to the form of the projectors.
This is however not true for the off-shell amplitudes with gluons
in the final state. There are several ways to deal with this problem.
First, Lipatov's effective action \cite{Lipatov:1995pn} and the resulting
Feynman rules \cite{Antonov:2004hh} can be used. This is because
the kinematics (\ref{eq:kAkB}) corresponds to quasi-multi-Regge kinematics
in the terminology of \cite{Antonov:2004hh} (for recent LHC-related calculations in that framework we refer e.g. to 
\cite{Nefedov:2013ywa,Saleev:2012np,Kniehl:2011hc}). A second approach developed
recently in \cite{vanHameren:2012if} is suitable for automatic calculation
of large final state multiplicities and uses a manifestly gauge invariant
method of embedding the off-shell process in a larger on-shell process
without compromising high-energy kinematics. Finally, there is one
more approach \cite{vanHameren:2012uj} suitable in a simplified situation described later in
this section.

\begin{figure}
\begin{centering}
\parbox{5cm}{A)\\\includegraphics[width=5cm]{fw3j_fig1}}~~~~~~~~~~~~\parbox{5cm}{B)\\\includegraphics[width=5cm]{fw3j_fig2}}
\par\end{centering}

\caption{\label{fig:factoriz}A) Factorization of a hadronic collision into
unintegrated PDFs (top and bottom blobs after 'squaring') and parton-level
sub-process (middle blob). B) The hard sub-process is defined by an
off-shell matrix elements with incoming off-shell gluon propagators
contracted with high-energy projectors (explained on the r.h.s). In
order to make this amplitude gauge invariant additional contributions
are needed (see the main text).}

\end{figure}

In the CCH approach, the transverse-momentum-dependent gluon densities
$\mathcal{F}_{g^{*}/H}$ were originally assumed to undergo the BFKL
evolution. As we have remarked above, we shall use the formula (\ref{eq:HEN_fact_1})
with $\mathcal{F}_{g*/H}$ incorporating more subtle effects, in particular
gluon saturation. In the present paper we shall use the nonlinear
BK equation, extended with a consistency constraint, a non singular
piece of the gluon splitting function and running strong coupling
constant \cite{Kutak:2003bd,Kutak:2004ym}. The strength of the non-linearity
is adjusted by a parameter that can be interpreted as a radius of
a hadron, either proton or nuclei (e.g. in Ref. \cite{Kutak:2012rf} it was applied for
a lead target). Let us remark, that since there is a conjecture that
integration of $\mathcal{F}_{g^{*}/H}$ over transverse momentum is
related to a collinear PDF, we shall often refer to $\mathcal{F}_{g^{*}/H}$
as the unintegrated gluon density. The relation between both quantities reads
\begin{equation}
\int dk^2_{T\, A}\,\mathcal{F}_{g^{*}/A}\left(x_{A},k_{T\, A}\right) = x_A f_{g/A}\left(x_A\right).
\end{equation}
Suppose now, that we deal with asymmetric kinematics, i.e. $x_{B}\gg x_{A}$,
 which is a characteristic feature of forward scattering (see the next section). 
Then the gluon originating
from hadron $B$ is probed near the mass-shell and $\mathcal{F}_{g^{*}/B}$
should be replaced by its collinear equivalent $f_{g/B}$. Moreover,
the valence quarks play important role. Thus, the proper
formula in such a setup is given by\begin{multline}
d\sigma_{AB\rightarrow X}=\int \frac{d^{2}k_{T\, A}}{\pi}\int\frac{dx_{A}}{x_{A}}\,\int dx_{B}\,\\
\sum_{b}\mathcal{F}_{g^{*}/A}\left(x_{A},k_{T\, A}\right)\, f_{b/B}\left(x_{B}\right)\, d\hat{\sigma}_{g^{*}b\rightarrow X}\left(x_{A},x_{B},k_{T\, A}\right),\label{eq:HENfact_2}\end{multline}
where $b$ runs over gluon and all the quarks that can contribute
to the production of multiparticle state $X$ (see also Ref. \cite{Deak:2009xt} and the appendix 
of \cite{Catani:1990eg} for the collinear limit in the high-energy factorization). In the formula
above, any scale dependence was suppressed. The off-shell gauge invariant
process $g^{*}b\rightarrow g\ldots g$ can be calculated along the
lines of Ref. \cite{vanHameren:2012uj}. The case with quarks is
actually straightforward, as in axial gauge any gauge contribution
due to Slavnov-Taylor identities vanishes.

Let us summarize our basic assumptions. We use $k_{T}$-factorized,
hybrid (i.e. collinear PDF is mixed with the unintegrated one, see
also \cite{Dumitru:2005gt} for CGC approach) form given in Eq. (\ref{eq:HENfact_2})
with the inclusion of non-liner effects in the evolution of the unintegrated
gluon PDF. The hard matrix elements are calculated fully off-shell
at tree-level; they are gauge invariant, and all of them are convoluted
with the same gluon density given in fundamental color representation, as given in  \cite{Kutak:2012rf}.

\section{Process definition and kinematics}

\label{sec:Kinem}

Let us now give a detailed description of the process we are interested
in. We want to study exclusive three jet events, namely,
\begin{equation}
A\left(p_{A}\right)B\left(p_{B}\right)\rightarrow J_{1}\left(p_{1}\right)J_{2}\left(p_{2}\right)J_{3}\left(p_{3}\right),\label{eq:process}
\end{equation}
where $A=\left\{ \mathrm{p^{+}},\mathrm{Pb}\right\} $, $B=\mathrm{p^{+}}$
and $J_{i}\left(p_{i}\right)$ denotes the jet with momentum $p_{i}$.
We work in the c.m. frame throughout the paper. This frame corresponds
to LAB frame for $\mathrm{p}^{+}\mathrm{p}^{+}$ collision, but not
for the $\mathrm{p}^{+}\mathrm{Pb}$ collisions. In our frame we define
\begin{equation}
p_{A}^{\mu}=\left(E,0,0,-E\right),\,\,\,\,\, p_{B}^{\mu}=\left(E,0,0,E\right),\label{eq:def_pApB}
\end{equation}
with $E=\sqrt{S}/2$ where $S$ is the total c.m. energy squared. In
the present paper we consider c.m. energies $\sqrt{S}=5.02\,\mathrm{TeV}$
and $\sqrt{S}=7.0\,\mathrm{TeV}$. 

Let us now discuss the kinematic cuts that are relevant to the physics
we would like to address. To this end let us decompose the final state
momenta as follows
\begin{equation}
p_{i}^{\mu}=\frac{\left|\vec{p}_{T\, i}\right|}{\sqrt{S}}\,\left(e^{\eta_{i}}p_{A}^{\mu}+e^{-\eta_{i}}p_{B}^{\mu}\right)+p_{T\, i}^{\mu},\label{eq:final_moms_decomp}\end{equation}
where $p_{T\, i}\cdot p_{A}=p_{T\, i}\cdot p_{B}=0$ and the rapidity
$\eta_{i}$ is defined as
\begin{equation}
\eta_{i}=\frac{1}{2}\,\ln\frac{p_{i}^{0}+p_{i}^{z}}{p_{i}^{0}-p_{i}^{z}}.\label{eq:rap_def}
\end{equation}
Further we note that \begin{equation}
p_{T\, i}^{\mu}=\left(0,\vec{p}_{T\, i},0\right)\label{eq:pTmuvecdef}\end{equation}
and
\begin{equation}
\vec{p}_{T\, i}=\left(\left|\vec{p}_{T\, i}\right|\sin\phi_{i},\left|\vec{p}_{T\, i}\right|\cos\phi_{i}\right).\label{eq:pTazimdecomp}
\end{equation}

Now let us come back to the kinematic cuts. First of all we assume that
\begin{equation}
\left|\vec{p}_{T\, i}\right|>p_{T\,\mathrm{cut}},\,\,\, i=1,2,3.\label{eq:pTcut}
\end{equation}
The actual values of the cuts shall be given in the following sections
when we discuss numerical results. Typically, we shall order the jets
with decreasing $\left|\vec{p}_{T\, i}\right|$ values, i.e. 
\begin{equation}
\left|\vec{p}_{T\,1}\right|>\left|\vec{p}_{T\,2}\right|>\left|\vec{p}_{T\,3}\right|.\label{eq:pT_order}
\end{equation}
Further restriction is given by a jet definition. Here we work with
anti-$k_{T}$ clustering algorithm \cite{Cacciari:2008gp} with radius
$R_{\mathrm{cut}}$, thus the final state momenta cannot be too close
in the $\phi-\eta$ space%
\footnote{Actually for tree-level parton-level processes it is equivalent to
a proper cut on the $\phi-\eta$ plane. %
}. In order to access the small $x$ region we have to impose
additional cuts. According to \eqref{eq:final_moms_decomp} and the
factorization formula (cf. Eq. (\ref{eq:kAkB})),
the longitudinal fractions of the hadrons' momenta $x_{A}$, $x_{B}$
that initiate the hard scattering are given by
\begin{equation}
x_{A}=\sum_{i}\frac{\left|\vec{p}_{T\, i}\right|}{\sqrt{S}}\, e^{\eta_{i}},\,\,\,\,\, x_{B}=\sum_{i}\frac{\left|\vec{p}_{T\, i}\right|}{\sqrt{S}}\, e^{-\eta_{i}}.\label{eq:xAxBrap}
\end{equation}
Thus, in order to select small, say, $x_{A}$ for the fixed $S$ and
$\left|\vec{p}_{T\, i}\right|$ we have to go to large rapidity values,
ideally for all the jets. In the same time $x_{B}$ would be large;
thus, this sort of kinematics is often referred to as \textit{asymmetric
kinematics}. If some of the jets appear in the central rapidity region,
we can still access small $x$ regime provided at least one of the jets
is in the forward region. Those issues shall be illustrated by a specific
calculation in Section \ref{sub:assymetry}. The forward region is defined
as
\begin{equation}
\eta_{f\,0}\leq \eta_{i} \leq\eta_{f\,1},\label{eq:fwregion_def}
\end{equation}
while the central region is defined as
\begin{equation}
\left|\eta_{j}\right|\leq\eta_{c}\label{eq:centrreg_def}
\end{equation}
with specific boundary values $\eta_{f\,0}$, $\eta_{f\,1}$, $\eta_{c}$
given later. Note, that we tag the forward jet in the positive rapidity
hemisphere only, both for proton-proton and proton-lead collisions.

There are also additional cuts that might be interesting for the studies
of unintegrated gluon densities. We may restrict the two leading jets
to be back-to-back-like. More precisely, we can define
\begin{equation}
p_{T\,12}=\left|\vec{p}_{T\,1}+\vec{p}_{T\,2}\right|<D_{\mathrm{cut}},\label{eq:btb_def}
\end{equation}
with $D_{\mathrm{cut}}$ parameter being not much smaller then $p_{T\,\mathrm{cut}}$.
Such a study is motivated by the fact, that for $D_{\mathrm{cut}}\rightarrow0$
the total transverse momentum of the initial state gluons is transferred
to the forward jet.

\section{Numerical results and discussion}

\label{sec:Results}

\subsection{Preliminary remarks}

\label{sub:Result_remarks}

The numerical calculations were performed using two new
Monte Carlo programs and were cross-checked against each other. The
first program is a $\mathtt{C++}$ code using the $\mathtt{foam}$ algorithm \cite{Jadach:2002kn}
and based on the method for off-shell matrix elements described in
Ref. \cite{vanHameren:2012uj}. The working-name of the program is
$\mathtt{LxJet}$%
\footnote{The program shall be publicly available.%
}. The second independent code is a $\mathtt{fortran}$ program based
on \cite{vanHameren:2012if}. Since we want to study some small $x$ properties
of the jet observables within high-energy factorization itself, we
do not interface the program with any parton shower in the present
calculation. The final state parton shower can be added using e.g. $\mathtt{pythia}$ \cite{Sjostrand:2007gs}
 and shall be done in the future. Another shortcoming comes from neglecting
multiple parton interactions (MPIs). In the contary to dijet production, where
MPIs lead to a change of the overall normalization of angle distributions \cite{Stasto:2011ru}, 
the situation for three-jet production is more complicated and is left for further study.

Let us now summarize the inputs we have used. For the unintegrated
parton densities $\mathcal{F}_{g^{*}/H}$ we take the ones described
in the previous section and fitted to HERA data in \cite{Kutak:2012rf}.
These include the nonlinear PDFs for proton, lead, and additionally
the proton PDF with linear evolution \cite{Kwiecinski:1997ee}. For
the collinear PDFs $f_{a}$ we take CTEQ10 NLO set \cite{Lai:2010vv}.
The consistent strong coupling constant is also taken from the same
source. Since our calculations are essentially tree-level as far as
the parton-level amplitude is concerned, there is a large dependence
on the choice of the scales. In order to estimate the theoretical
uncertainty we do the following standard procedure. First we set the
renormalization and collinear factorization scales to be equal
$\mu_{f}=\mu_{r}\equiv\mu$ and choose
\begin{equation}
\mu=E_{1}+E_{2}+E_{3}.\label{eq:scale_ensum}
\end{equation}
Our error estimate is then given by the band constructed from the two outputs
with the two choices of the scale (including statistical errors):
$\mu/2$ and $2\mu$. In all calculations we choose the radius of
the anti-$k_{T}$ algorithm to be $R_{\mathrm{cut}}=0.5$. 

We consider two rapidity configurations:
\begin{itemize}
\item \textit{central-forward region}: we demand that the two hardest jets (with
indices $1$,$2$) are in the central region defined by $\eta_{c}=2.8$,
while the softest jet (with index $3$) is in the forward region defined
by $\eta_{f\,0}=3.2$, $\eta_{f\,1}=4.7$,
\item \textit{forward region}; all three jets are within the region defined
by $\eta_{f\,0}=3.2$, $\eta_{f\,1}=4.9$.
\end{itemize}

\subsection{Asymmetry distributions}

\label{sub:assymetry}

In order to check if the region of the longitudinal fractions $x_{A}$,
$x_{B}$ we access is consistent with our assumptions leading to Eq.
(\ref{eq:HENfact_2}) let us we define the following variable
\begin{equation}
x_{\mathrm{as}}=\frac{\left|x_{A}-x_{B}\right|}{x_{A}+x_{B}}.\label{eq:x_as}
\end{equation}
It has the support in $\left[0,1\right]$ and measures the asymmetry
of the event (for $x_{\mathrm{as}}\rightarrow1$ we have totally asymmetric
events). We expect that within our kinematic cuts the cross section
is dominated by asymmetric events, being thus in agreement with Eq.
(\ref{eq:HENfact_2}). This point shall be verified by explicit calculations
below.

\begin{figure}
\begin{centering}
\includegraphics[width=0.5\textwidth]{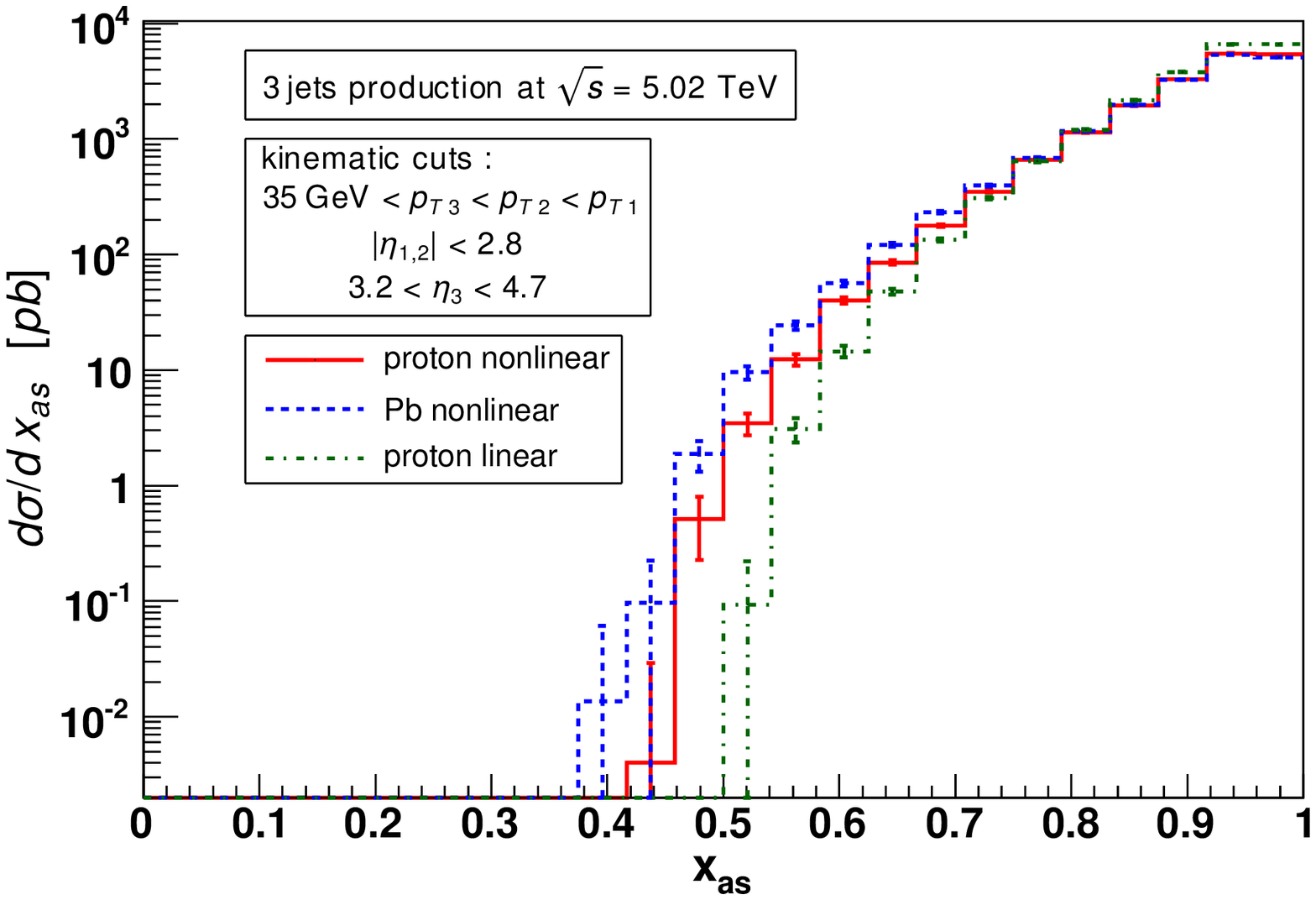}\includegraphics[width=0.5\textwidth]{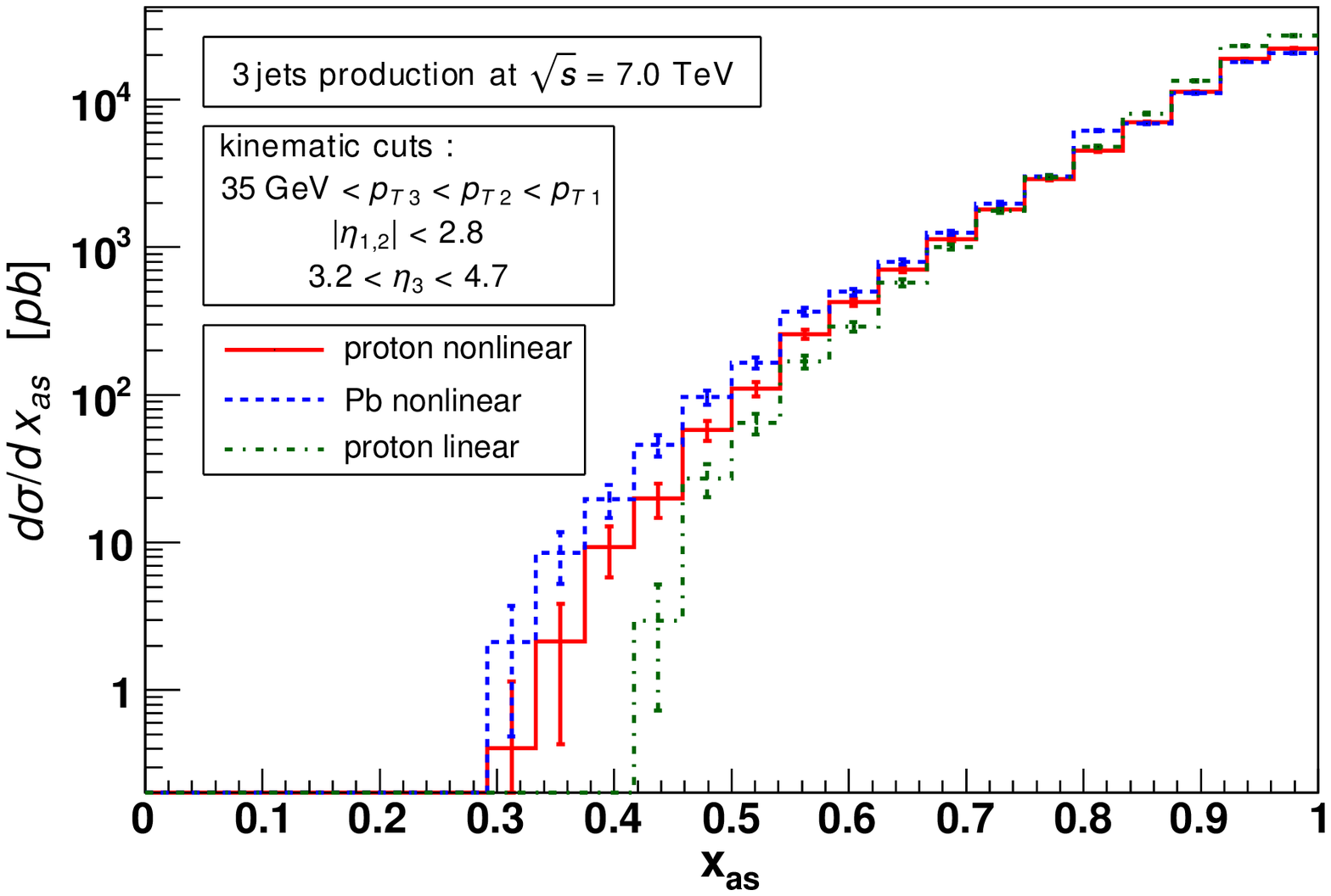}
\par\end{centering}

\caption{\label{fig:xassymetry_fw}The differential cross section as a function
of asymmetry variable $x_{\mathrm{as}}$ for the forward-central rapidity
region and two different c.m. energies: left for $5.02\,\mathrm{TeV}$,
right for $7.0\,\mathrm{TeV}$.}
\end{figure}

\begin{figure}
\begin{centering}
\includegraphics[width=0.5\textwidth]{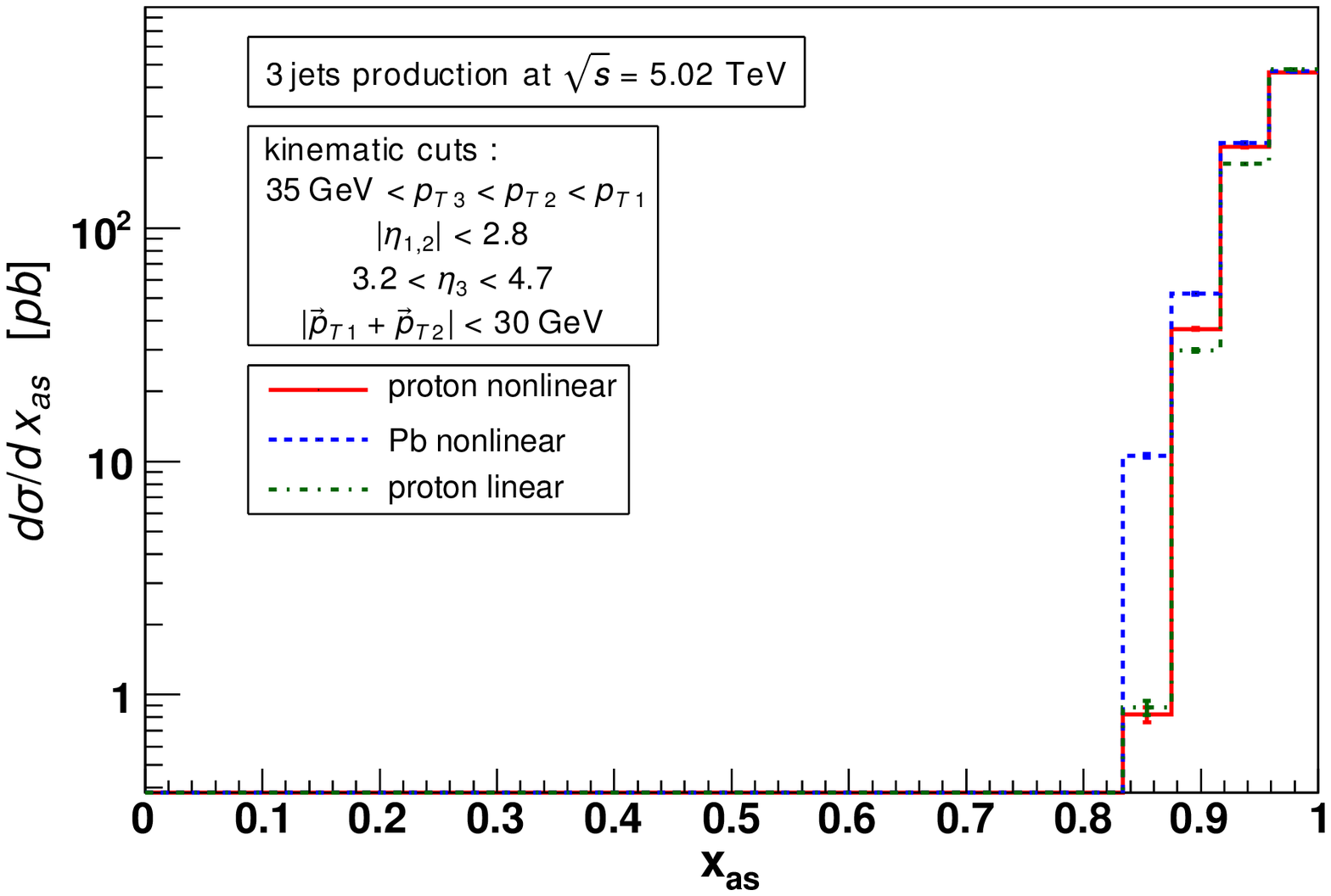}\includegraphics[width=0.5\textwidth]{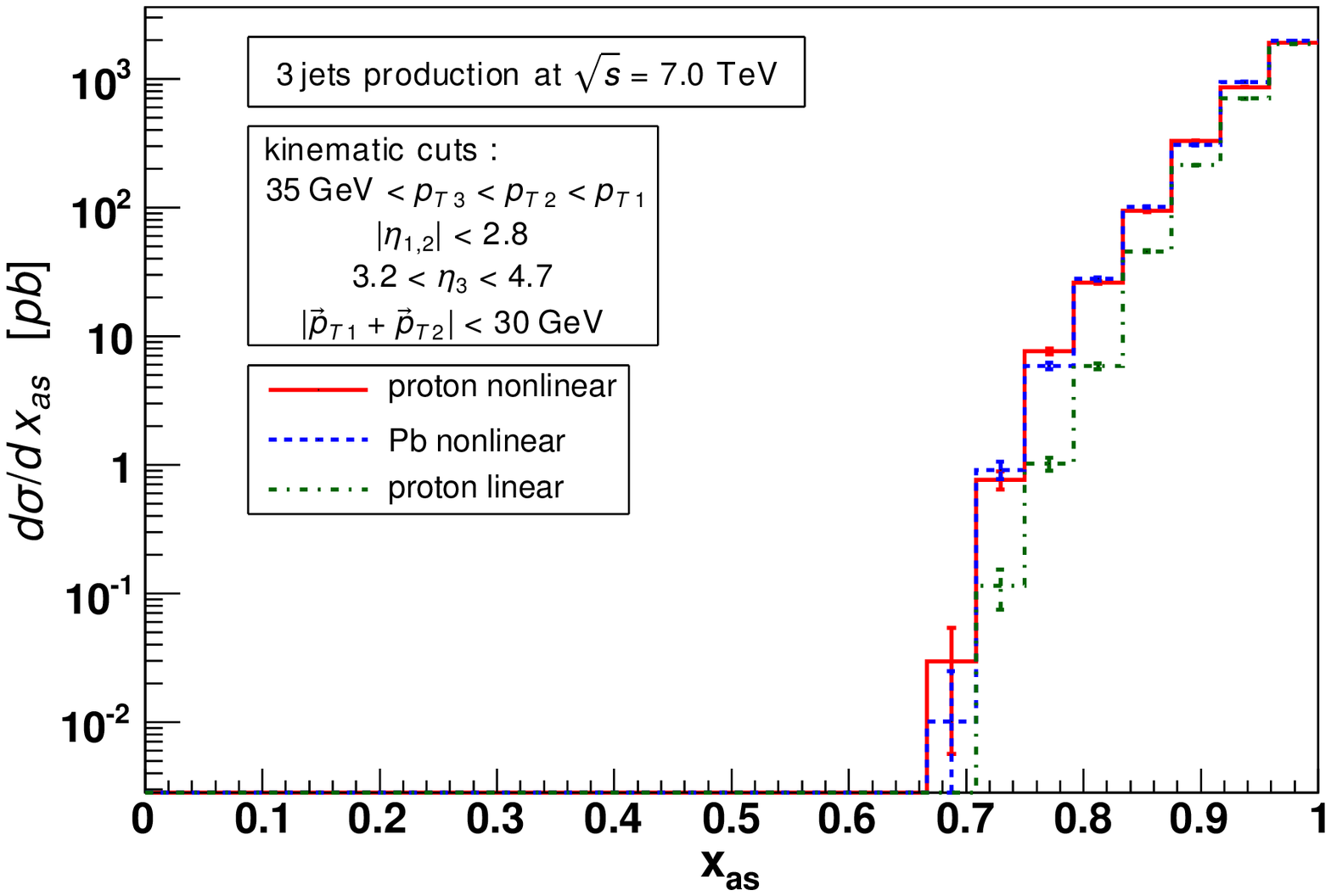}
\par\end{centering}

\caption{\label{fig:xassymetry_btb}The differential cross section as a function
of asymmetry variable $x_{\mathrm{as}}$ for forward-central rapidity
region with back-to-back cut $D_{\mathrm{cut}}=30\,\mathrm{GeV}$
and two different c.m. energies: left for $5.02\,\mathrm{TeV}$, right
for $7.0\,\mathrm{TeV}$.}
\end{figure}

\begin{figure}
\begin{centering}
\includegraphics[width=0.5\textwidth]{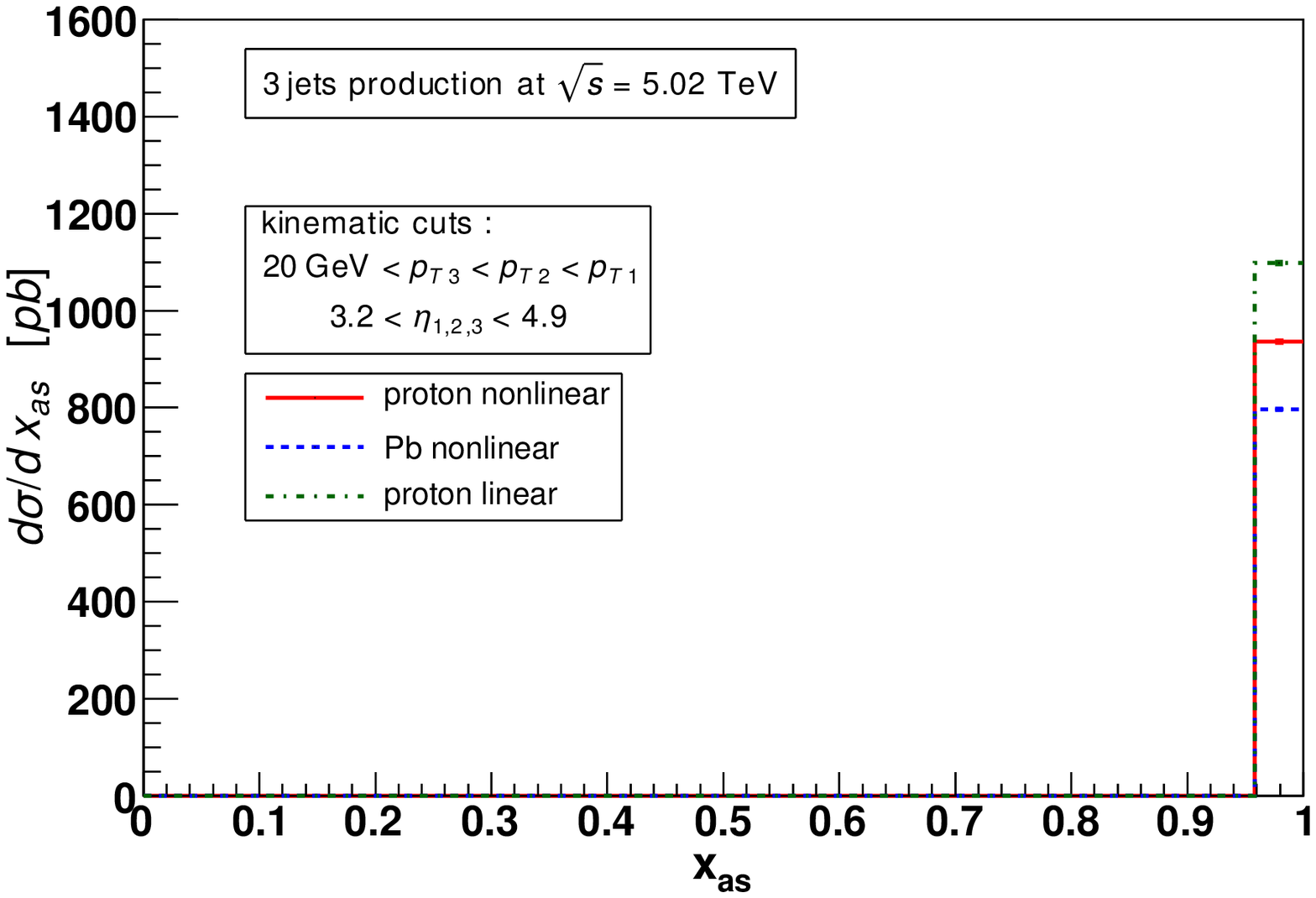}\includegraphics[width=0.5\textwidth]{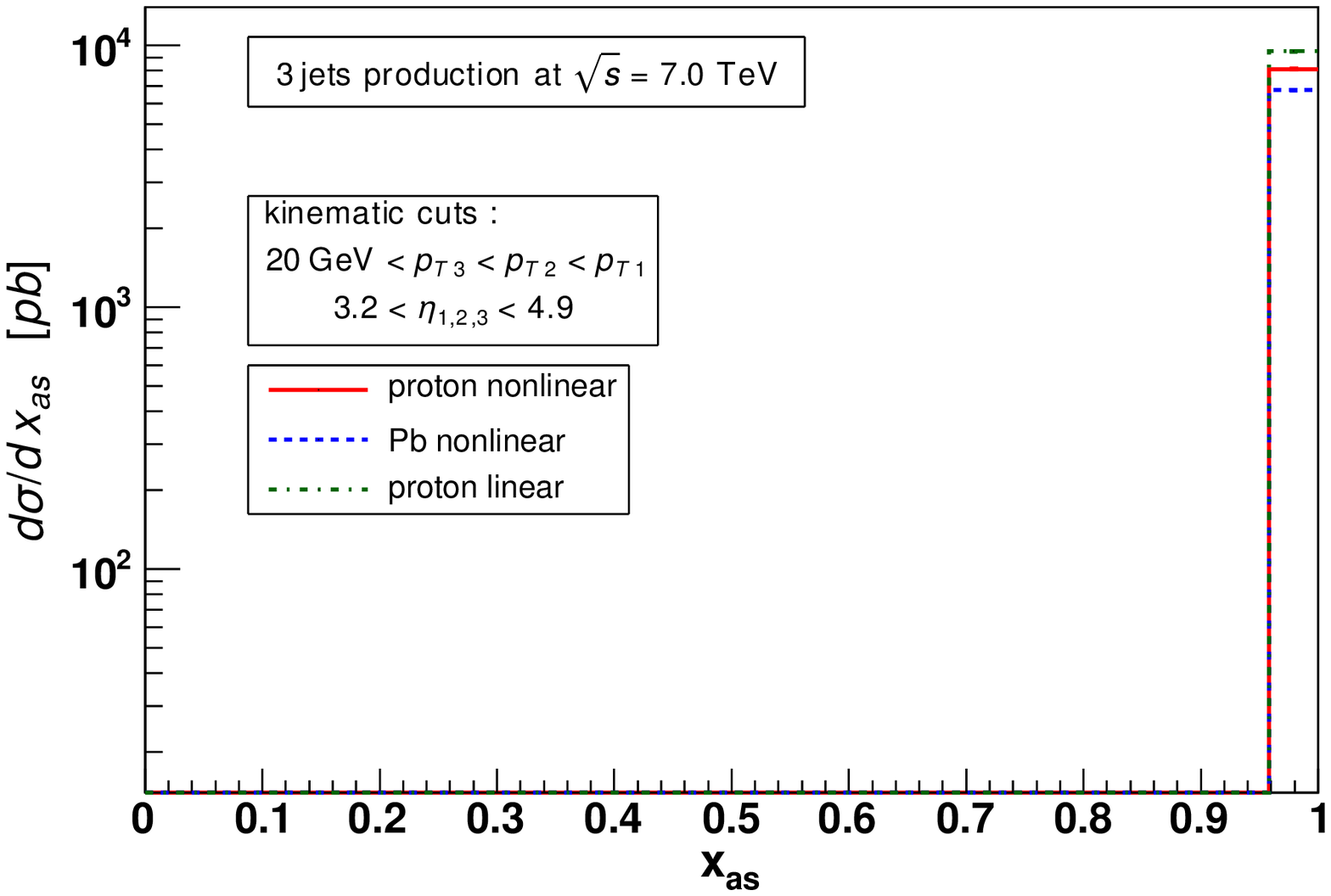}
\par\end{centering}

\caption{\label{fig:xassymetry_ffw}The differential cross section as a function
of asymmetry variable $x_{\mathrm{as}}$ for purely forward rapidity
region and two different c.m. energies: left for $5.02\,\mathrm{TeV}$,
right for $7.0\,\mathrm{TeV}$.}
\end{figure}

In the figures \ref{fig:xassymetry_fw}-\ref{fig:xassymetry_ffw}
we present differential cross sections in $x_{\mathrm{as}}$ for three
different scenarios: forward-central rapidity region, forward-central
region with two leading jets being close to back-to-back, and the purely
forward region. The rapidity regions were defined in the previous
section. Further details are given in the plots. We see that the collisions
in the forward region (Fig. \ref{fig:xassymetry_ffw}) are completely
asymmetric as one should expect. However, most of the events
in forward-central region are also asymmetric, as seen in Fig. \ref{fig:xassymetry_fw}.
We have observed, that -- as far as forward-central collisions are
concerned -- lowering the $p_{T\,\mathrm{cut}}$ spoils the asymmetry
of the events; thus one cannot go to as low $p_{T\,\mathrm{cut}}$
as for purely forward collisions.

\subsection{Azimuthal decorrelations}

\label{sub:Results_corr}

\subsubsection{Central-forward jets}

Let us now present the results for azimuthal decorrelations for central-forward
jet configuration. There are many azimuthal observables that can be
studied within this context. In this paper we study distributions
in the azimuthal angle between the leading jet (with index $1$) and
the softest jet (with index $3$) 
\begin{equation}
\phi_{13}=\left|\phi_{1}-\phi_{3}\right|,\,\,\,\,\phi_{13}\in\left[0,2\pi\right).\label{eq:phi13_def}
\end{equation}
Note, that this angle is always calculated in one direction and is
not just the smallest angle between the jets. This is important; if
we assume that the direction of the leading jet divides the azimuthal
plane to two half-planes, the events with the forward jet lying on
the left half-plane and right-half-plane (with the same smallest angle
to the leading jet) are not symmetric. Let us note that in the collinear
factorization at leading order the momentum conservation requires that $\pi/2<\phi_{13}<3\pi/2$.
Thus the shapes given in the plots discussed below are a characteristic
feature of the high-energy factorization.

\begin{figure}
\begin{centering}
\includegraphics[width=0.5\textwidth]{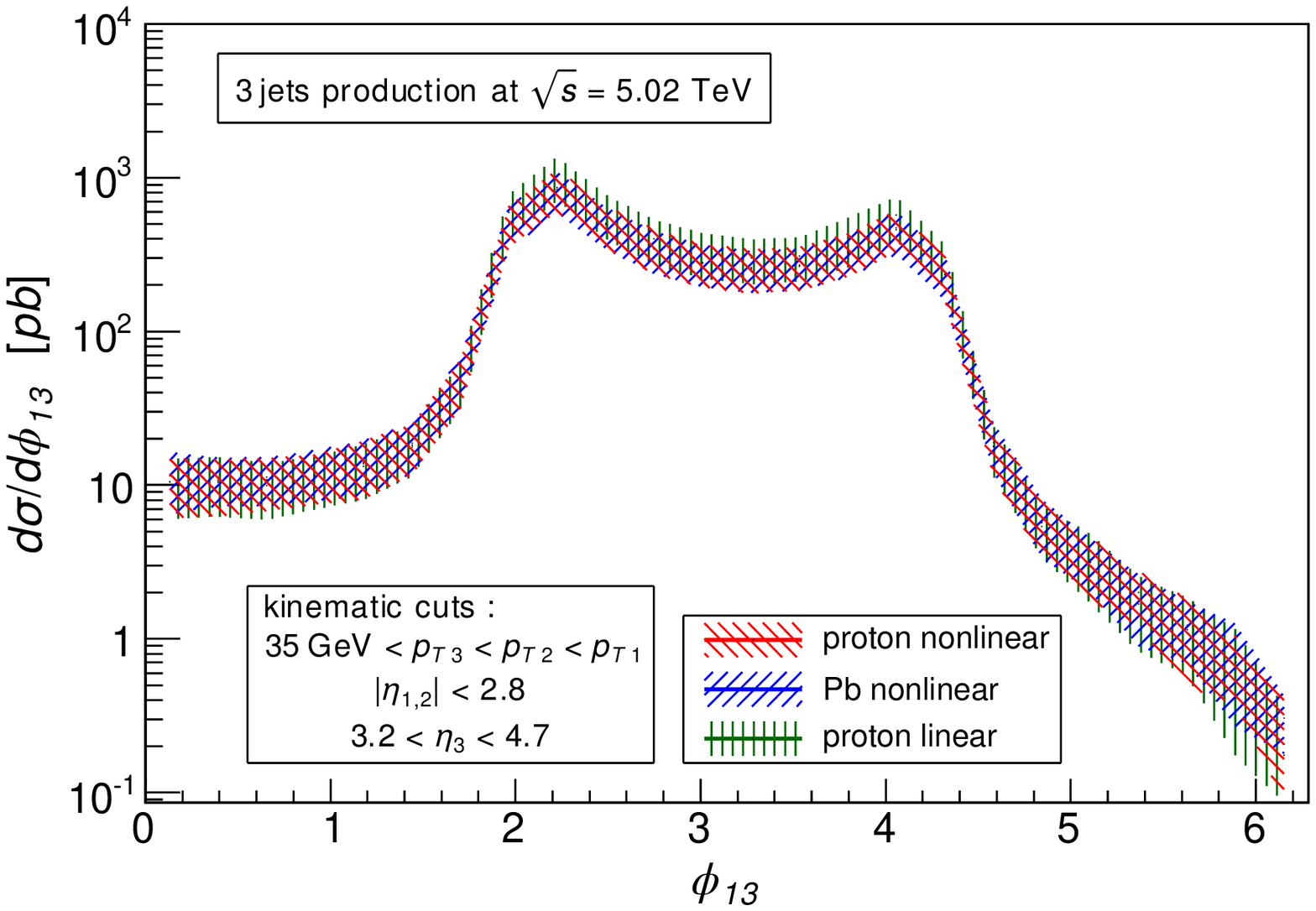}\includegraphics[width=0.5\textwidth]{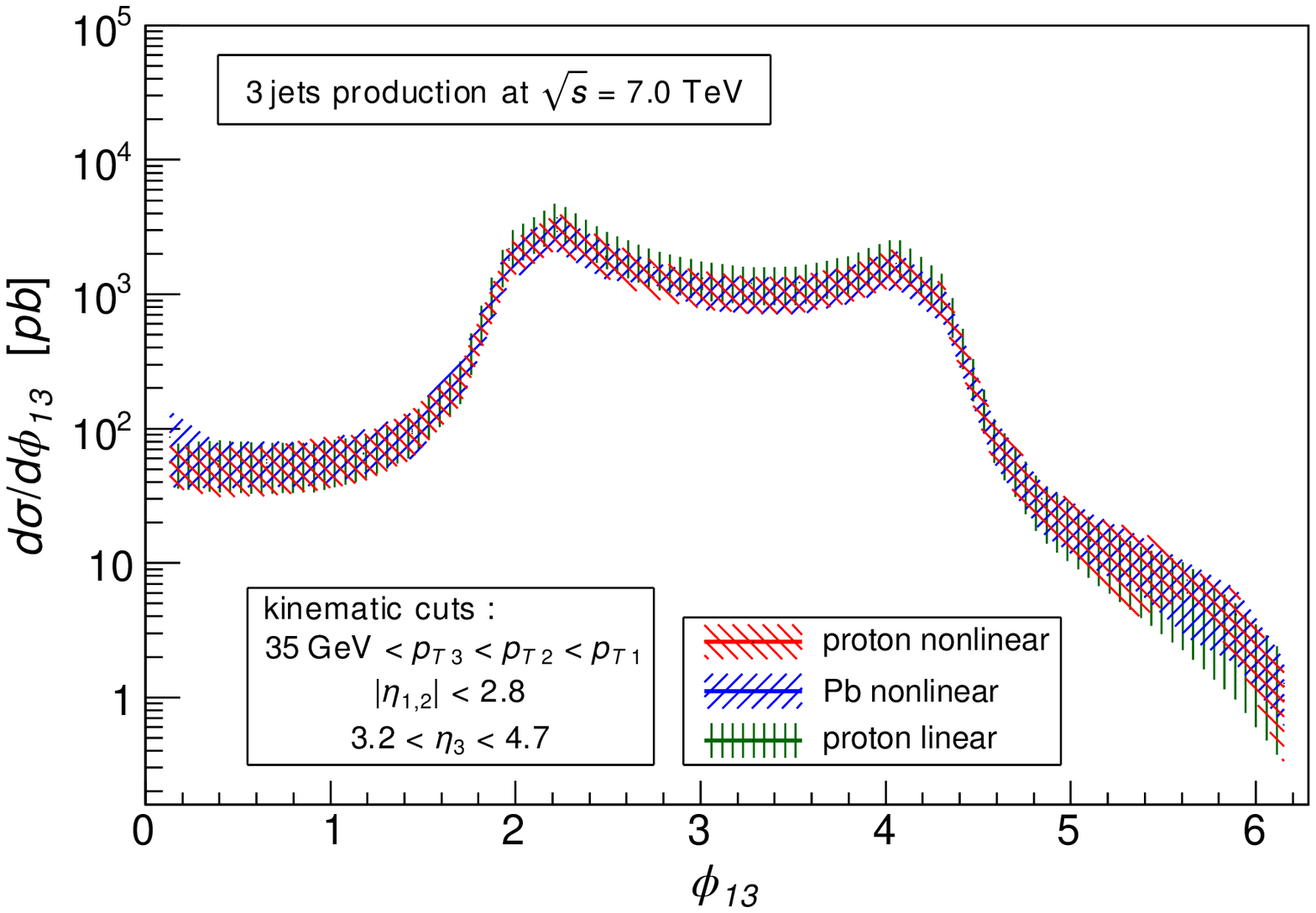}
\par\end{centering}

\caption{\label{fig:fw_azimcorr_bands} Differential cross section in difference
of the azimuthal angles between the leading and forward jets. The
band represents the theoretical uncertainty due to scale variation
and statistical errors. The left plot corresponds to c.m. energy $5.02\,\mathrm{TeV}$,
 the right to $7.0\,\mathrm{TeV}$.}
\end{figure}

\begin{figure}
\begin{centering}
\includegraphics[width=0.5\textwidth]{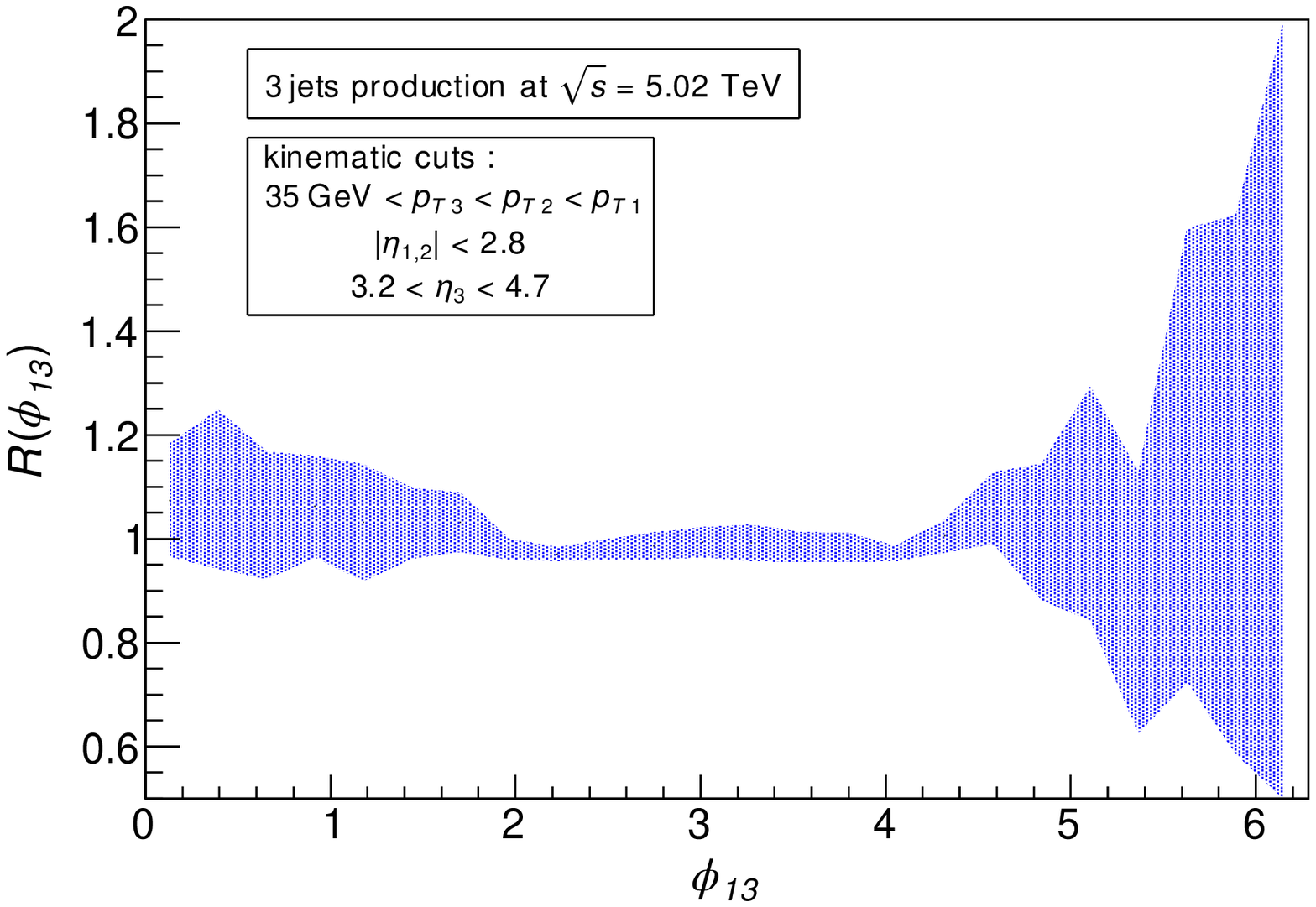}\includegraphics[width=0.5\textwidth]{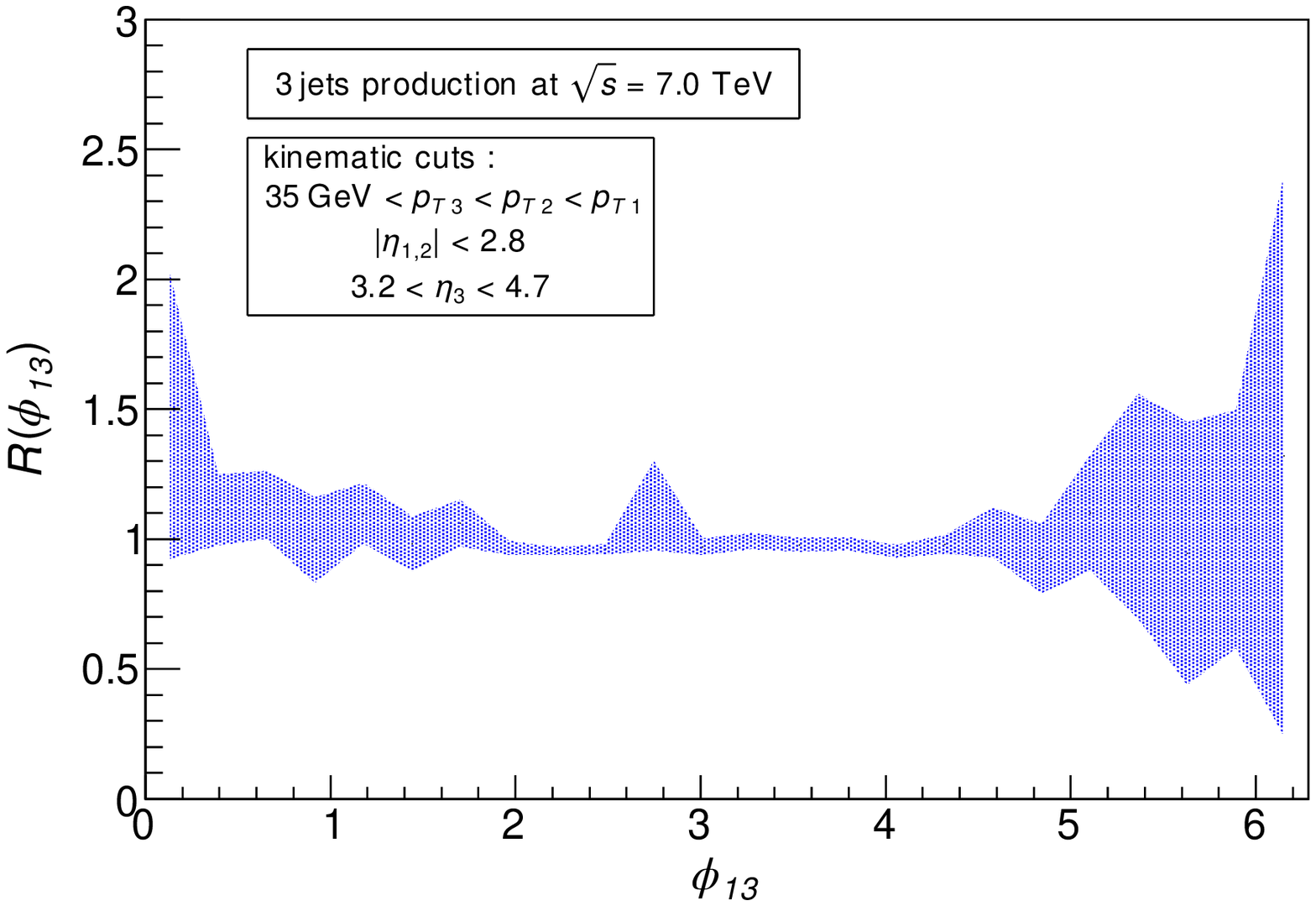}
\par\end{centering}

\caption{\label{fig:fw_azimcorr_ratios} The nuclear modification factor as
a function of difference of the azimuthal angles between the leading
and forward jets. The band represents the theoretical uncertainty
due to scale variation and statistical errors. The left plot corresponds
to c.m. energy $5.02\,\mathrm{TeV}$, the right to $7.0\,\mathrm{TeV}$.}
\end{figure}

\begin{figure}
\begin{centering}
\includegraphics[width=0.5\textwidth]{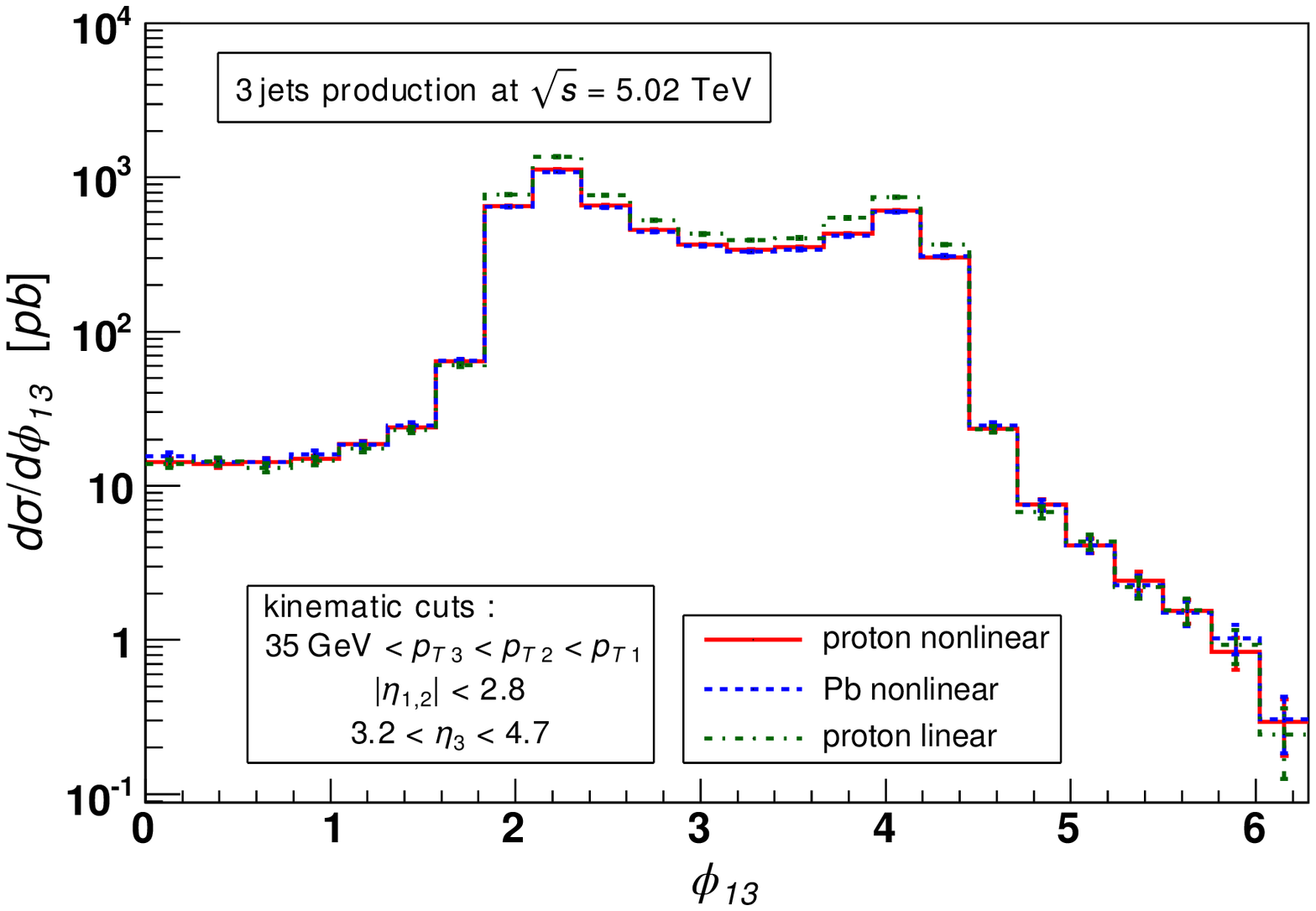}\includegraphics[width=0.5\textwidth]{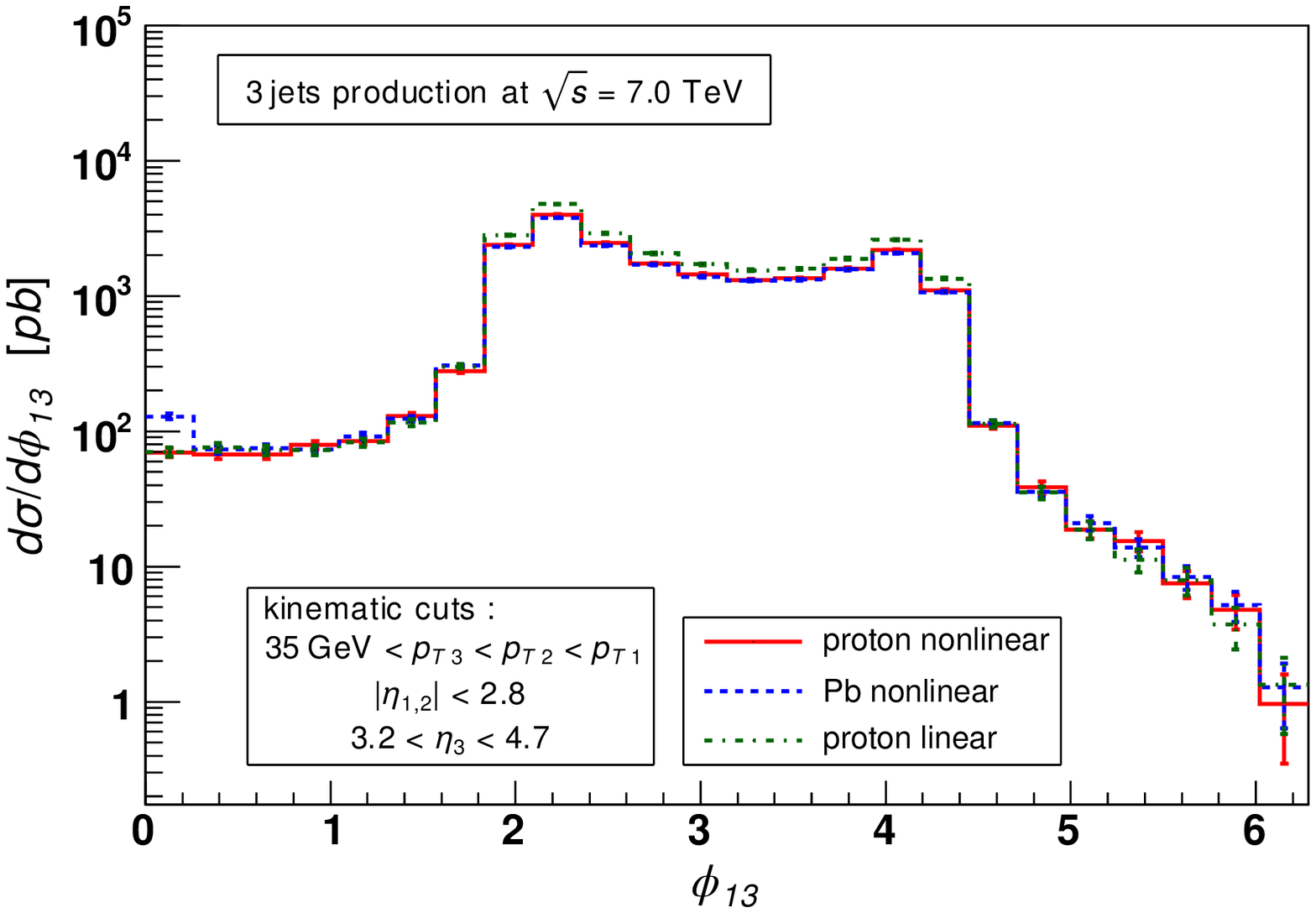}
\par\end{centering}

\caption{\label{fig:fw_azimcorr_exampl}Differential cross section in difference
of the azimuthal angles between the leading and forward jets for a
particular choice of the scale $\mu/2$. The left column corresponds to
c.m. energy $5.02\,\mathrm{TeV}$, the right to $7.0\,\mathrm{TeV}$.}
\end{figure}

In Figs. \ref{fig:fw_azimcorr_bands}-\ref{fig:fw_azimcorr_exampl}
we present a sample of our result for the differential cross section
in the variable $\phi_{13}$. We observe that indeed the whole region
$\left(0,2\pi\right]$ is covered by events, however the ``collinear''
region $\pi/2<\phi_{13}<3\pi/2$ dominates. The results for the nonlinear
evolution described in the Section \ref{sec:HighEnFact} for proton
and lead, as well as the BFKL with sub leading corrections are similar
and the nuclear modification ratio is consistent with unity, as seen
in Fig. \ref{fig:fw_azimcorr_ratios}.

\begin{figure}
\begin{centering}
\includegraphics[width=0.5\textwidth]{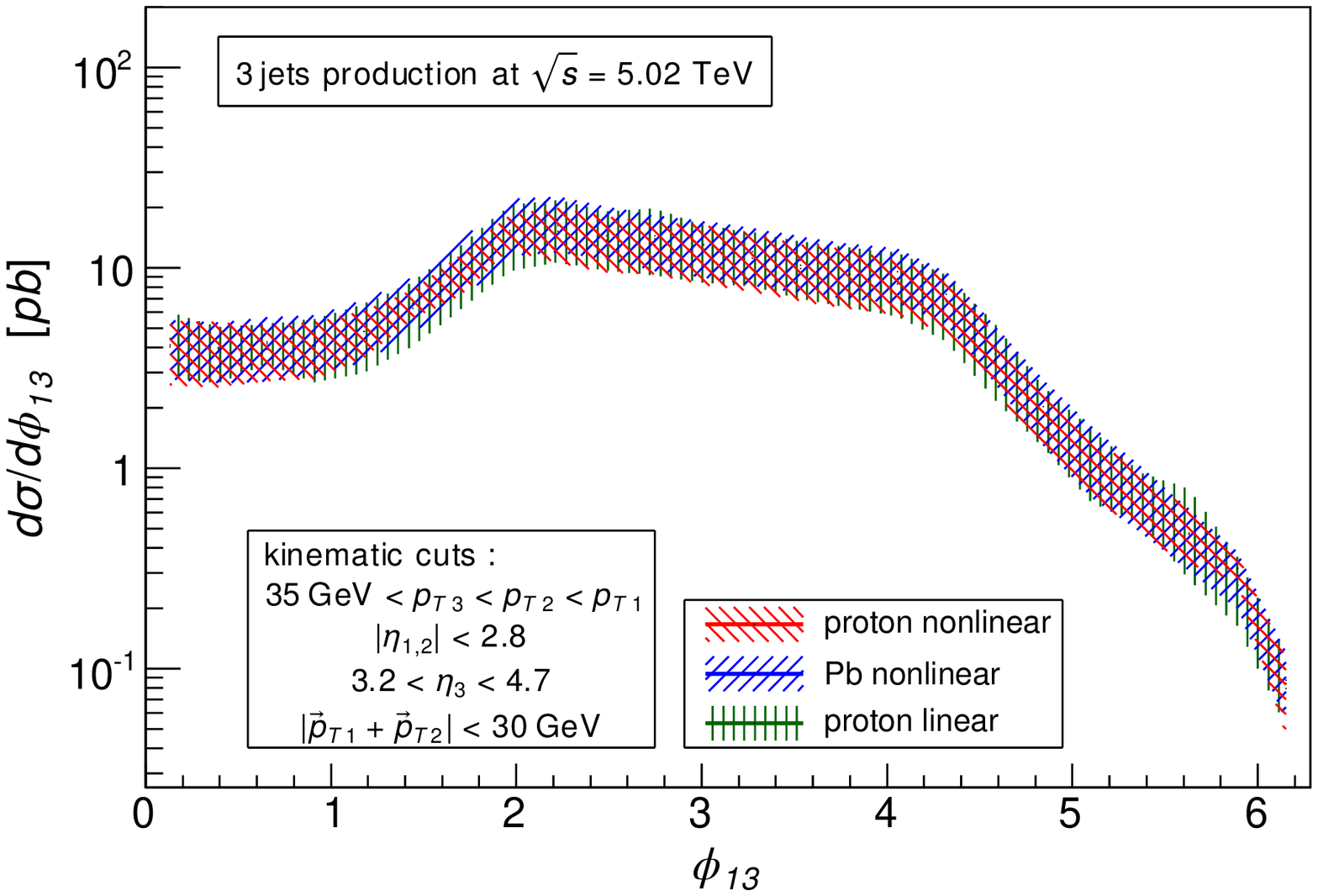}\includegraphics[width=0.5\textwidth]{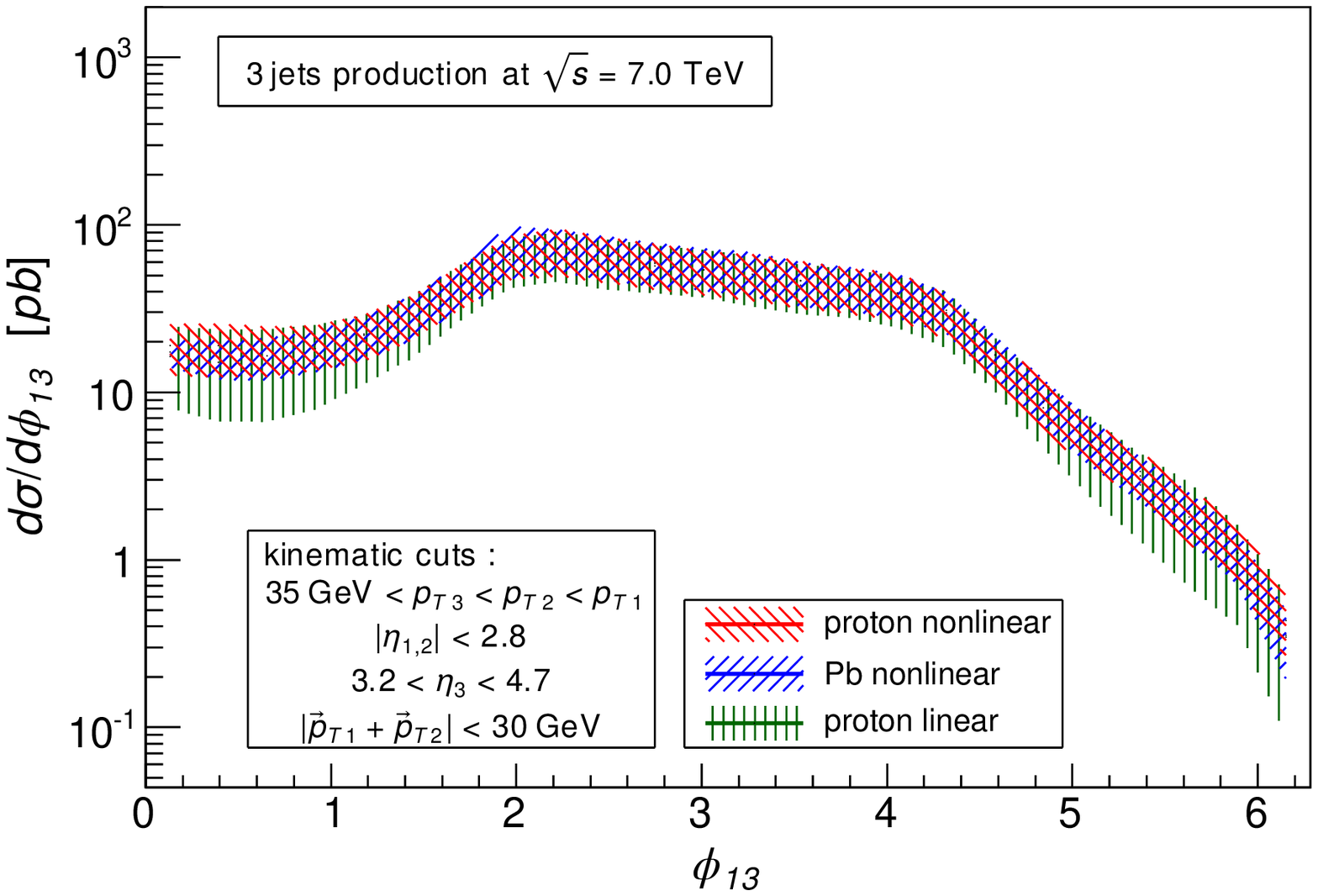}
\par\end{centering}

\caption{\label{fig:btb_azimcorr_bands} Differential cross section in difference
of the azimuthal angles between the leading and forward jets with
the additional restriction that the two leading jets are back-to-back-like.
The band represents the theoretical uncertainty due to scale variation
and statistical errors. The left plot corresponds to c.m. energy $5.02\,\mathrm{TeV}$,
the right to $7.0\,\mathrm{TeV}$.}
\end{figure}

\begin{figure}
\begin{centering}
\includegraphics[width=0.5\textwidth]{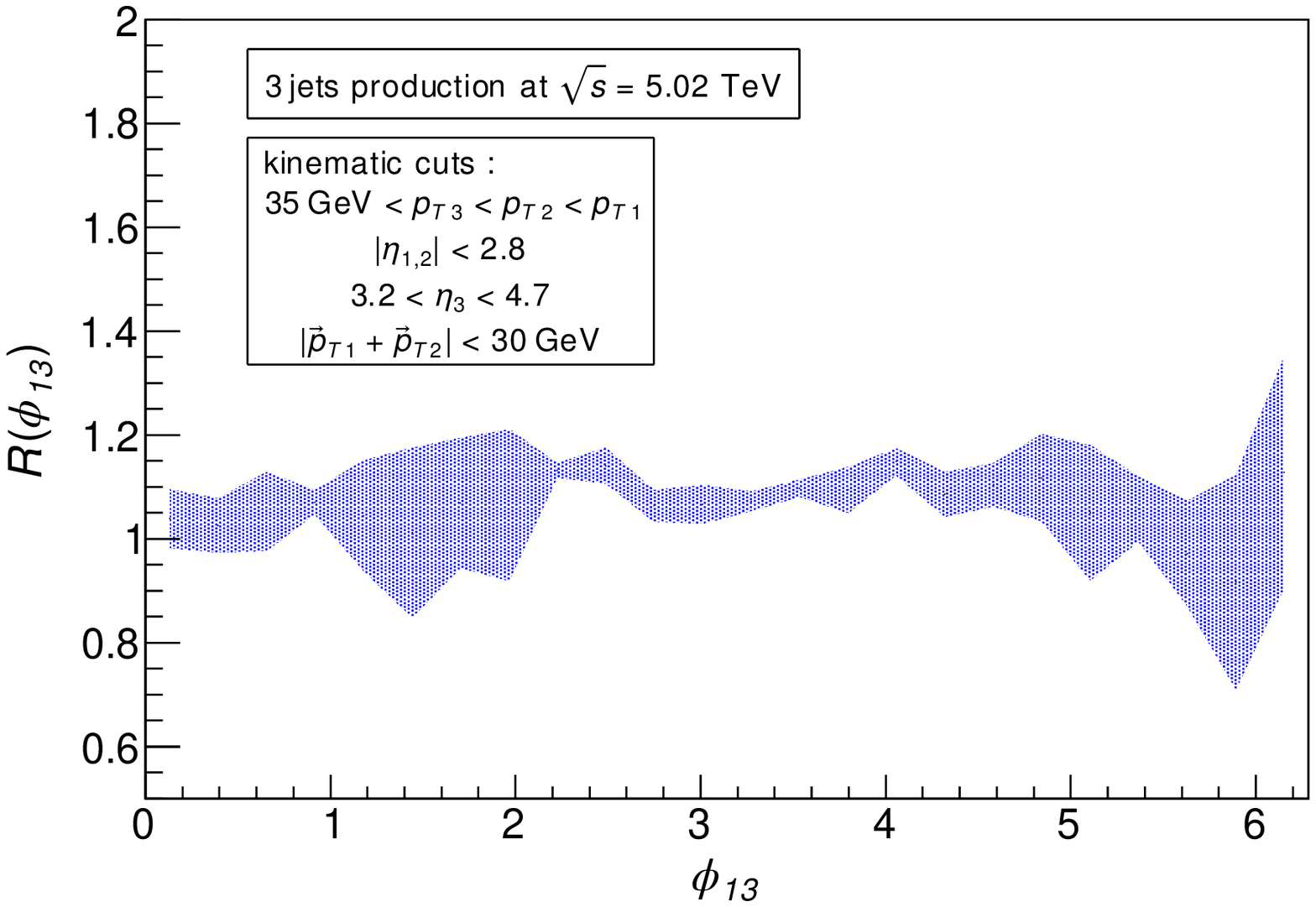}\includegraphics[width=0.5\textwidth]{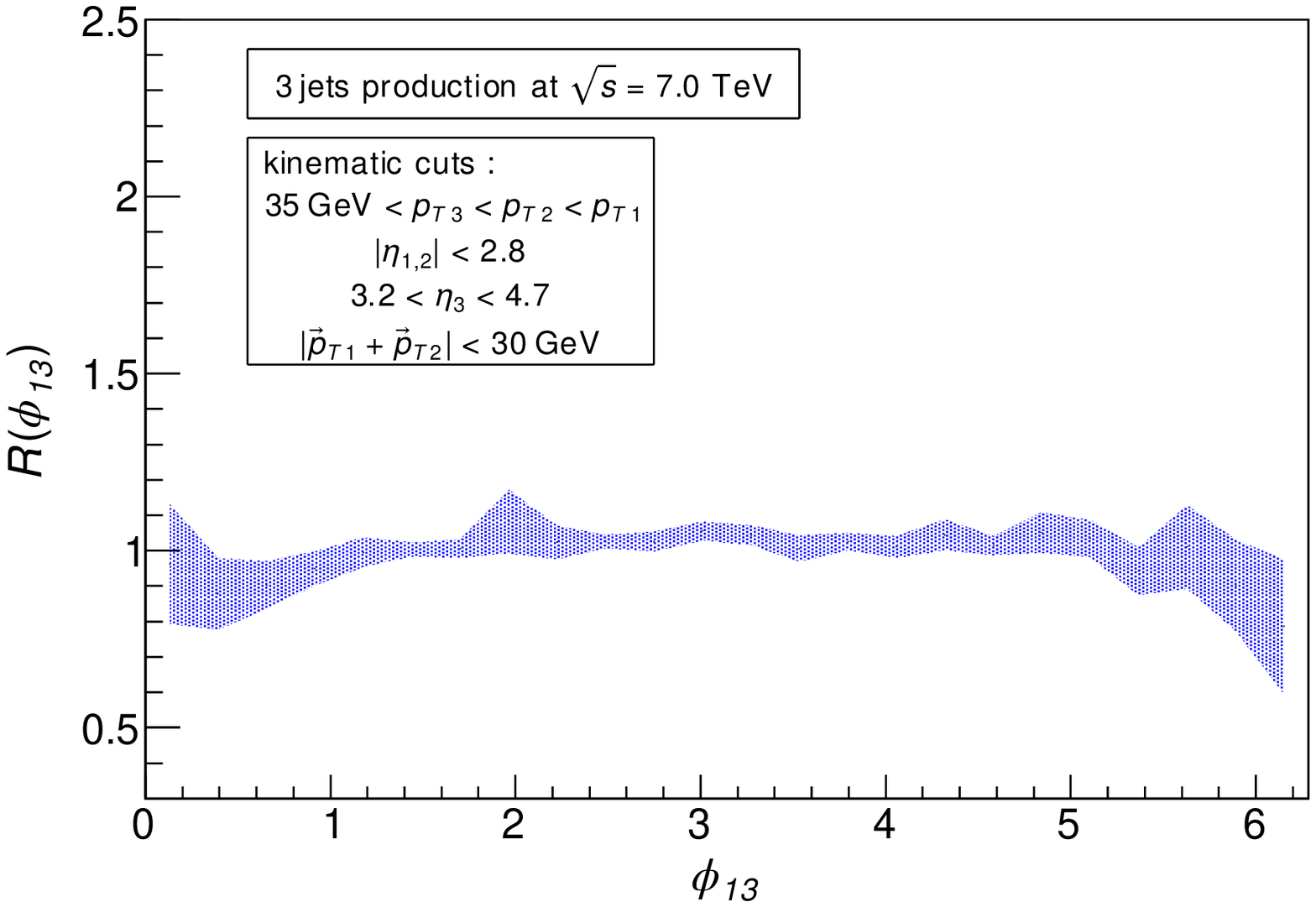}
\par\end{centering}

\caption{\label{fig:btb_azimcorr_ratios} The nuclear modification factor as
a function of difference of the azimuthal angles between the leading
and forward jets, with the additional restriction that the two leading
jets are back-to-back-like. The band represents the theoretical uncertainty
due to scale variation and statistical errors. The left plot corresponds
to c.m. energy $5.02\,\mathrm{TeV}$, the right to $7.0\,\mathrm{TeV}$.}
\end{figure}

\begin{figure}
\begin{centering}
\includegraphics[width=0.5\textwidth]{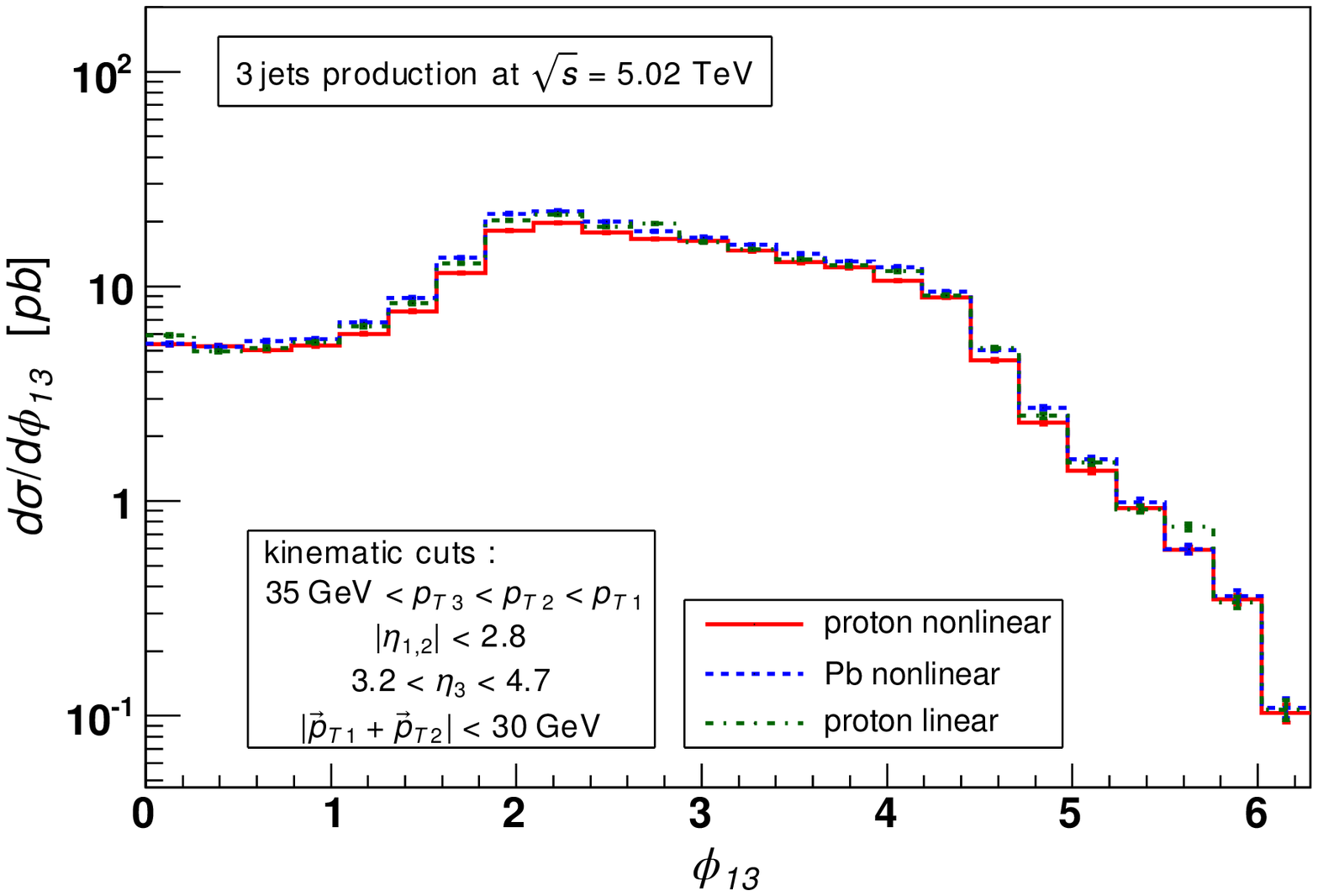}\includegraphics[width=0.5\textwidth]{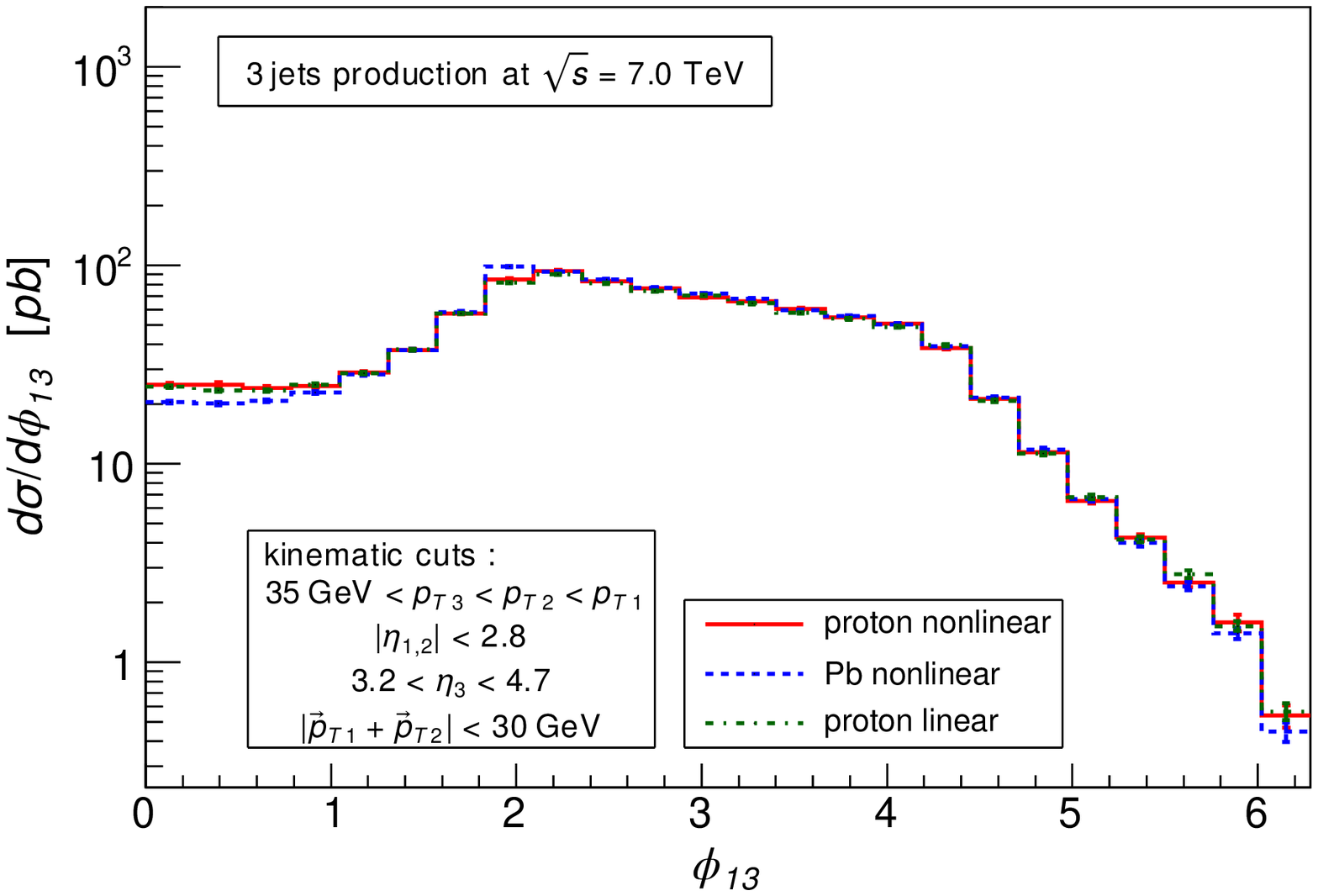}
\par\end{centering}

\begin{centering}
\includegraphics[width=0.5\textwidth]{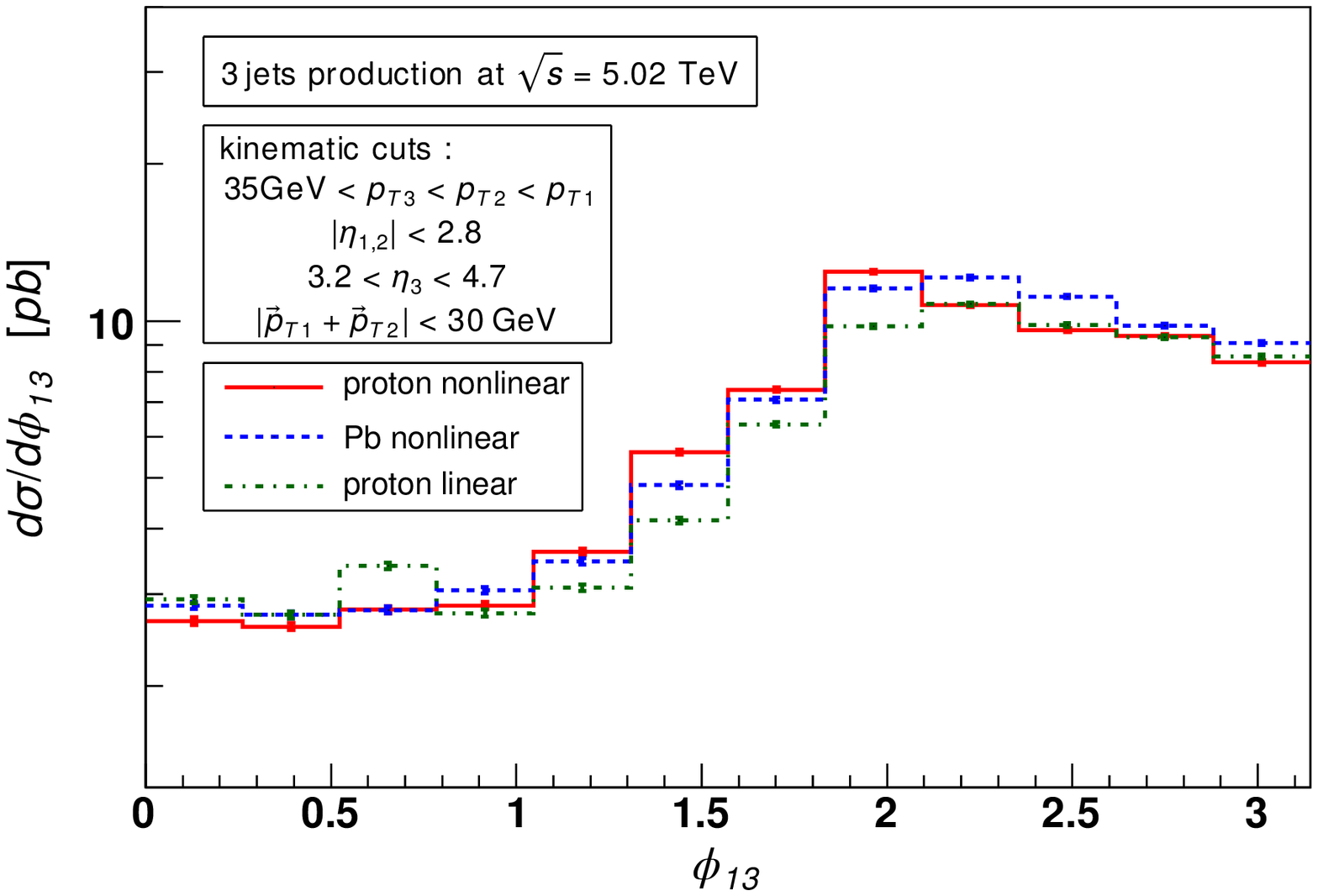}\includegraphics[width=0.5\textwidth]{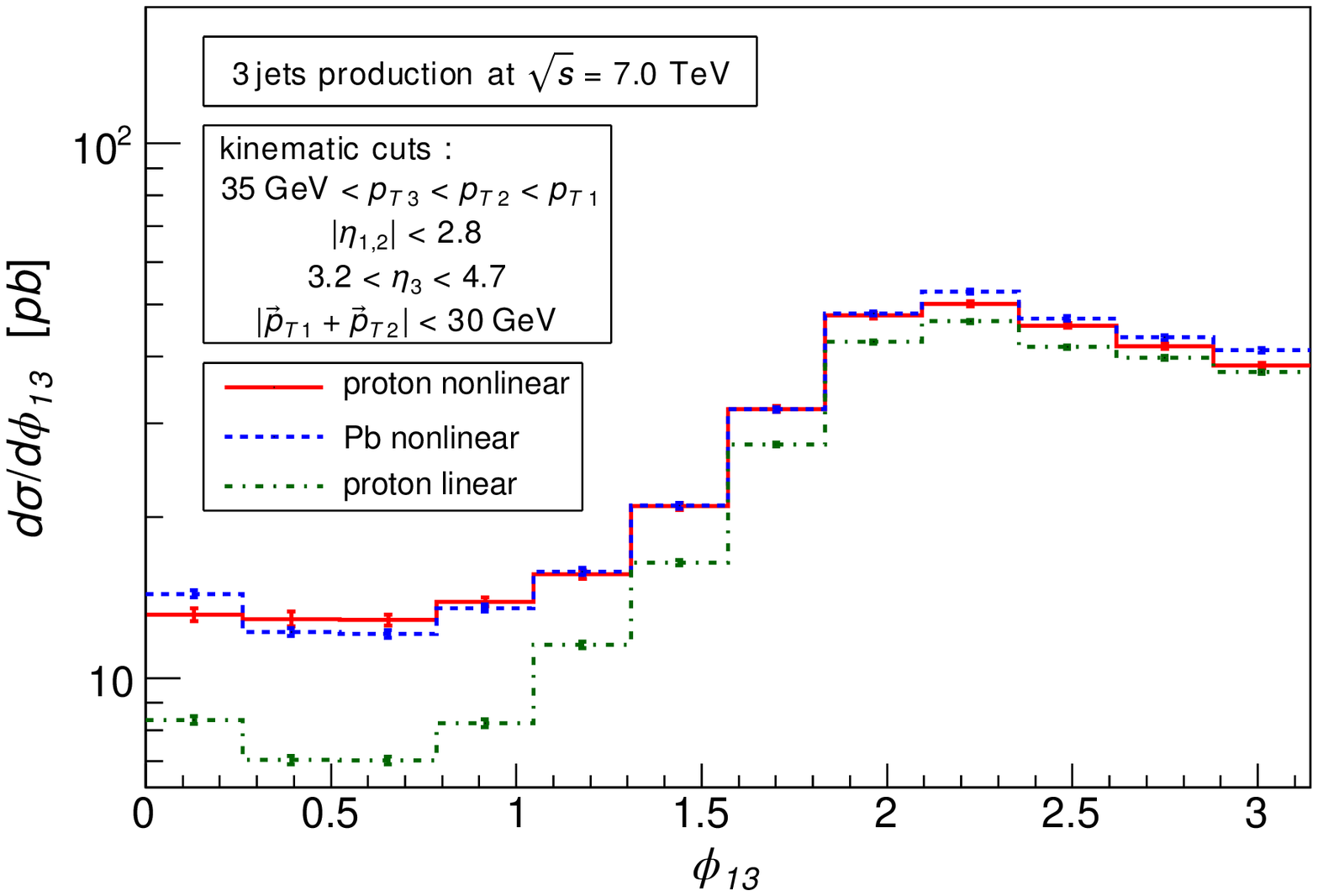}
\par\end{centering}

\caption{\label{fig:btb_azimcorr_exampl}Differential cross section in difference
of the azimuthal angles between the leading and forward jets, with
the additional restriction that the two leading jets are back-to-back-like.
The left column corresponds to c.m. energy $5.02\,\mathrm{TeV}$, the right
to $7.0\,\mathrm{TeV}$. The top plots are made for the scale $\mu/2$
while the bottom plots zoom the top plots for $0<\phi_{13}<\pi$ and
are made for the scale $2\mu$.}
\end{figure}

Let us now turn to the case when the two leading jets are restricted
to be back-to-back-like. In the present calculation we choose $D_{\mathrm{cut}}=30\,\mathrm{GeV}$.
We have checked empirically that in order to observe any significant
difference comparing to forward-central case discussed above we should
use $D_{\mathrm{cut}}<p_{T\,\mathrm{cut}}$. Our results are presented
in Figs. \ref{fig:btb_azimcorr_bands}-\ref{fig:btb_azimcorr_exampl}.
The main properties of the distributions are the following. First,
the relative magnitude between the ``collinear'' region and the
``non-collinear'' region is decreased comparing to the previous
case. This indicates that we enter the region which is small $x$ sensitive.
We see (right of Figs. \ref{fig:btb_azimcorr_bands}, \ref{fig:btb_azimcorr_exampl})
that for the c.m. energy of $7$ TeV and large energy scale the distributions
are different for linear and nonlinear evolution and the difference
is most significant in the ``noncollinear'' region. The nuclear
modification factor (Fig. \ref{fig:btb_azimcorr_ratios}) is, however, still
consistent with unity.

\subsubsection{Forward jets}

\begin{figure}
\begin{centering}
\includegraphics[width=0.5\textwidth]{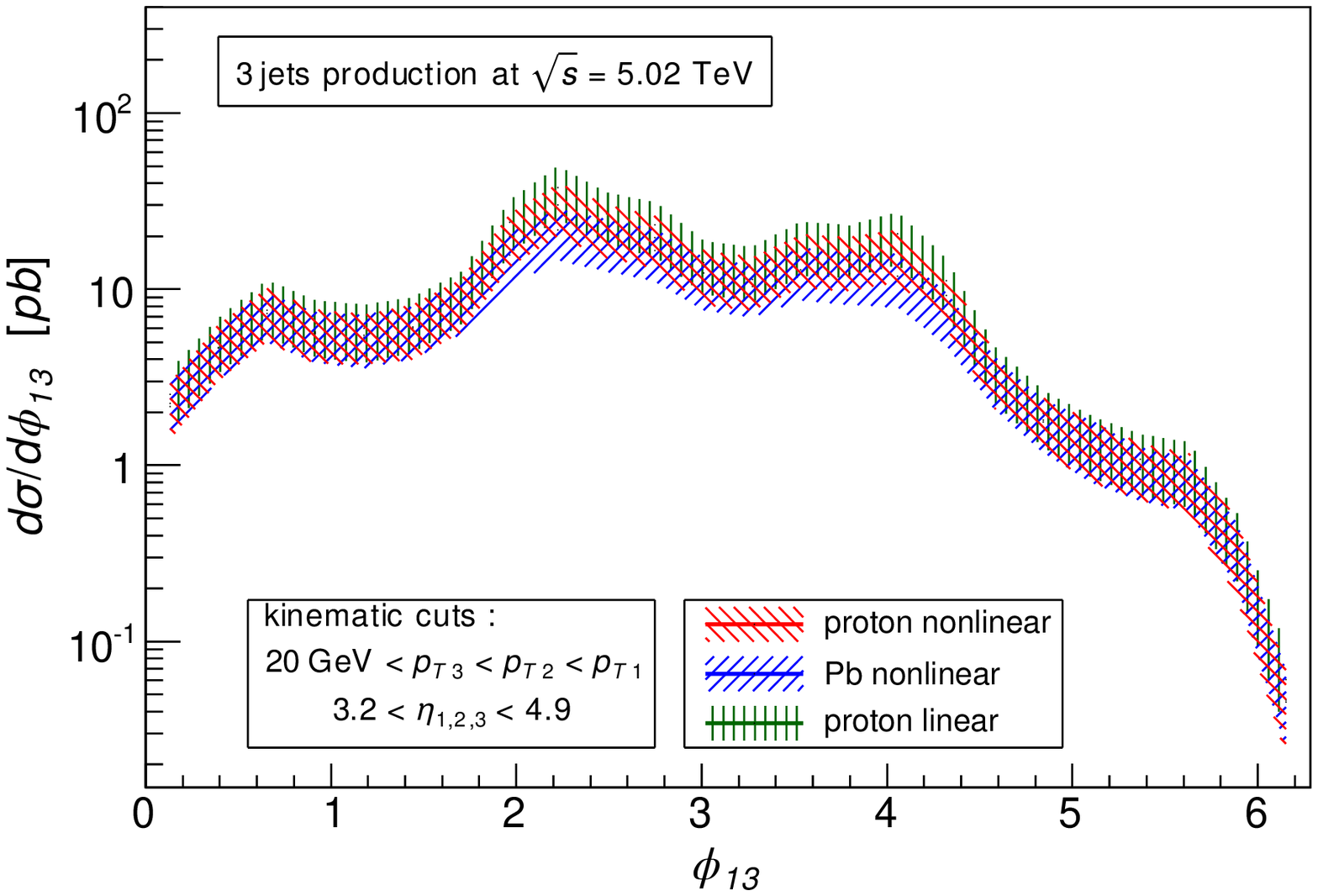}\includegraphics[width=0.5\textwidth]{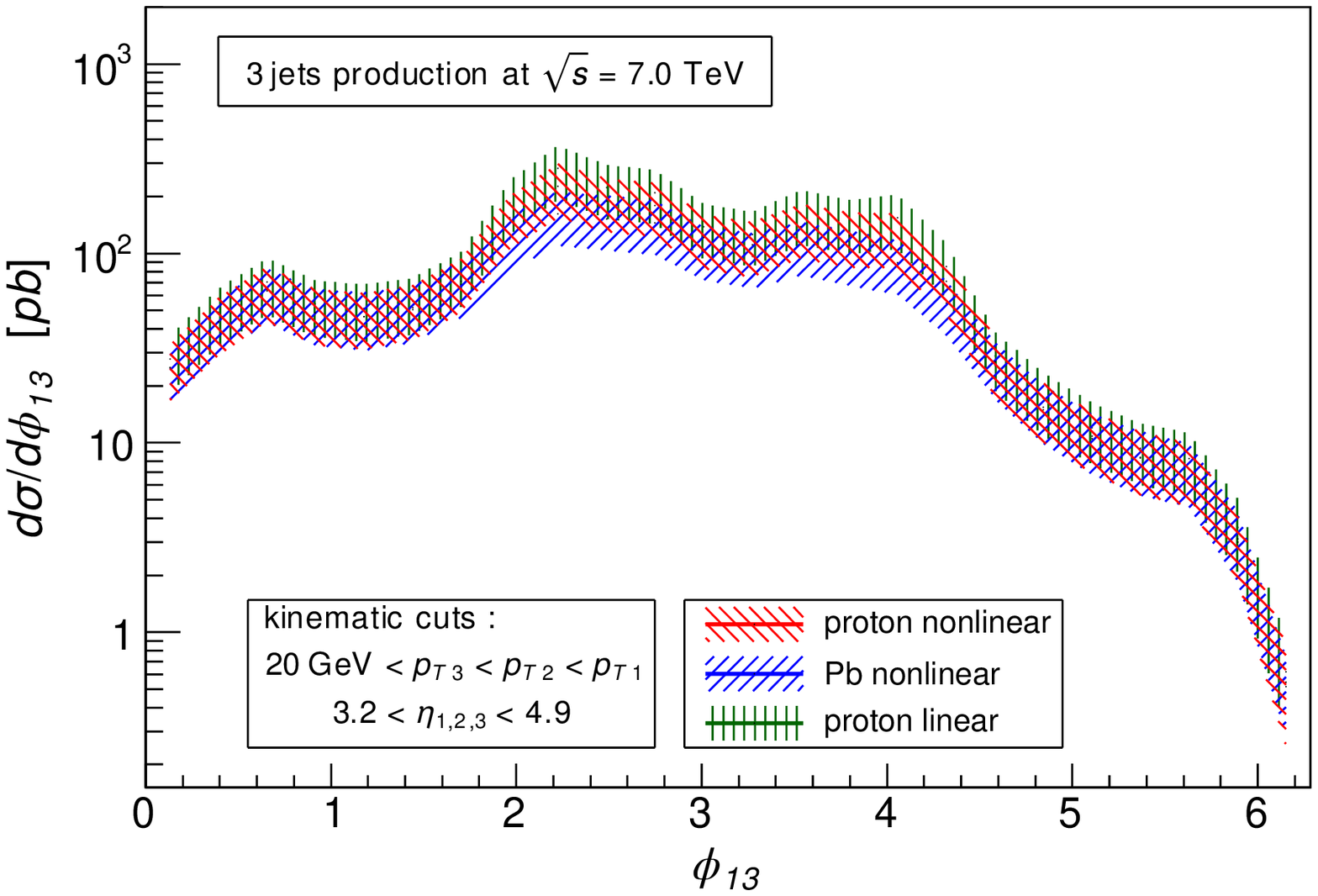}
\par\end{centering}

\caption{\label{fig:fww_azimcorr_bands} Differential cross section in difference
of the azimuthal angles between the leading and forward jets for the
forward rapidity region. The band represents the theoretical uncertainty
due to scale variation and statistical errors. The left plot corresponds
to c.m. energy $5.02\,\mathrm{TeV}$, the right to $7.0\,\mathrm{TeV}$.}
\end{figure}

\begin{figure}
\begin{centering}
\includegraphics[width=0.5\textwidth]{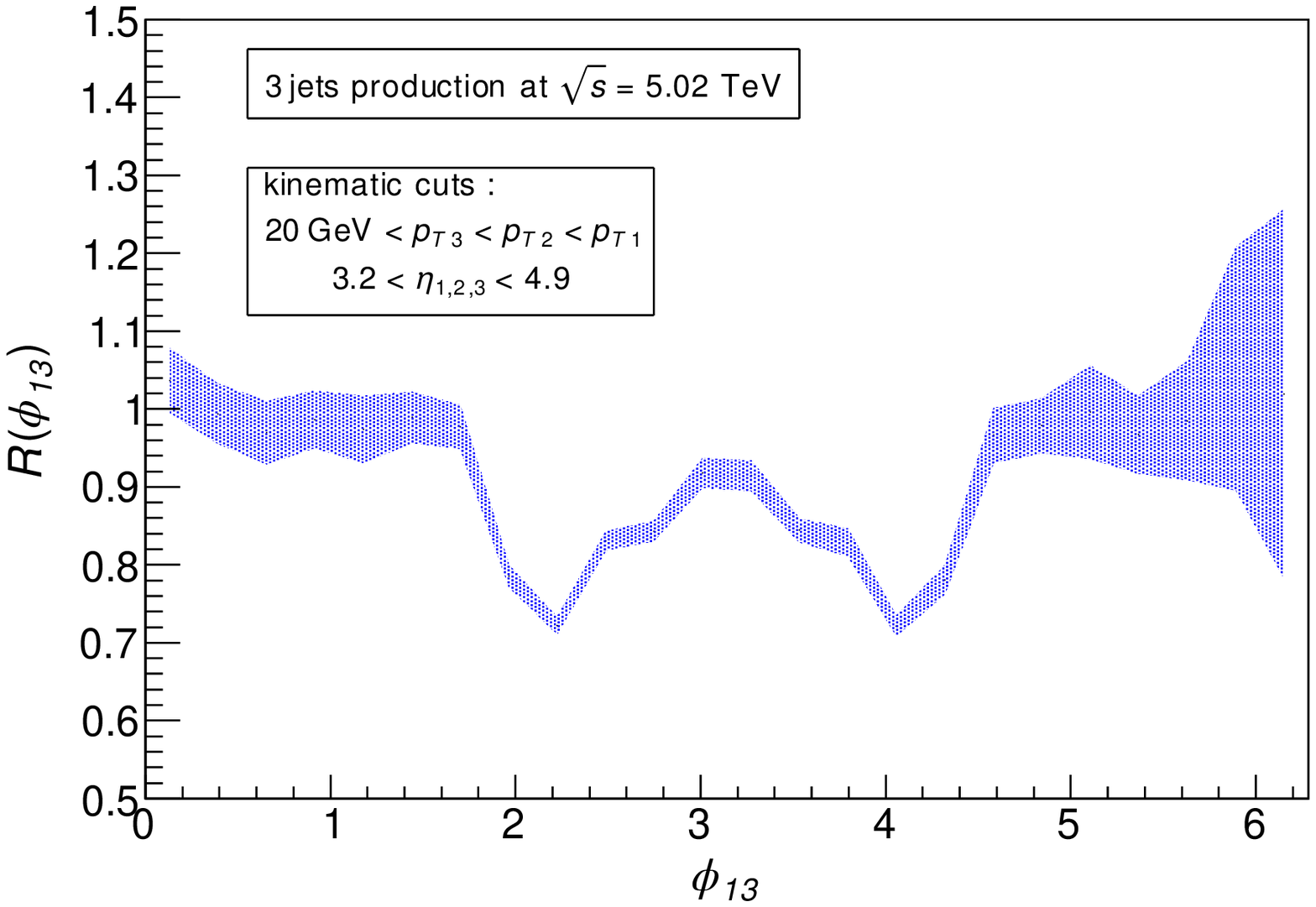}\includegraphics[width=0.5\textwidth]{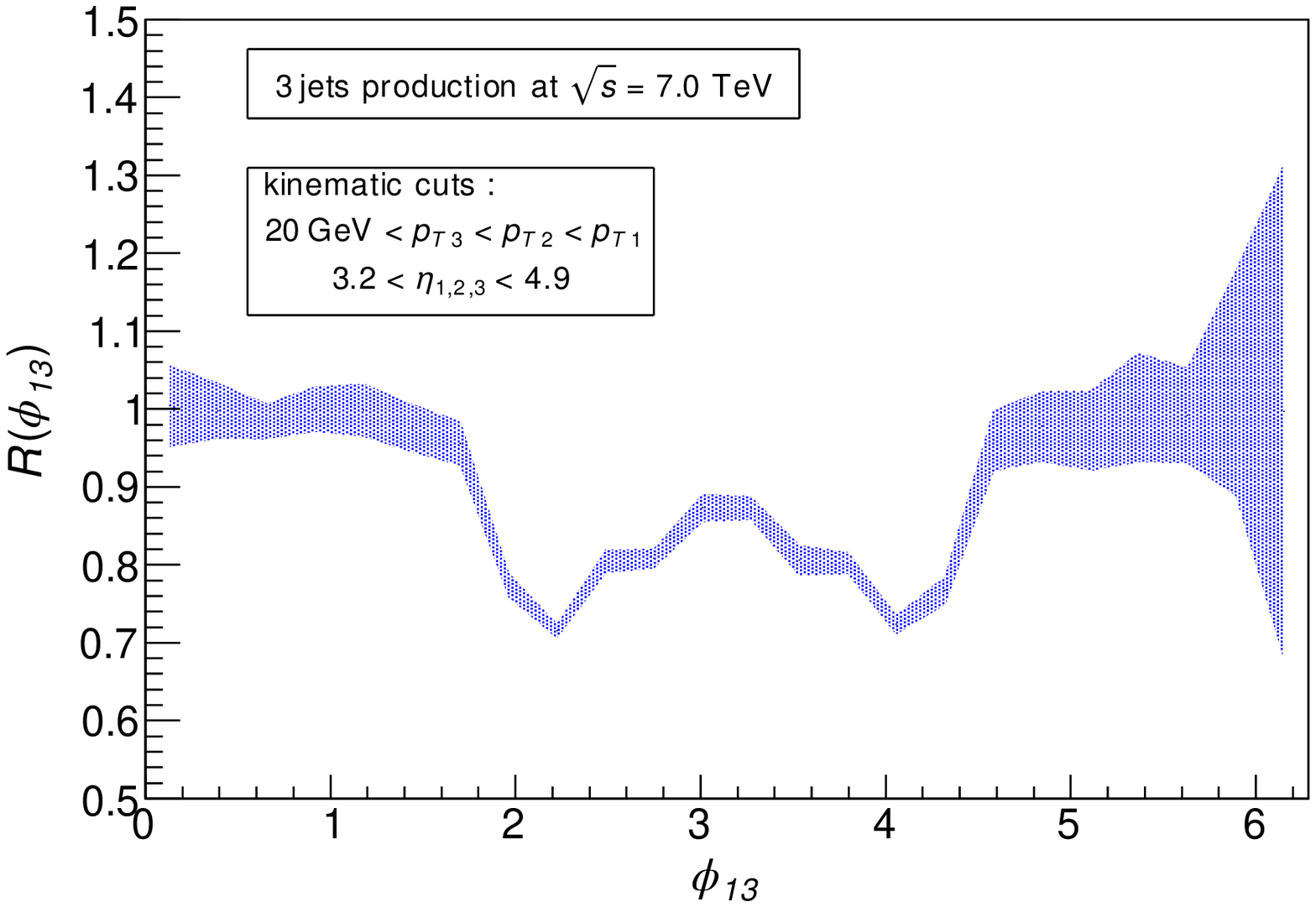}
\par\end{centering}

\caption{\label{fig:ffw_azimcorr_ratios} The nuclear modification factor as
a function of difference of the azimuthal angles between the leading
and forward jets for the forward rapidity region. The band represents
the theoretical uncertainty due to scale variation and statistical
errors. The left plot corresponds to c.m. energy $5.02\,\mathrm{TeV}$,
the right to $7.0\,\mathrm{TeV}$.}
\end{figure}

\begin{figure}
\begin{centering}
\includegraphics[width=0.5\textwidth]{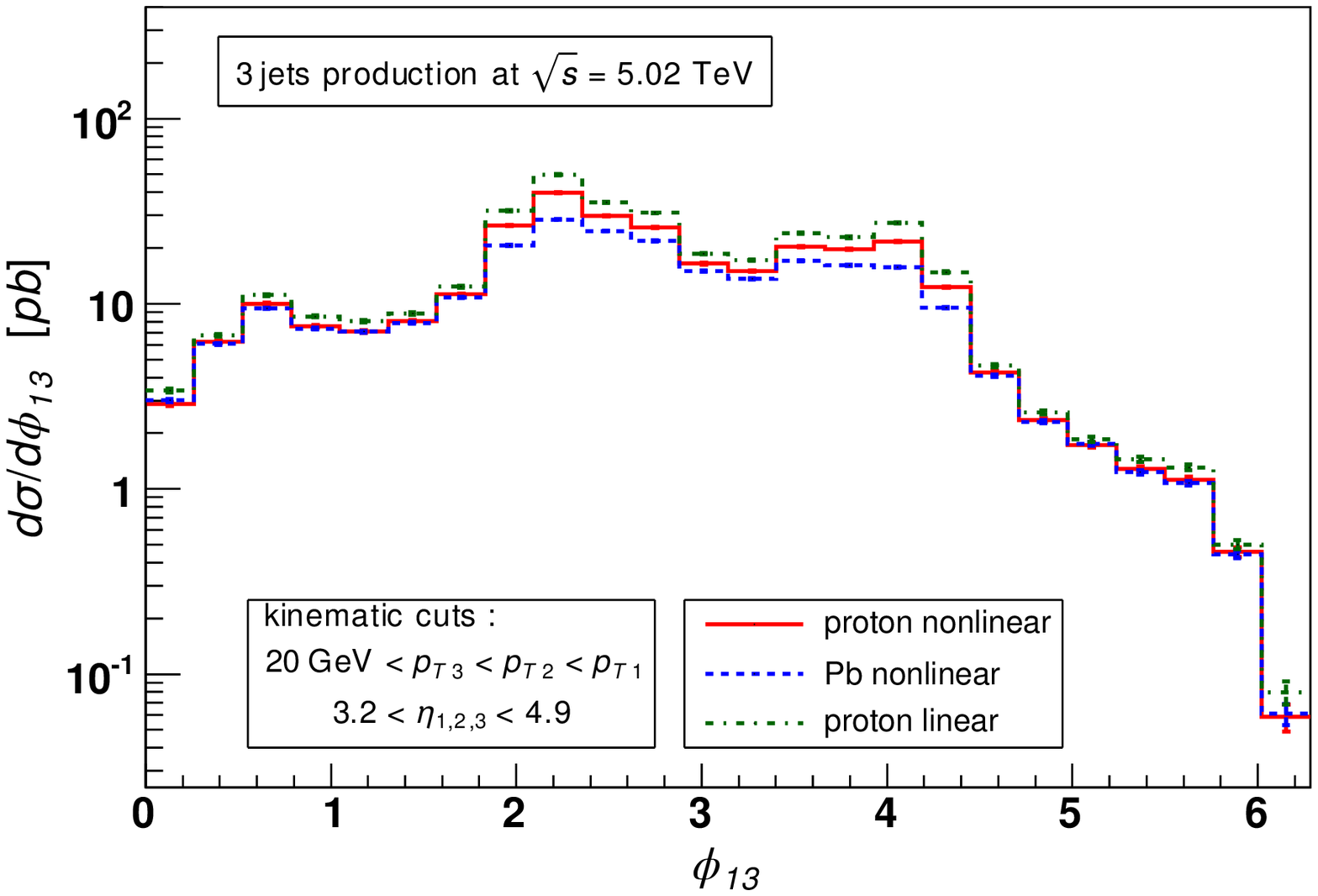}\includegraphics[width=0.5\textwidth]{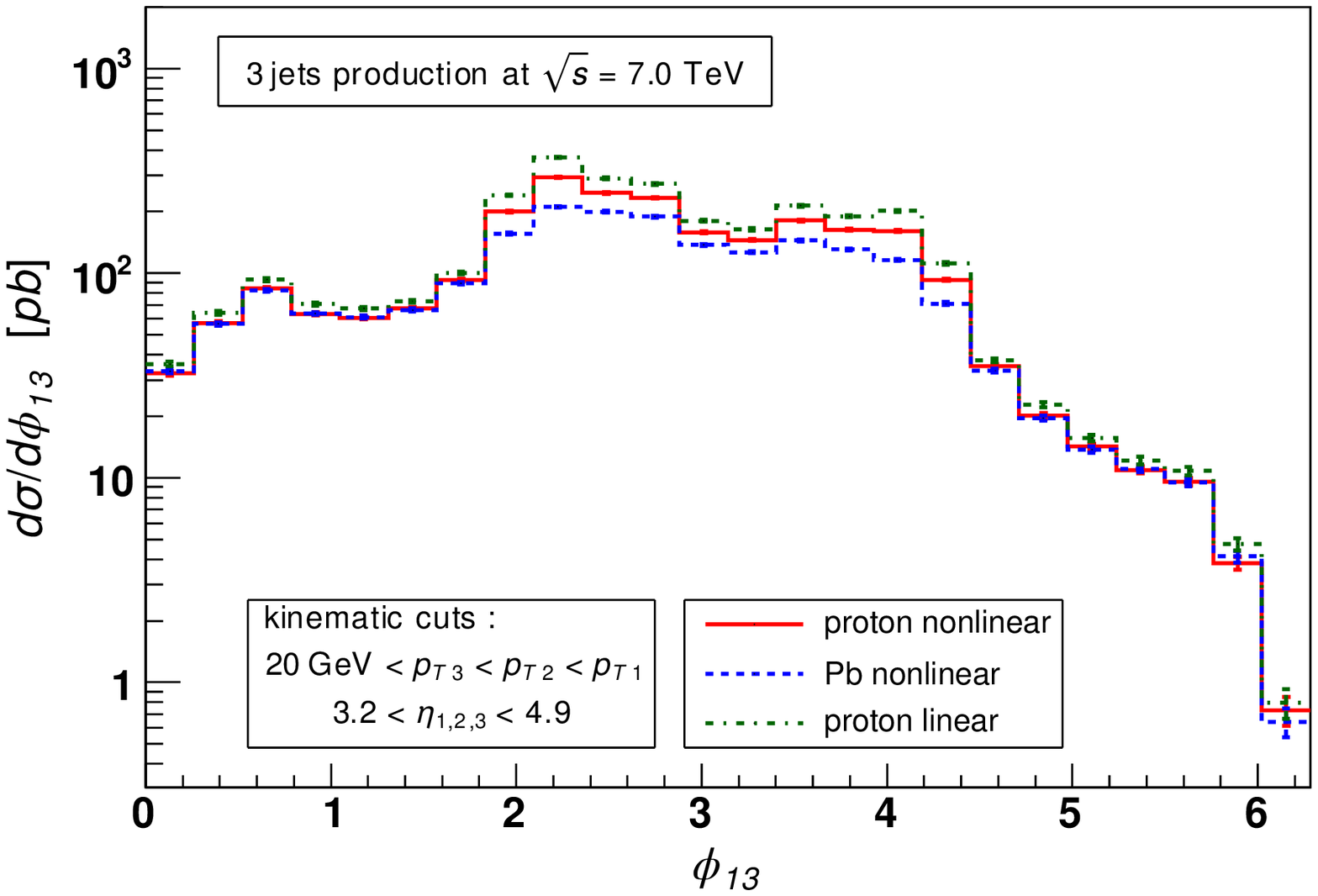}
\par\end{centering}

\begin{centering}
\includegraphics[width=0.5\textwidth]{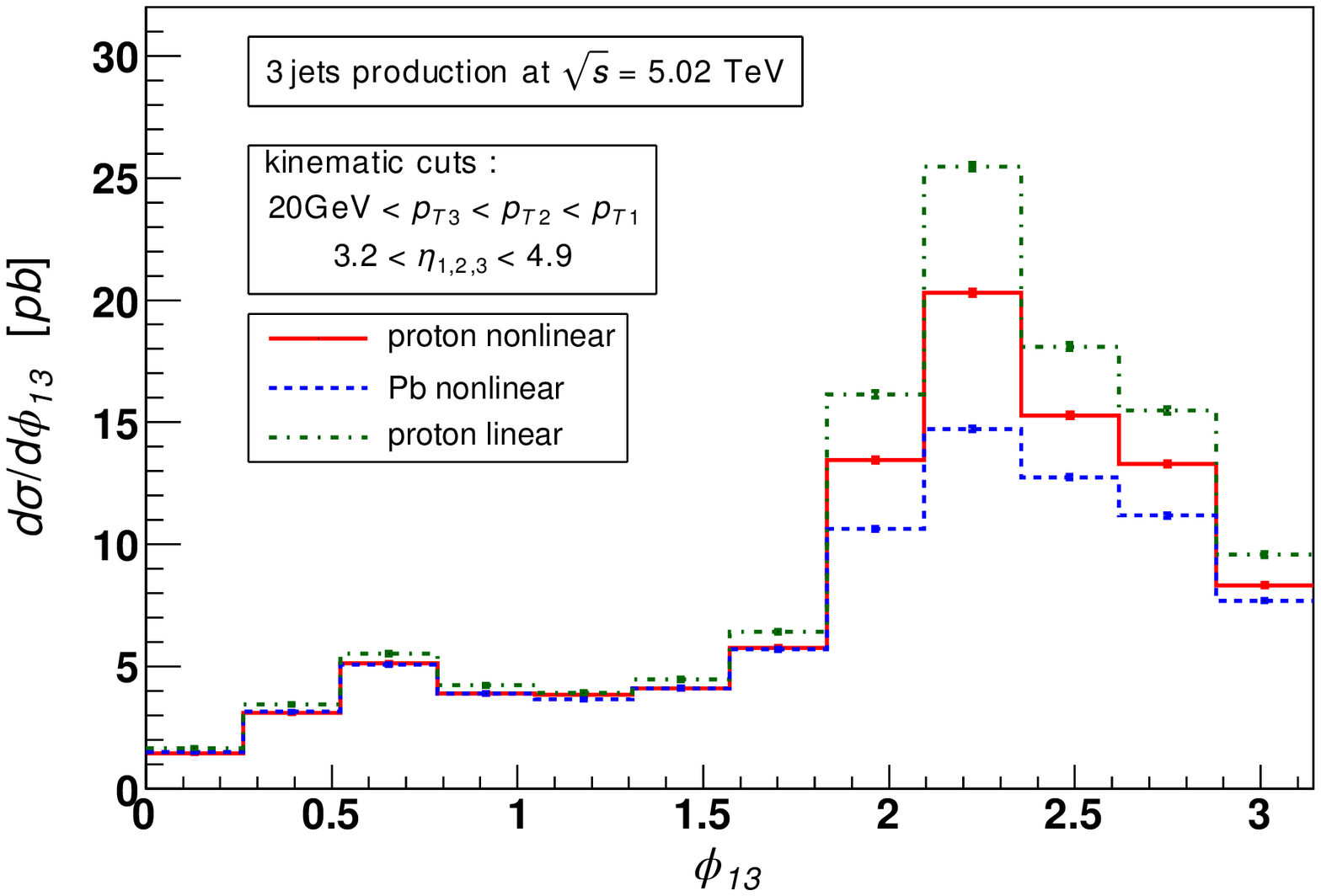}\includegraphics[width=0.5\textwidth]{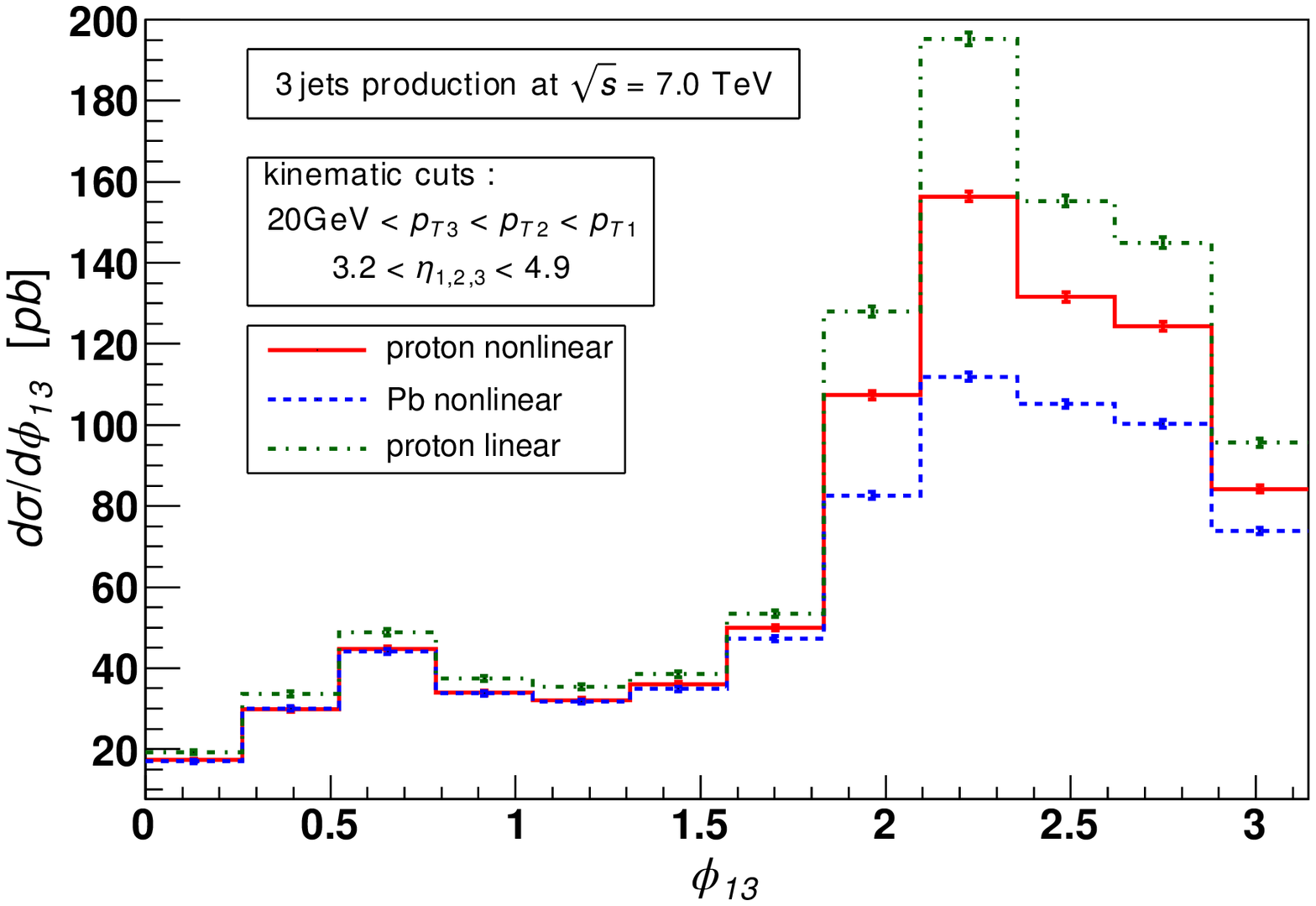}
\par\end{centering}

\caption{\label{fig:ffw_azimcorr_exampl}Differential cross section in difference
of the azimuthal angles between the leading and forward jets for the
forward rapidity region. The left column corresponds to c.m. energy $5.02\,\mathrm{TeV}$,
the right to $7.0\,\mathrm{TeV}$. The top plots are made for the scale
$\mu/2$ while the bottom for the scale $2\mu$ and $0<\phi_{13}<\pi$.}
\end{figure}

Let us move to the case where all of the jets are in the forward rapidity
region. As discussed in Subsection \ref{sub:assymetry}, in this region
we can safely go to relatively low values of $p_{T\,\mathrm{cut}}$;
thus we set $p_{T\,\mathrm{cut}}=20\,\mathrm{GeV}$. We present the
results in Figs. \ref{fig:fww_azimcorr_bands}-\ref{fig:ffw_azimcorr_exampl}.
We observe significant differences between the three scenarios (nonlinear
proton, nonlinear Pb and linear proton) in the middle region of $\phi_{13}$
distributions, i.e. in the ``collinear'' region. It is also nicely
illustrated by the nuclear modification ratios (Fig. \ref{fig:ffw_azimcorr_ratios})
which now have two dips in that region, indicating that the region
is sensitive to the nonlinear effects. The qualitative behavior is
the same for both considered c.m. energies.

\subsection{Unbalanced jet transverse momentum}

\label{sub:Results_unbalpT}

Let us now switch to an analysis of the cross section as a function of the
following quantity
\begin{equation}
\Delta p_{T}=\left|\vec{p}_{T\,1}+\vec{p}_{T\,2}+\vec{p}_{T\,3}\right|,\label{eq:DpT}
\end{equation}
which in the prescription given by the Eq. (\ref{eq:HENfact_2}) corresponds
to the transverse momentum of the off-shell gluon, i.e., $\Delta p_{T}=\left|\vec{k}_{T\, A}\right|$.

Let us remark, that the distributions we are going to present can
be much more affected by the final state parton shower, than the decorrelation
distribution presented above. Nevertheless, it is very interesting
to study the influence of different evolution equations for the multiparticle
production purely within the high-energy factorization.

\subsubsection{Central-forward jets}

\begin{figure}
\begin{centering}
\includegraphics[width=0.5\textwidth]{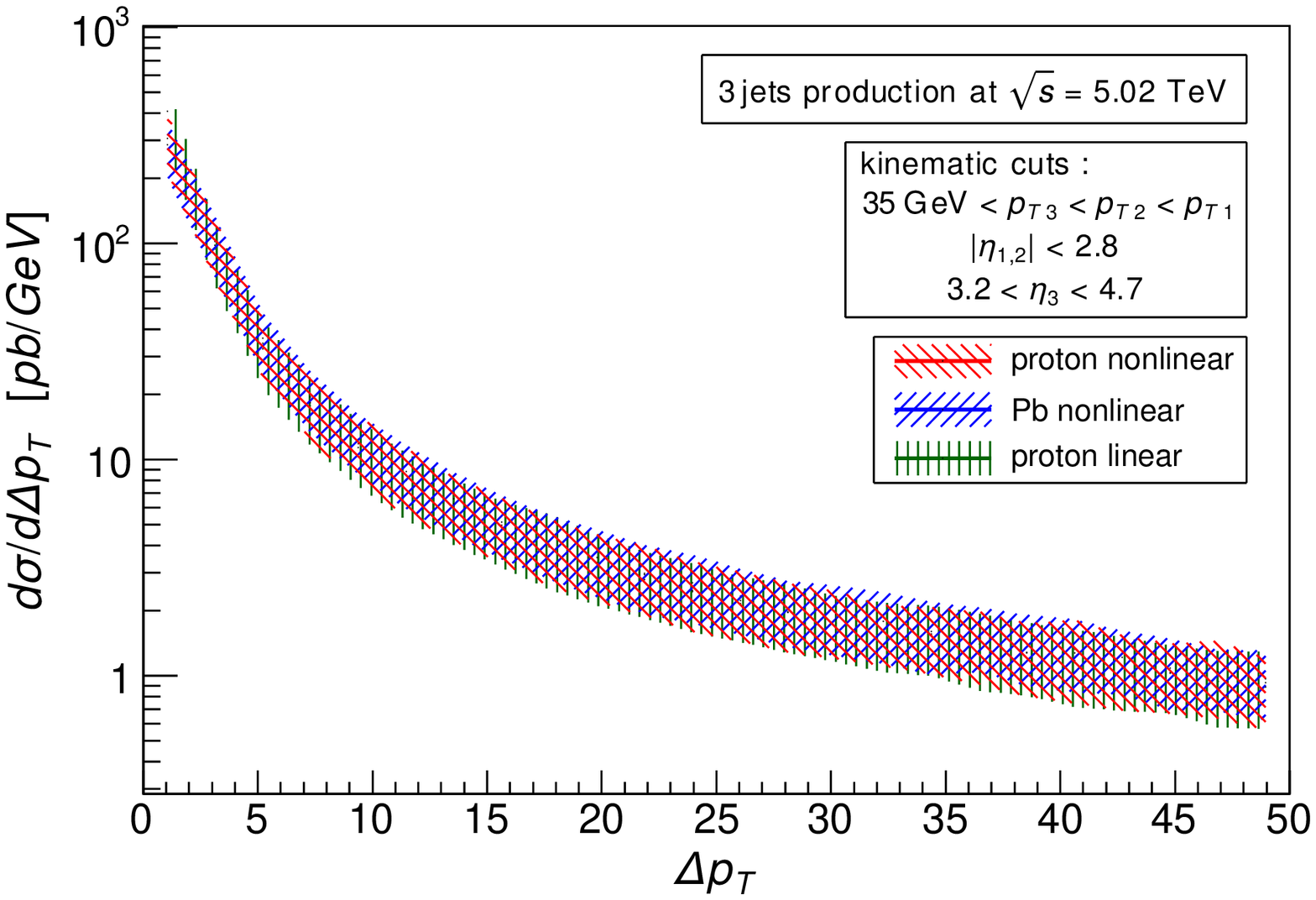}\includegraphics[width=0.5\textwidth]{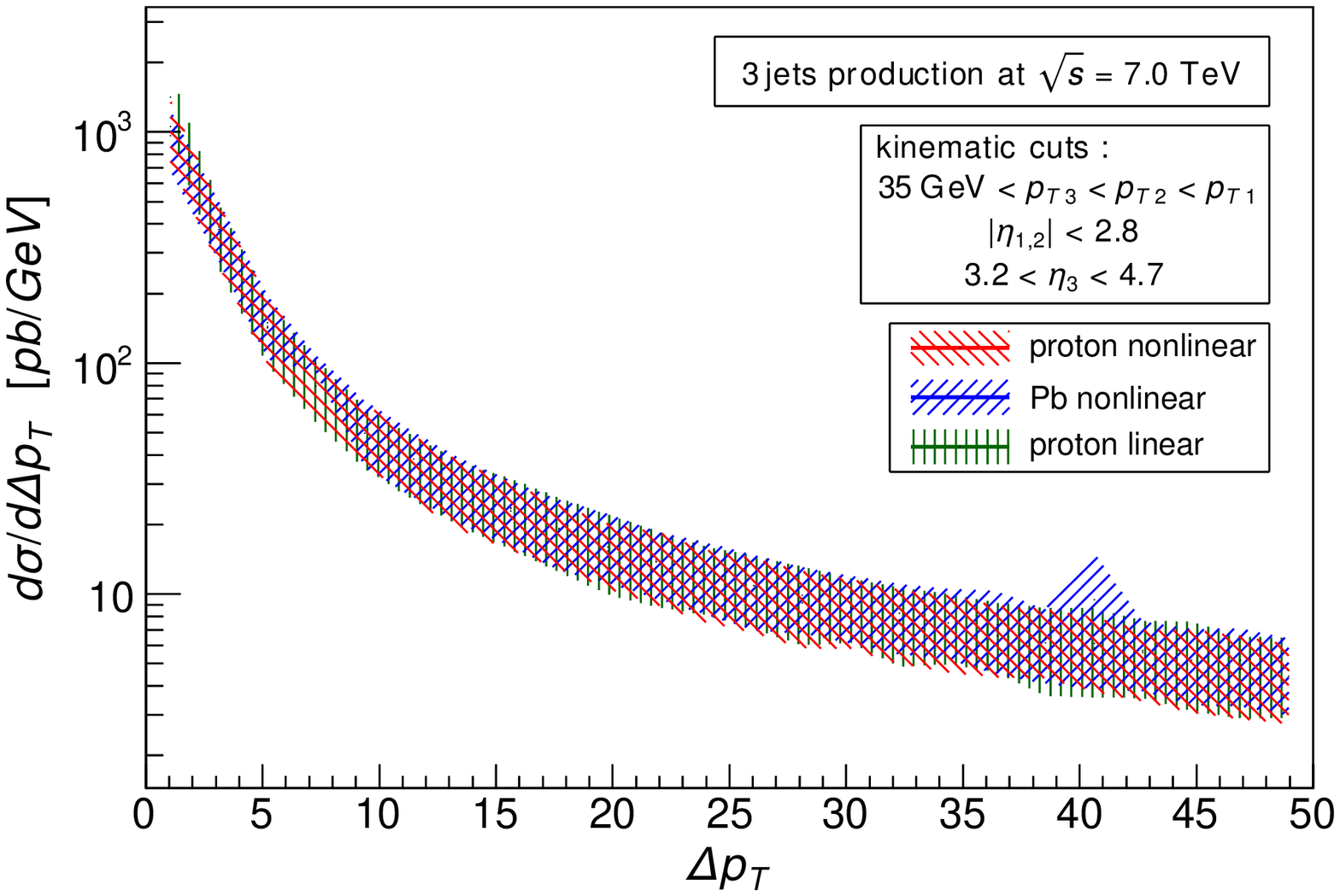}
\par\end{centering}

\begin{centering}
\includegraphics[width=0.5\textwidth]{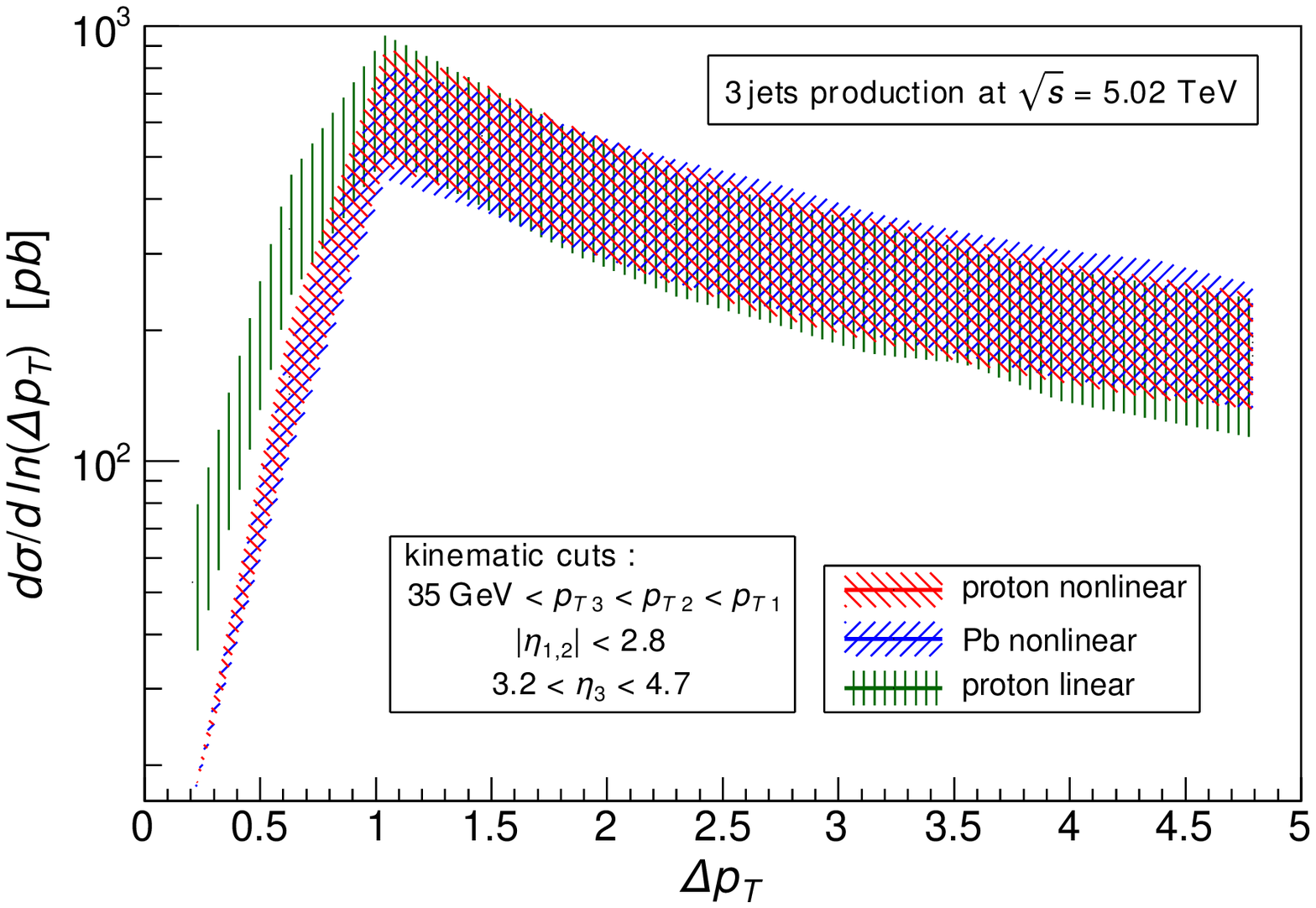}\includegraphics[width=0.5\textwidth]{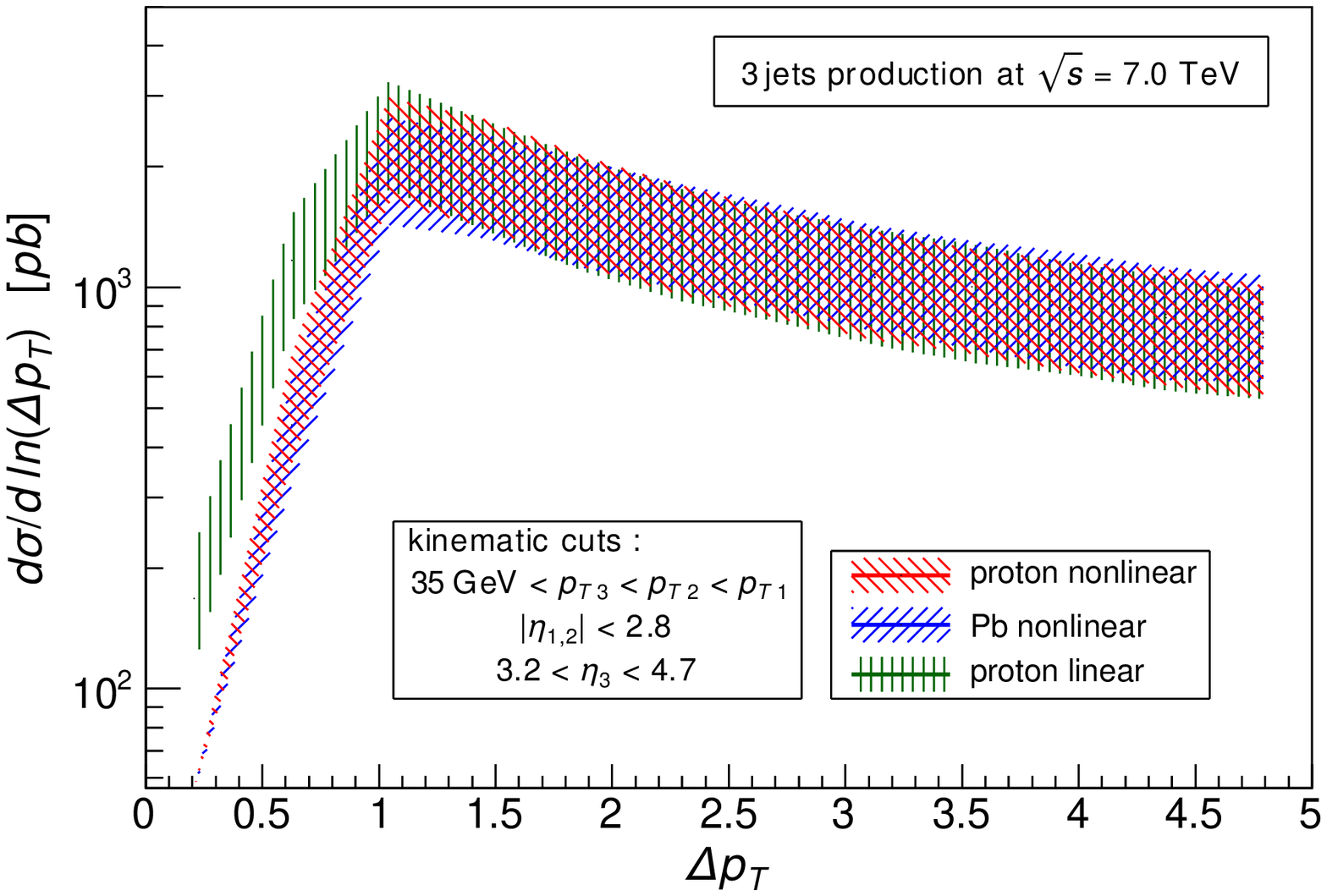}
\par\end{centering}

\caption{\label{fig:fw_kT_band} Differential cross section in the unbalanced
$p_{T}$ for central-forward jets. The band represents the theoretical
uncertainty due to scale variation and statistical errors. The left plots
correspond to c.m. energy $5.02\,\mathrm{TeV}$, the right to $7.0\,\mathrm{TeV}$.
The bottom plots zoom the low $\Delta p_{T}$ region (note the distributions
are differential in $\ln\left(\Delta p_{T}\right)$ there).}
\end{figure}

\begin{figure}
\begin{centering}
\includegraphics[width=0.5\textwidth]{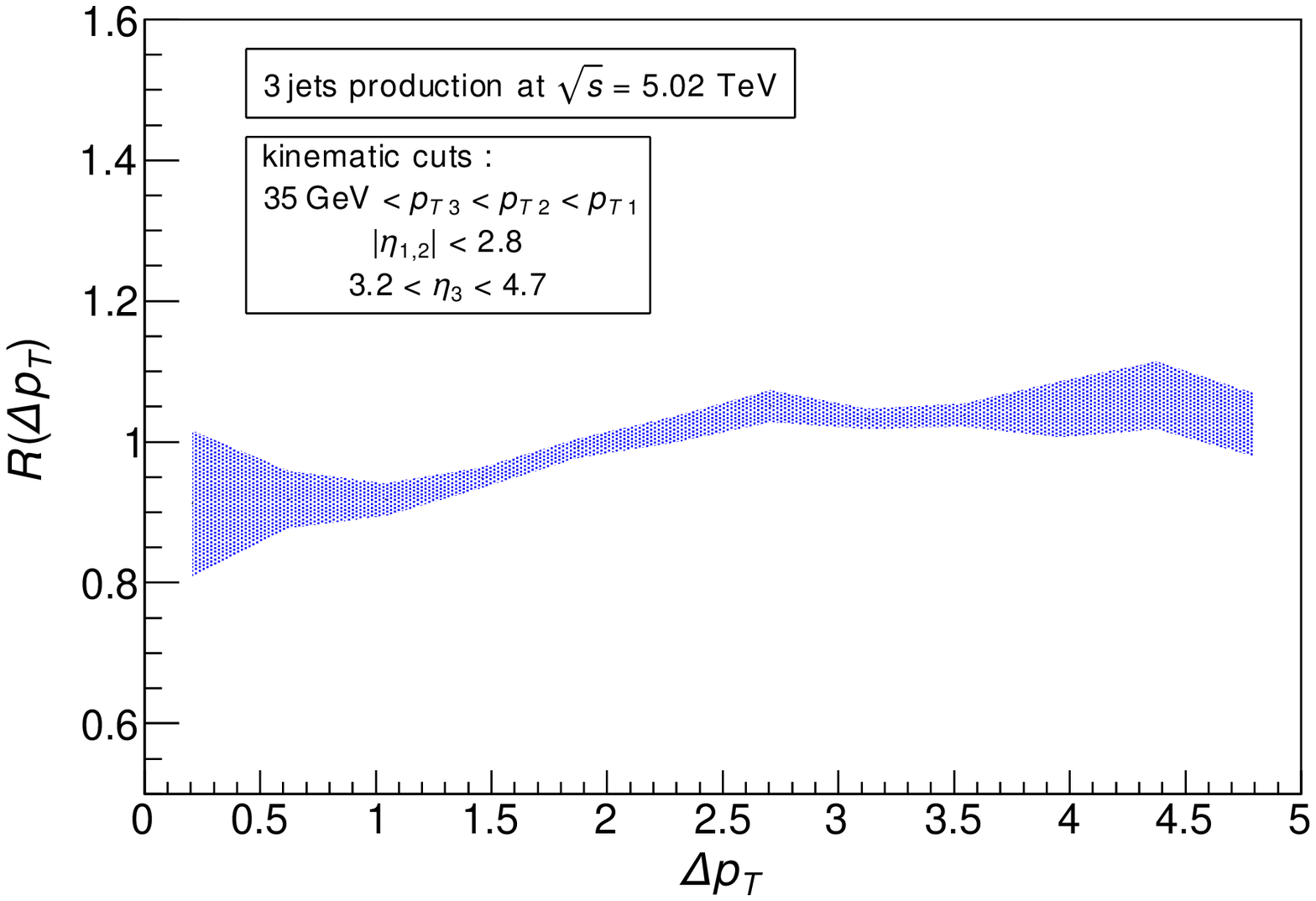}\includegraphics[width=0.5\textwidth]{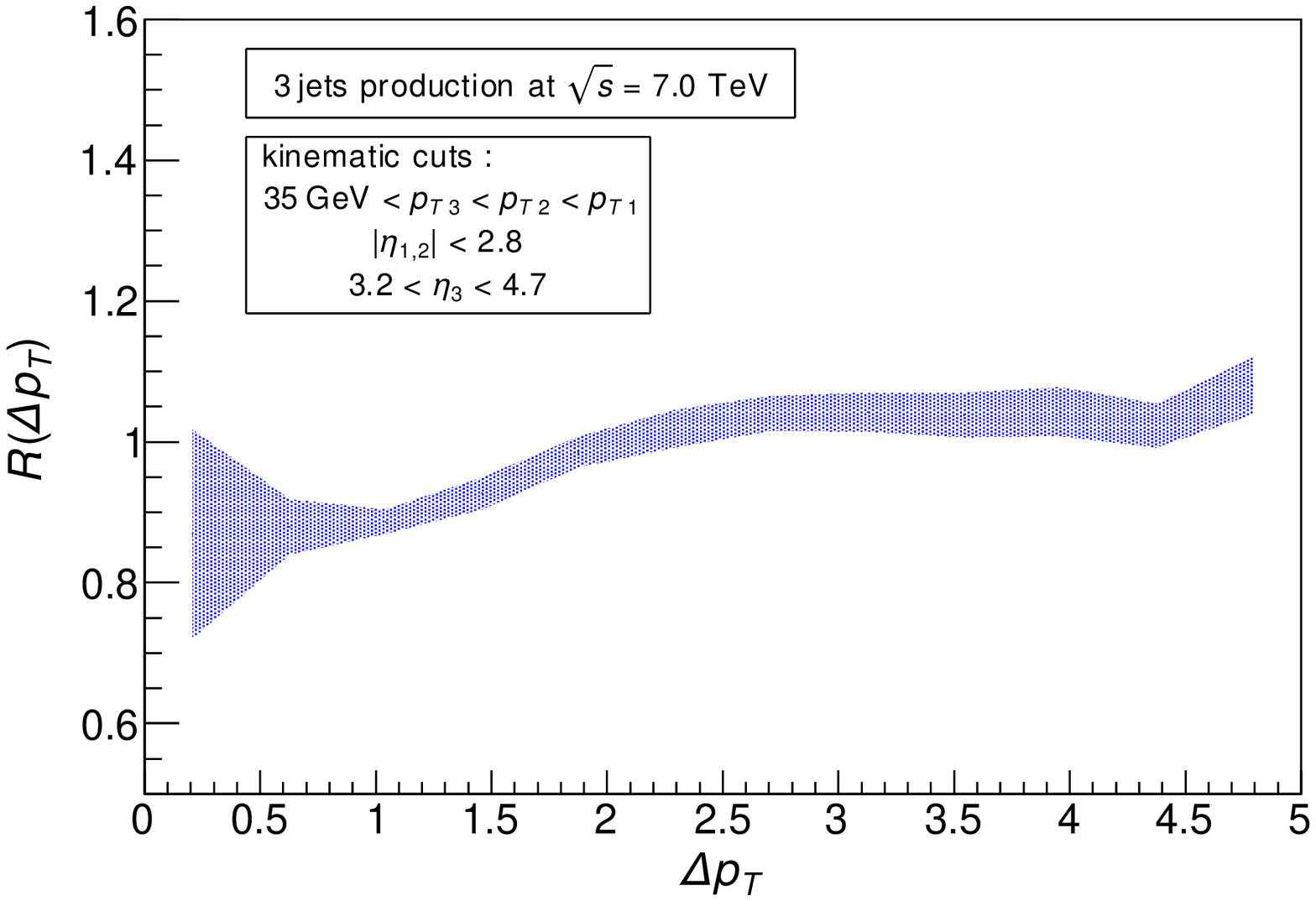}
\par\end{centering}

\caption{\label{fig:fw_kT_ratio} The nuclear modification factor for central-forward
jets as a function of the unbalanced $p_{T}$ for region close to
zero. The band represents the theoretical uncertainty due to scale
variation and statistical errors. The left plot corresponds to c.m. energy
$5.02\,\mathrm{TeV}$, the right to $7.0\,\mathrm{TeV}$.}
\end{figure}

\begin{figure}
\begin{centering}
\includegraphics[width=0.5\textwidth]{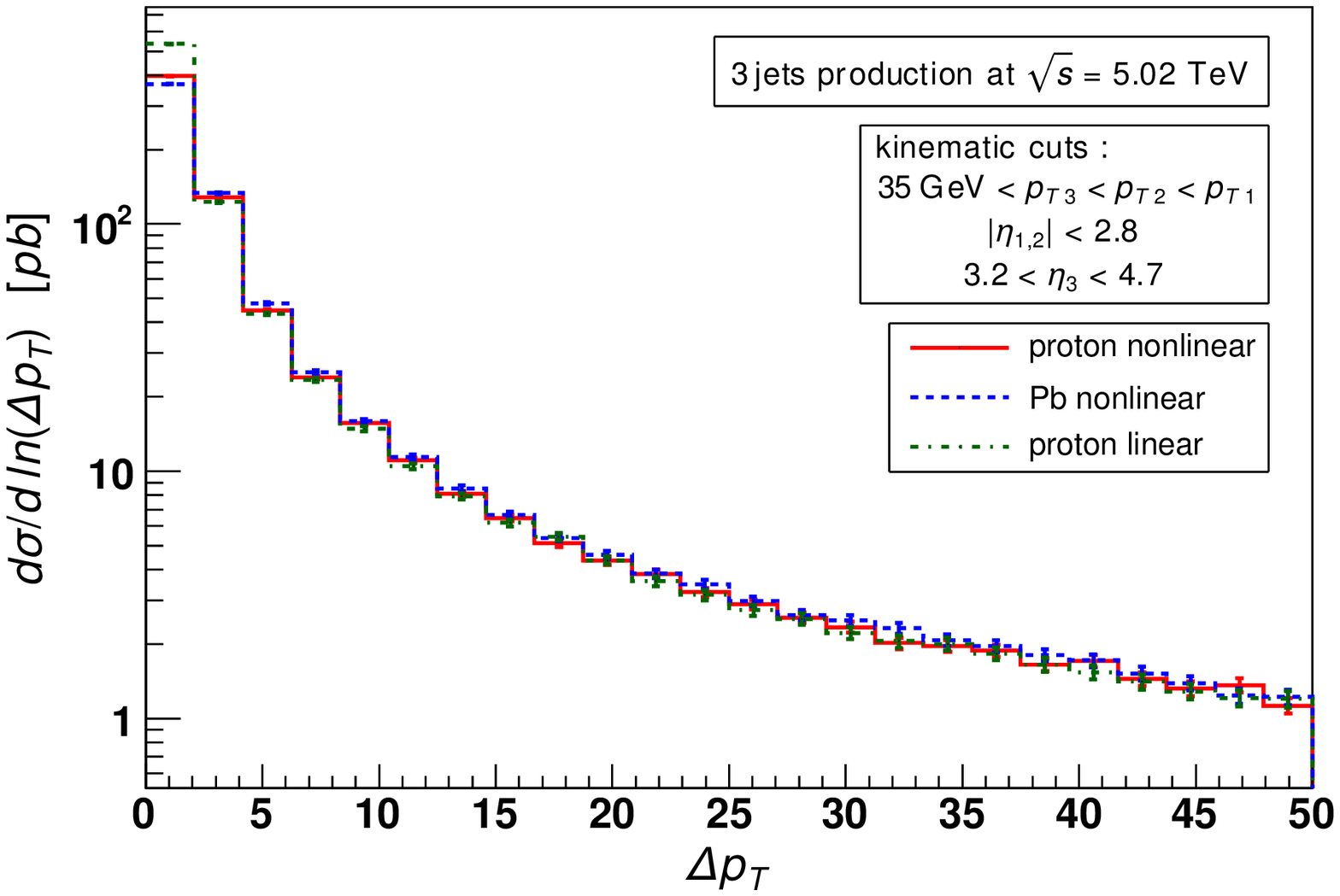}\includegraphics[width=0.5\textwidth]{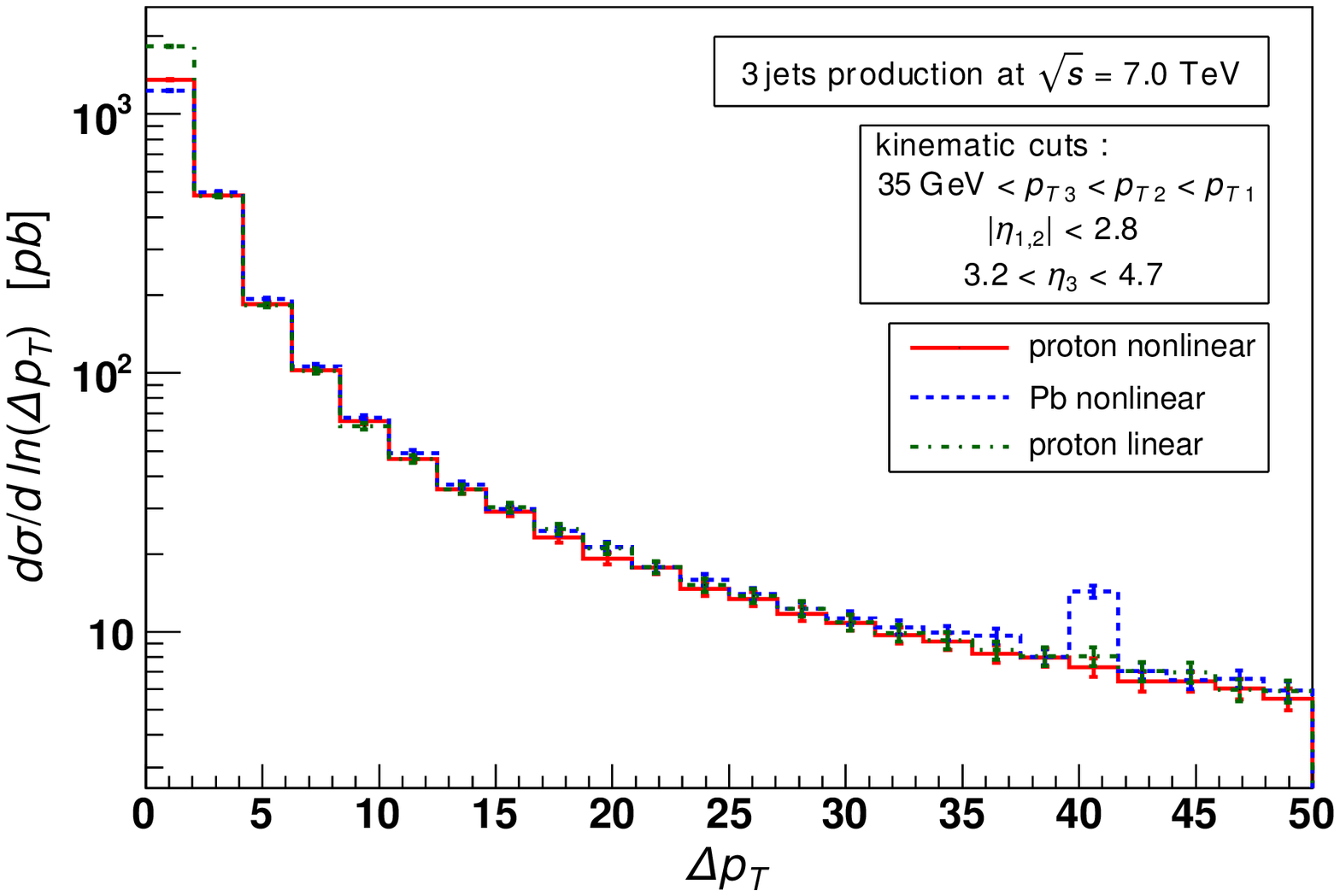}
\par\end{centering}

\begin{centering}
\includegraphics[width=0.5\textwidth]{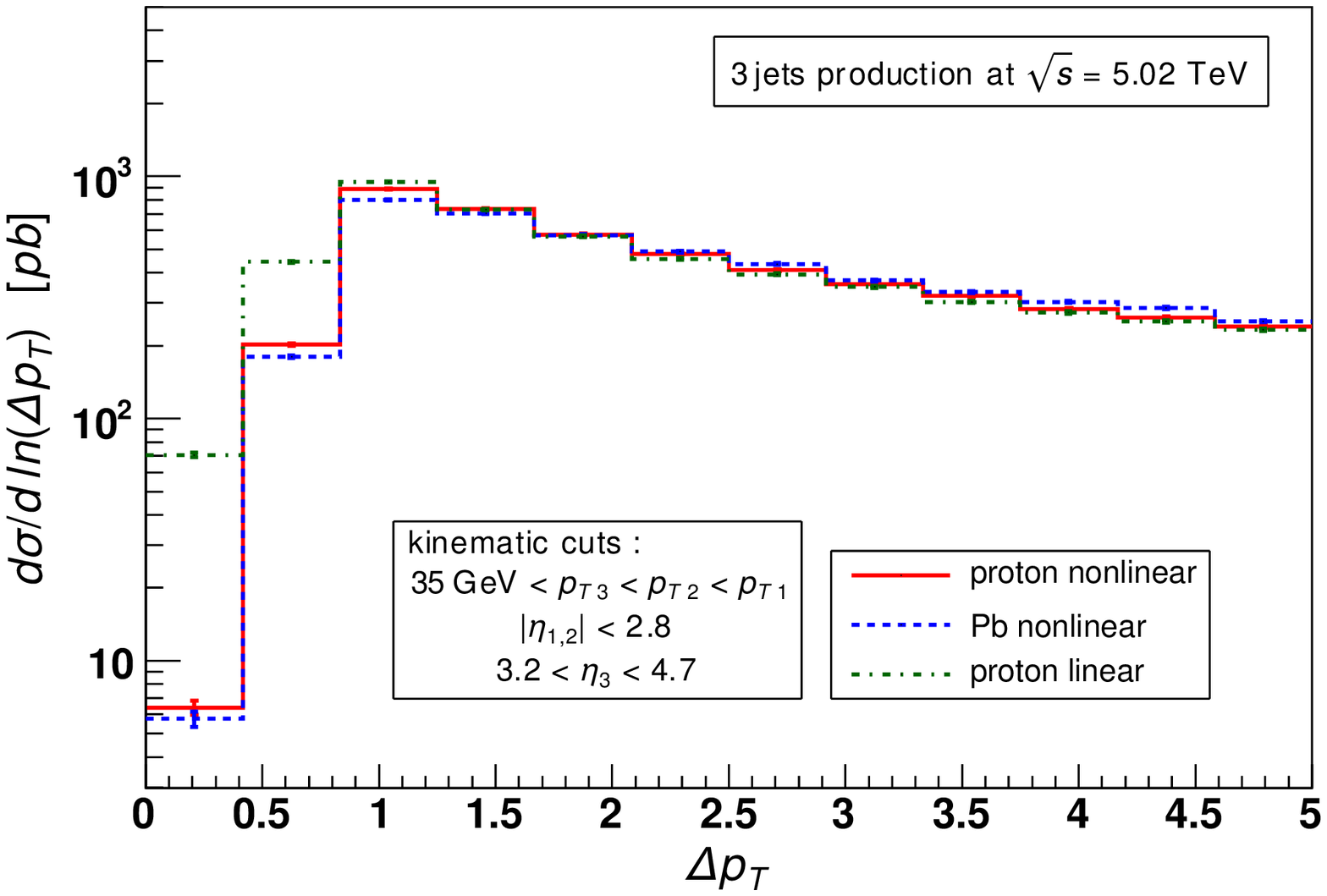}\includegraphics[width=0.5\textwidth]{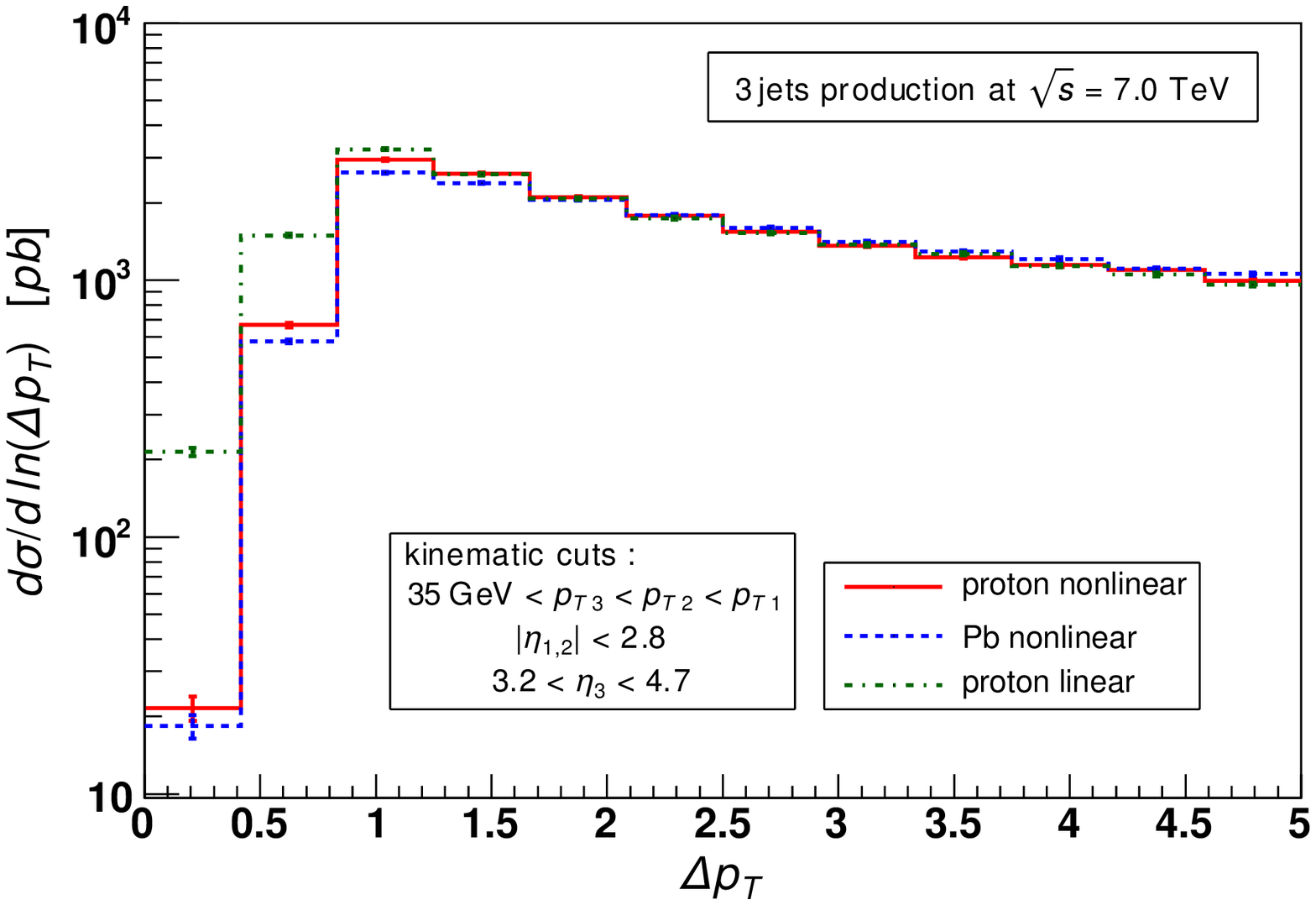}
\par\end{centering}

\caption{\label{fig:fw_kT_exampl}Differential cross section for central-forward
jets in the unbalanced $p_{T}$ for a particular choice of the scale
$\mu/2$. The left column corresponds to c.m. energy $5.02\,\mathrm{TeV}$,
the right to $7.0\,\mathrm{TeV}$. The bottom plots zoom the top plots
for low $\Delta p_{T}$ region but are calculated for the scale $2\mu$ (note the distributions are in $\ln\left(\Delta p_{T}\right)$
there).}
\end{figure}

We present the results in Figs. \ref{fig:fw_kT_band}, \ref{fig:fw_kT_ratio},
\ref{fig:fw_kT_exampl}. The first immediate observation is that the
distributions possess a maximum around $1\,\mathrm{GeV}$ (bottom
of Figs. \ref{fig:fw_kT_band}, \ref{fig:fw_kT_exampl}) which corresponds to the maximum of
the unintegrated gluon densities (see Ref. \cite{Kutak:2012rf}) used
in the calculations. The region below $1\,\mathrm{GeV}$ is sensitive
to the different evolutions; the most significant difference is between
linear and non-linear evolution, however the nuclear modification
factor is slightly suppressed for $\Delta p_{T}<2.5\,\mathrm{GeV}$.

The scenario with the two leading jets being back-to-back-like is
not especially interesting for the unbalanced transverse momentum
distributions due to the kinematics involved. It turns out that the region
of $\Delta p_{T}$ that is smaller then a few $\mathrm{GeV}$ is kinematically
forbidden. Although back-to-back forward-central jets probe actually the high
transverse momenta in the unintegrated gluon density, we did not see
any conclusive features of tails in our distributions.

\subsubsection{Forward jets}

\begin{figure}
\begin{centering}
\includegraphics[width=0.5\textwidth]{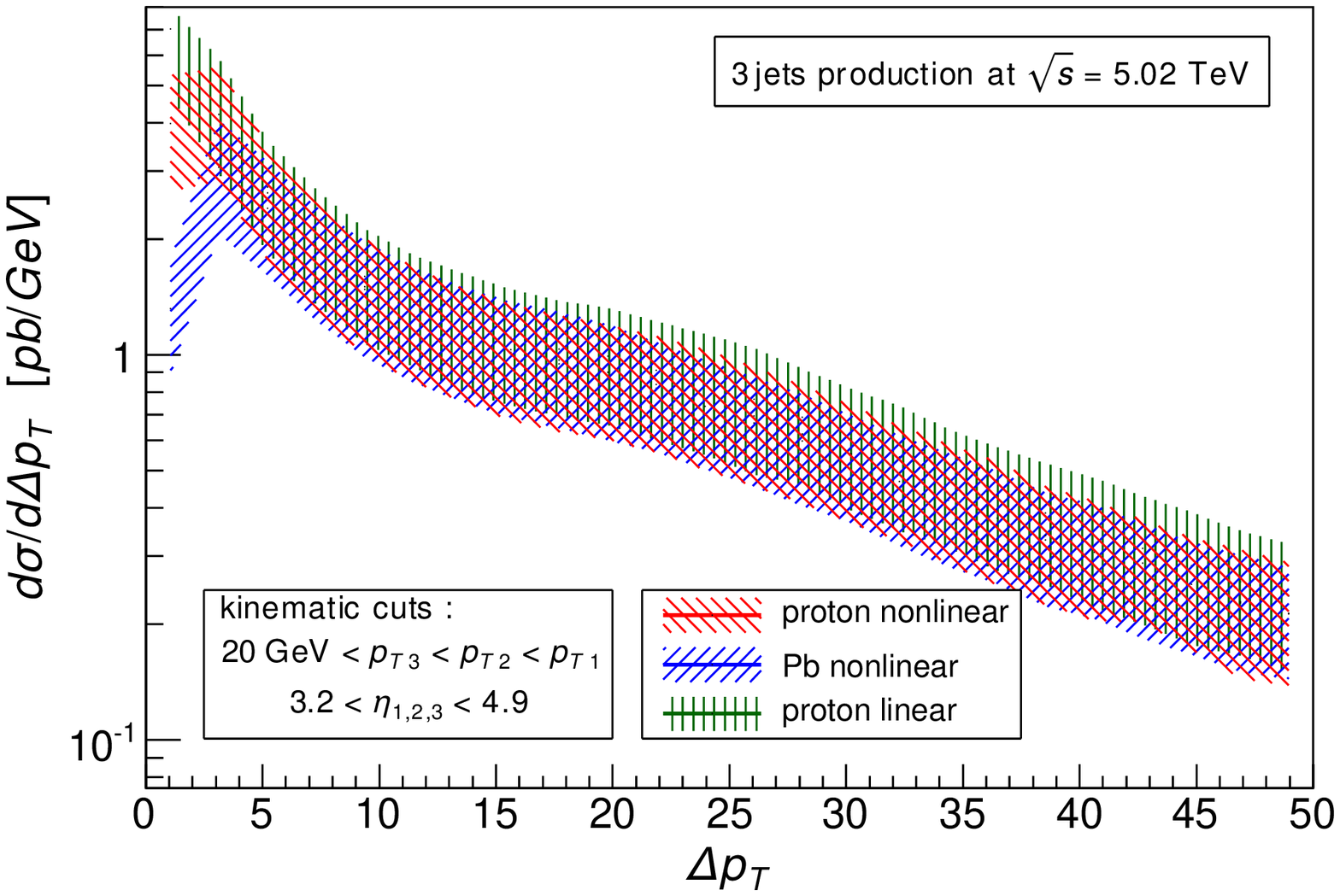}\includegraphics[width=0.5\textwidth]{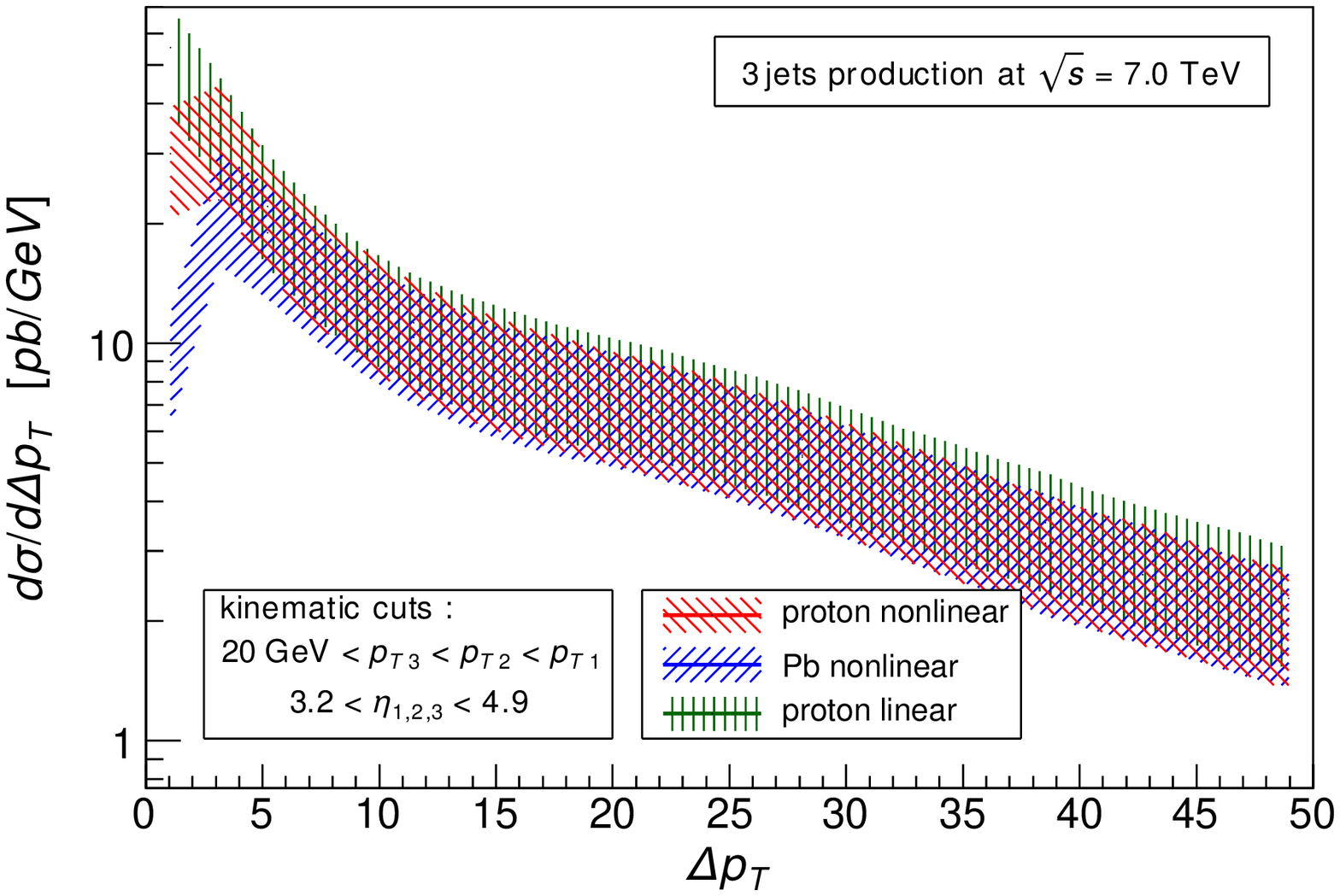}
\par\end{centering}

\begin{centering}
\includegraphics[width=0.5\textwidth]{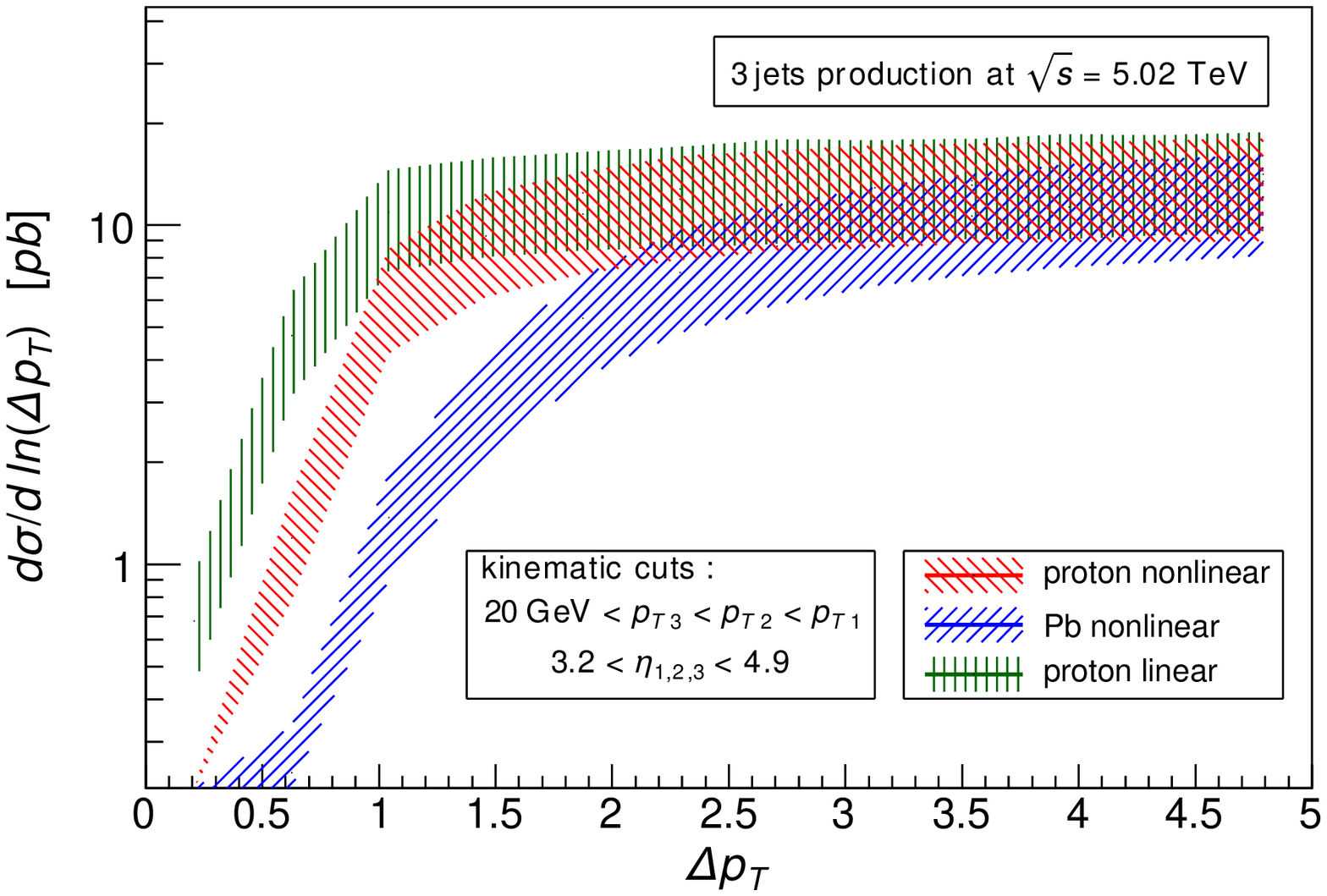}\includegraphics[width=0.5\textwidth]{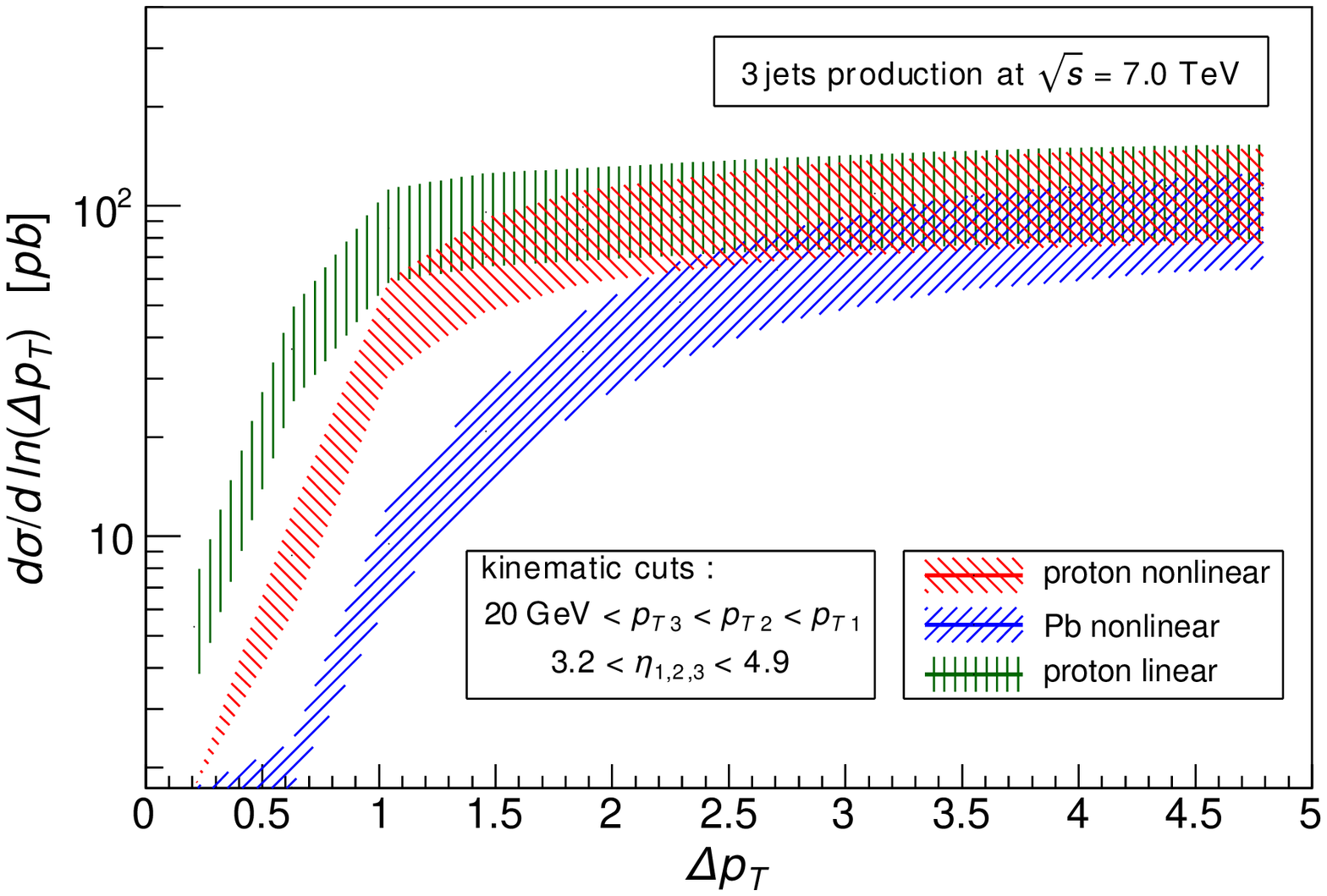}
\par\end{centering}

\caption{\label{fig:ffw_kT_band} Differential cross section in the unbalanced
$p_{T}$ for forward jets. The band represents the theoretical uncertainty
due to scale variation and statistical errors. The left plots correspond
to c.m. energy $5.02\,\mathrm{TeV}$, the right to $7.0\,\mathrm{TeV}$.
The bottom plots zoom the low $\Delta p_{T}$ region (note the distributions
are in $\ln\left(\Delta p_{T}\right)$ there).}
\end{figure}

\begin{figure}
\begin{centering}
\includegraphics[width=0.5\textwidth]{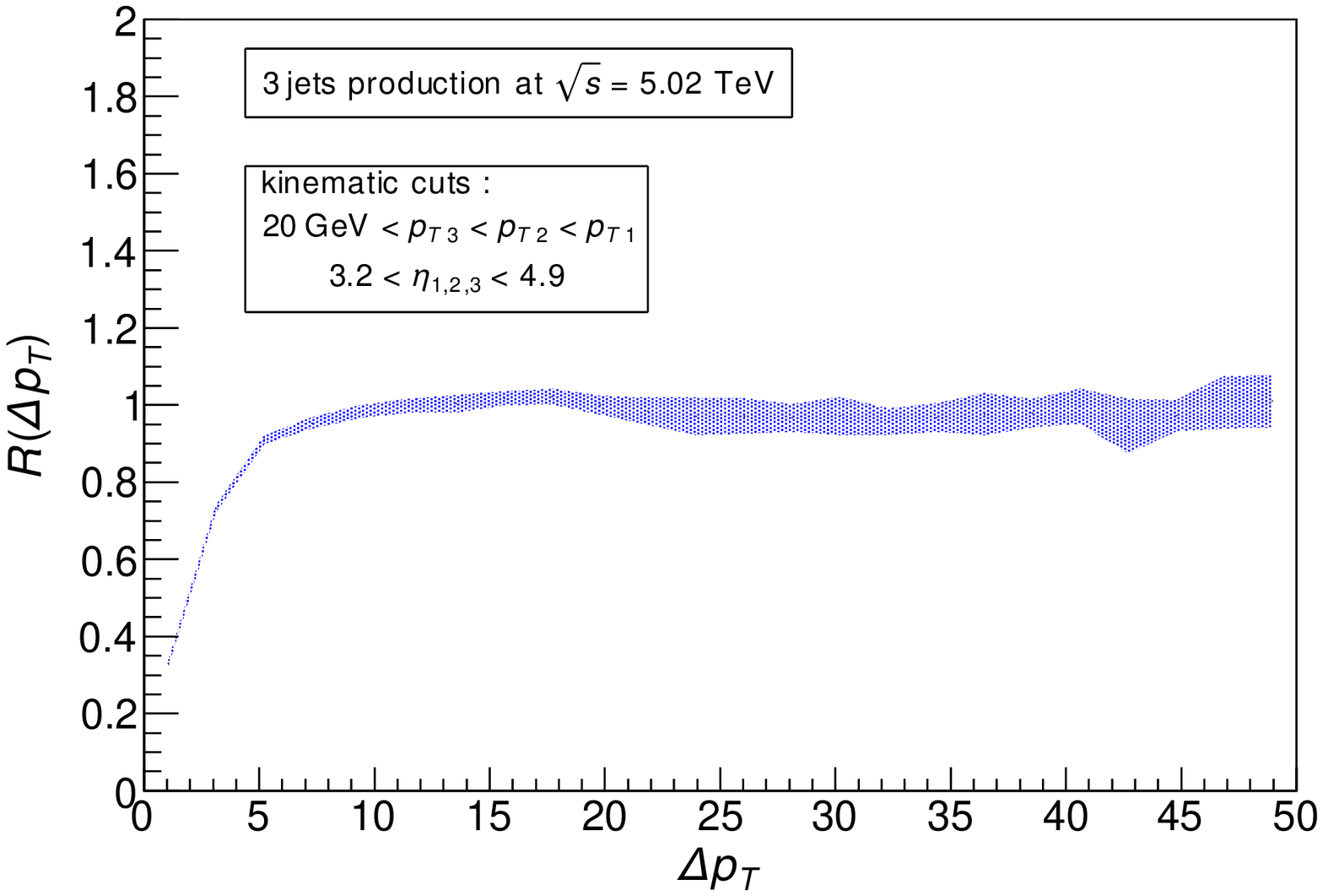}\includegraphics[width=0.5\textwidth]{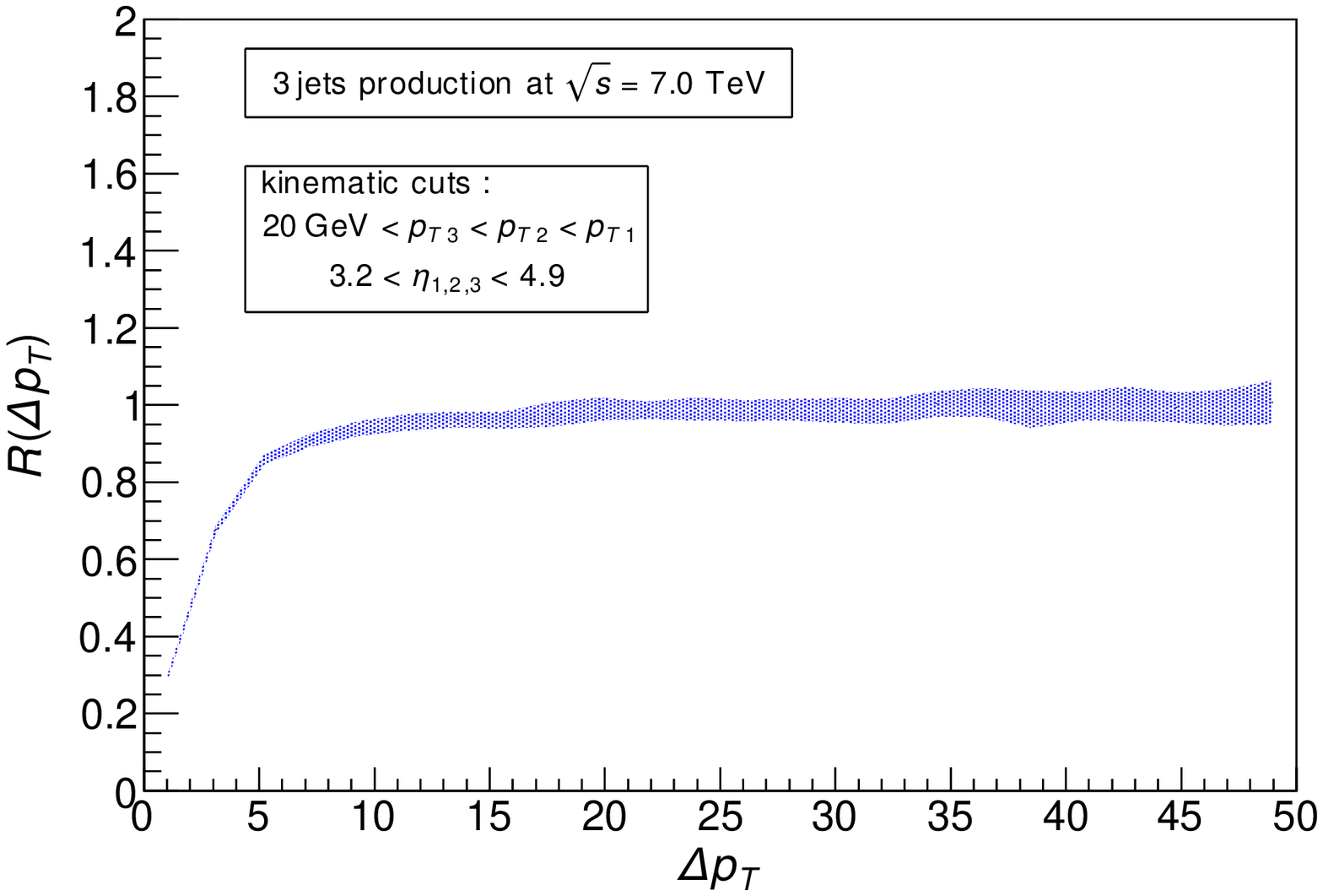}
\par\end{centering}

\begin{centering}
\includegraphics[width=0.5\textwidth]{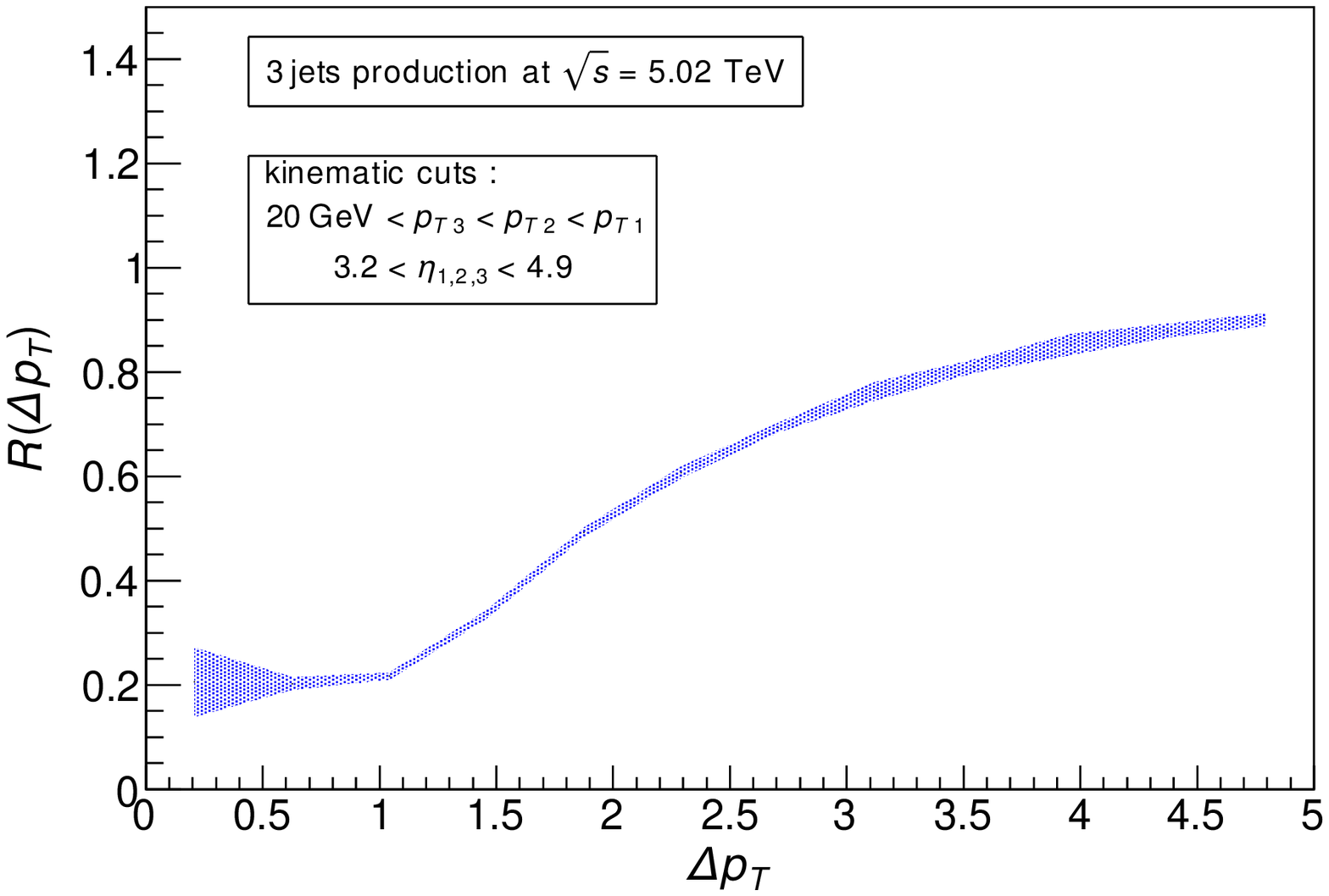}\includegraphics[width=0.5\textwidth]{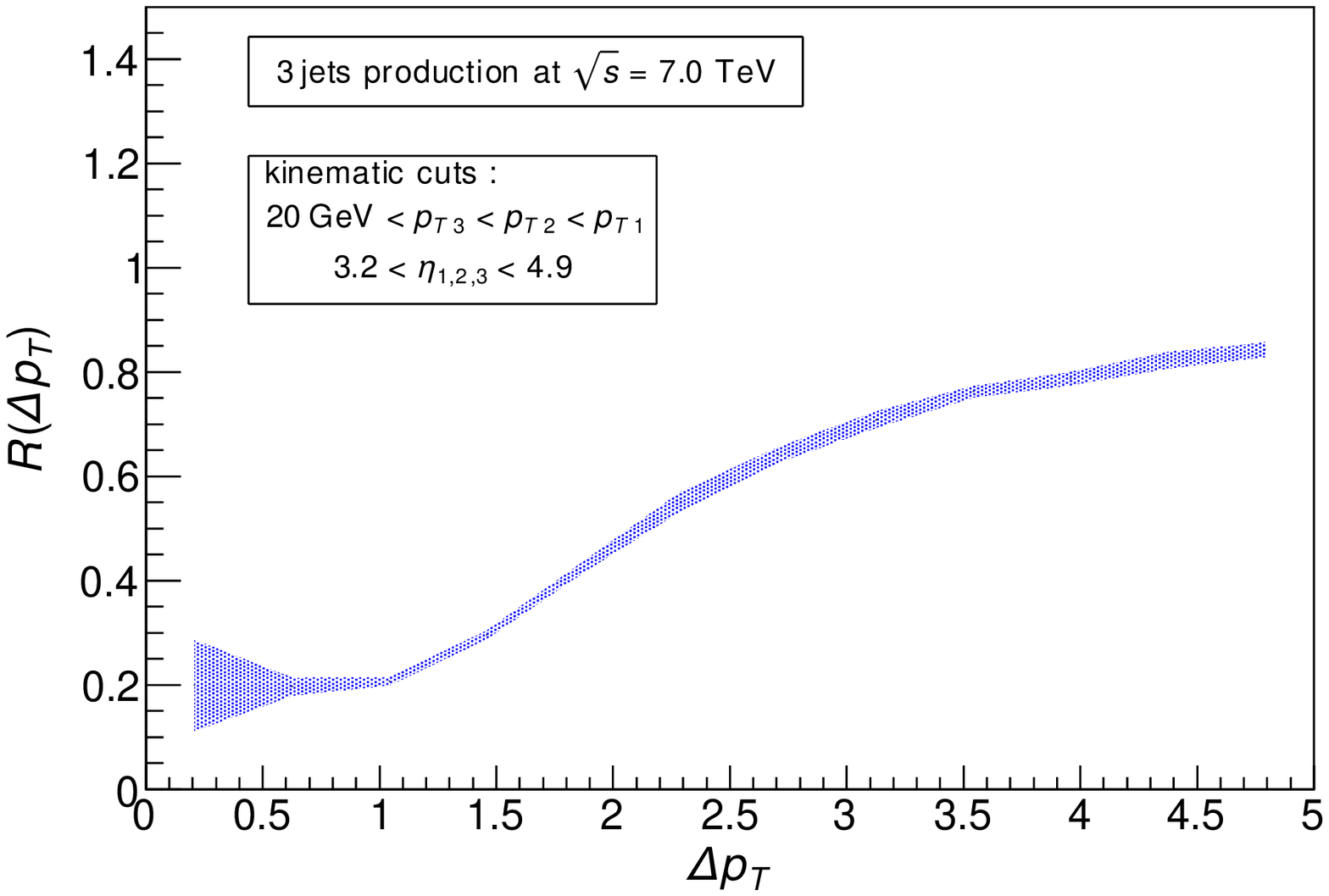}
\par\end{centering}

\caption{\label{fig:ffw_kT_ratio} The nuclear modification factor for forward
jets as a function of the unbalanced $p_{T}$. The band represents the theoretical uncertainty due to scale
variation and statistical errors. The left plots correspond to c.m. energy
$5.02\,\mathrm{TeV}$, the right to $7.0\,\mathrm{TeV}$. The bottom plots
zoom the region close to zero.}
\end{figure}

\begin{figure}
\begin{centering}
\includegraphics[width=0.5\textwidth]{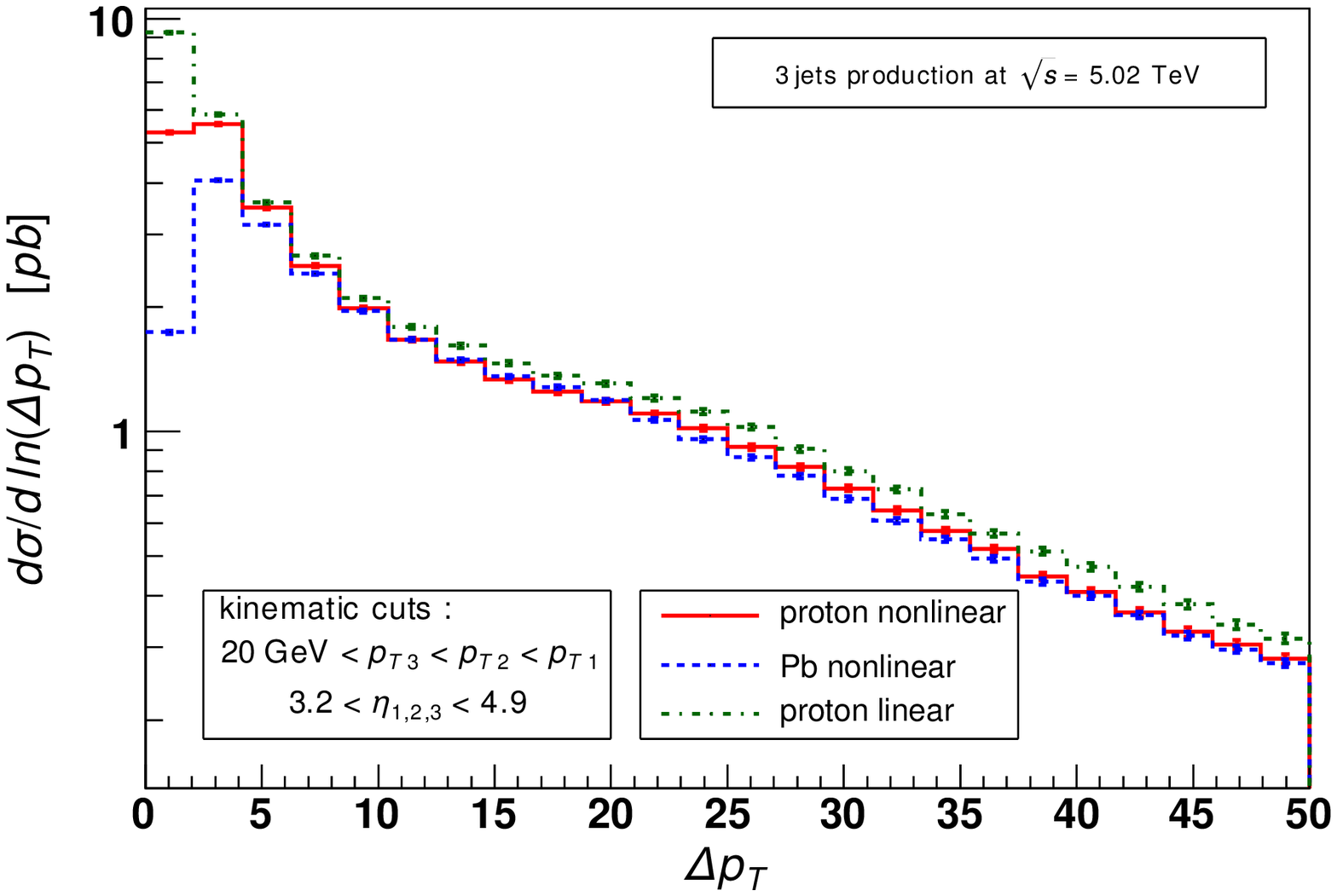}\includegraphics[width=0.5\textwidth]{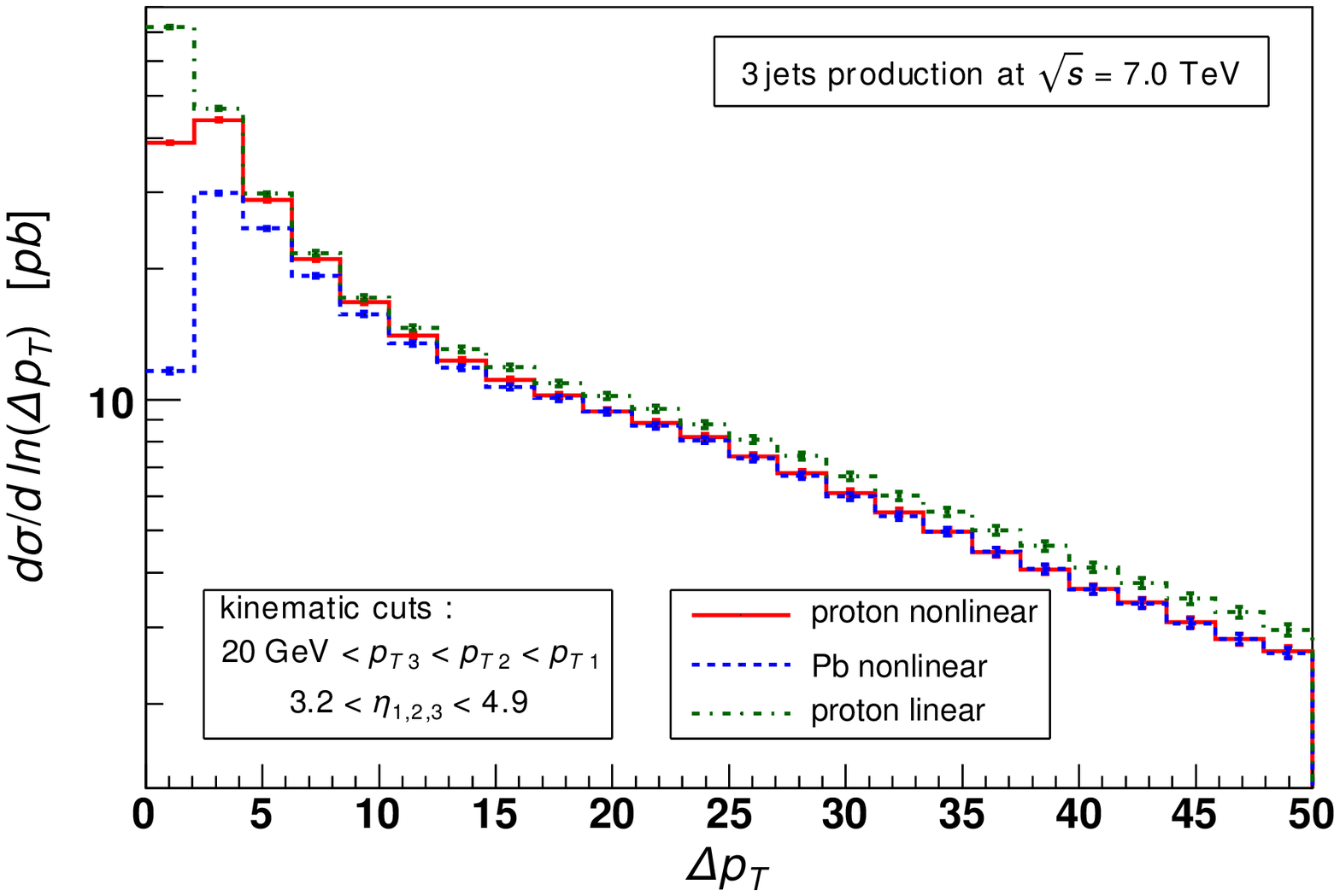}
\par\end{centering}

\begin{centering}
\includegraphics[width=0.5\textwidth]{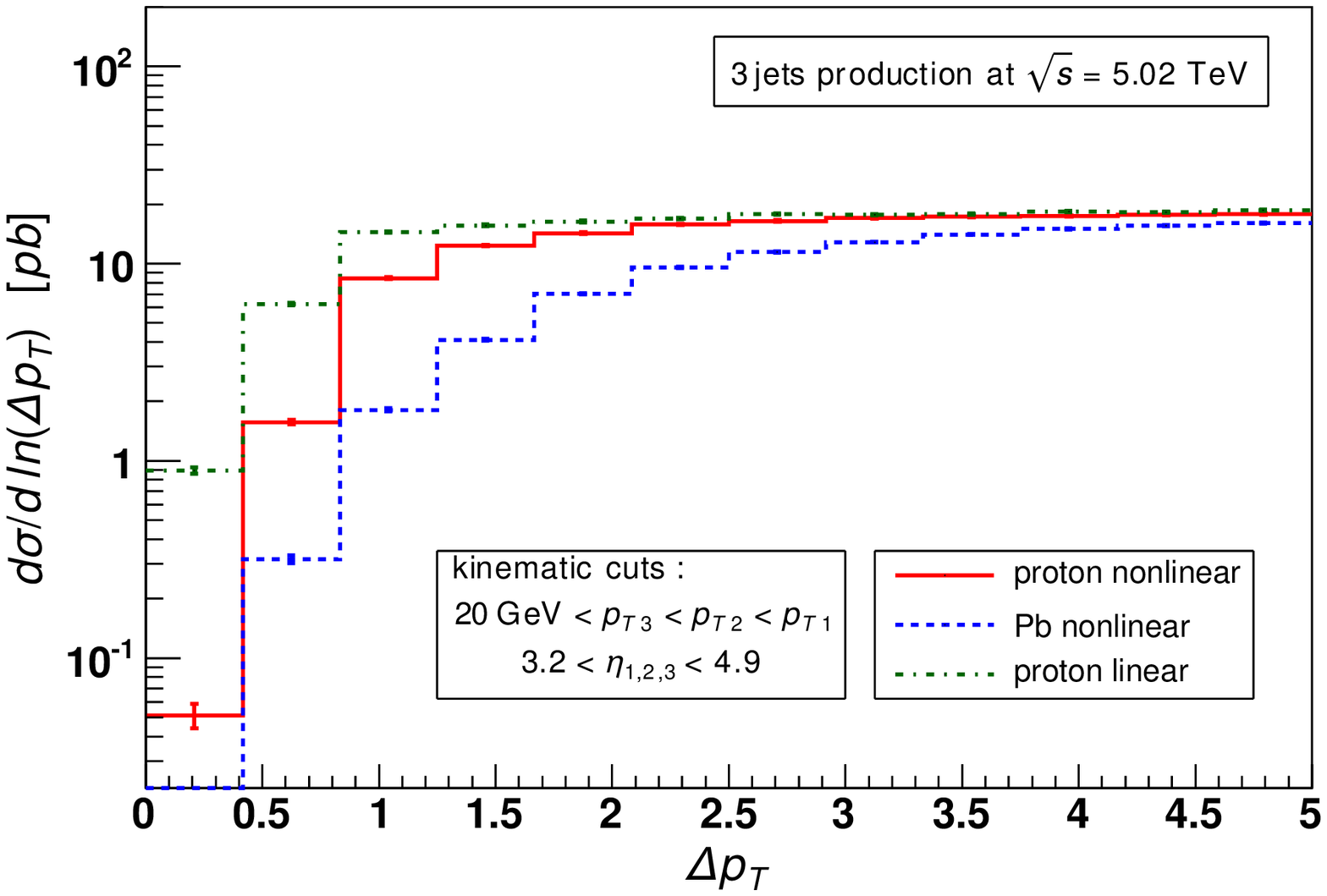}\includegraphics[width=0.5\textwidth]{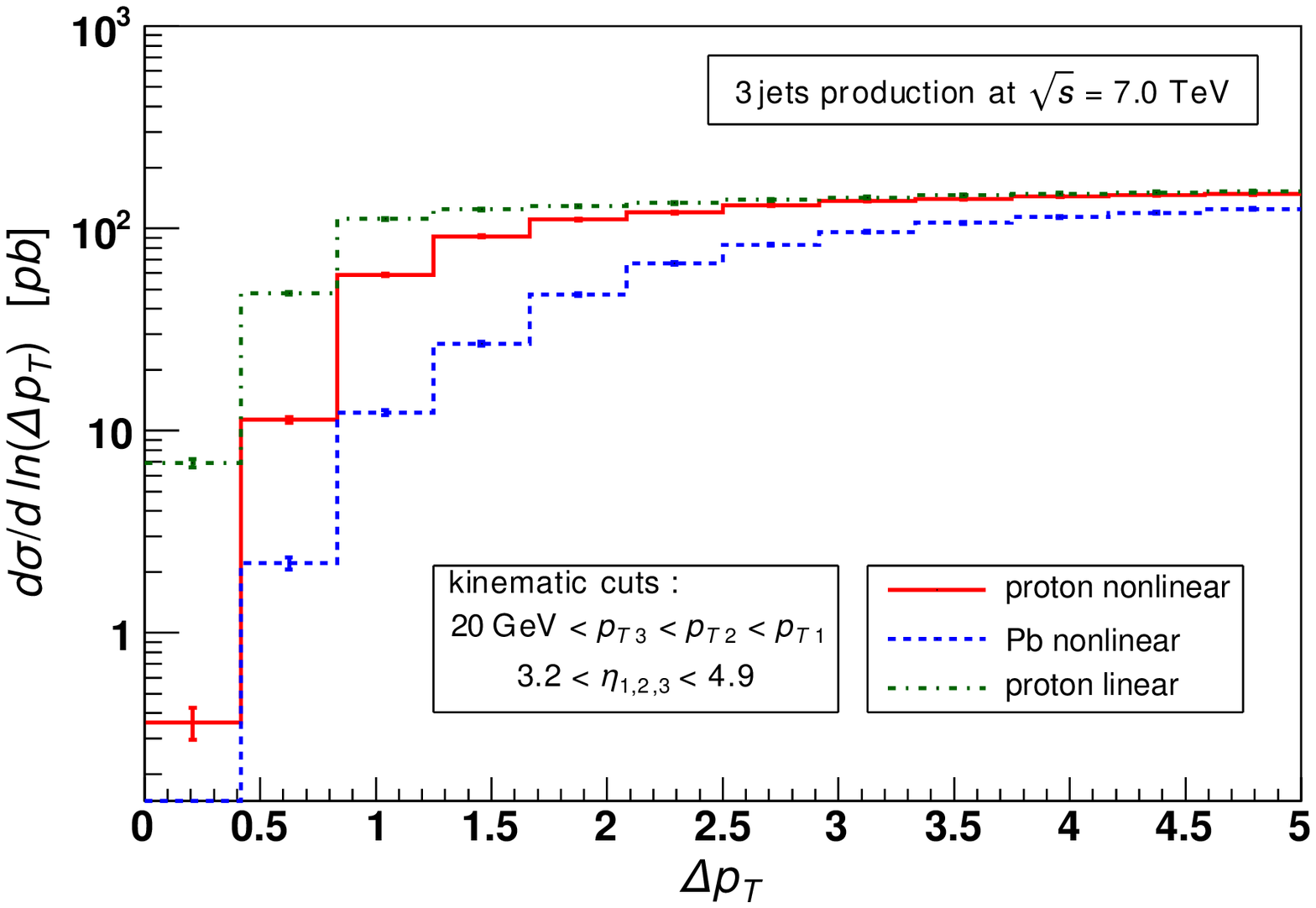}
\par\end{centering}

\caption{\label{fig:ffw_kT_exampl}Differential cross section for forward jets
in the unbalanced $p_{T}$ for a particular choice of the scale $\mu/2$.
The left column corresponds to c.m. energy $5.02\,\mathrm{TeV}$, the right
to $7.0\,\mathrm{TeV}$. The bottom plots zoom the top plots for low
$\Delta p_{T}$ region (note the distributions are in $\ln\left(\Delta p_{T}\right)$
there).}
\end{figure}

Finally we turn to the forward rapidity region. Again we note the significant
differences in the distributions for different evolution scenarios (Figs. \ref{fig:ffw_kT_band}-\ref{fig:ffw_kT_exampl}).
They are most prominent in the low unbalanced transverse momentum
region $\Delta p_{T}<5\,\mathrm{GeV}$ (see the bottom plots in Figs.
\ref{fig:ffw_kT_band}-\ref{fig:ffw_kT_exampl}). We note the suppression
of the distributions with nonlinear evolution, with, however, lead being
suppressed much more. This is also reflected in the nuclear modification
ratios as is evident from Fig. \ref{fig:ffw_kT_ratio}. Interestingly,
the nuclear modification factor is not so much sensitive to the scale
variation.

\section{Summary}

\label{sec:Summary}

In the present work we have studied three jet production at LHC for
proton-proton and proton-lead collisions. As we pointed out the trijet
final state is an ideal tool to perform scan of small $x$ unintegrated
gluon density and to discriminate between different evolution scenarios.
In our work we have used two independent Monte Carlo programs which implement
high-energy $k_{T}$-factorization with a single off-shell gluon and
gauge invariant matrix elements. We have studied three scenarios for
the evolution of the unintegrated gluon density: nonlinear evolution
for proton and lead according to Refs. \cite{Kutak:2003bd,Kutak:2004ym}
and its linear version. From numerous observables that can be constructed
for a three jet process, we have chosen azimuthal decorrelations and
the unbalanced transverse momentum of the jets. We considered two
rapidity regions accessible experimentally: forward-central region
and purely forward region. In addition we have considered the situation,
when the two central jets are back-to-back-like. Our findings can be briefly
summarized as follows. Forward-central collisions with relatively
high cut on the transverse jet momenta are not sensitive to different
kinds of gluon evolution, although they reflect some key features of the high-energy kinematics.
 The situation changes, when the two leading
jets are approximately back-to-back as the distributions start to
be sensitive to the region of a relatively large transverse momentum
in the unintegrated gluon density. For the case of forward scattering,
we observe significant difference between all three kinds of evolution,
in particular the shape of the nuclear modification factors (Figs.
\ref{fig:ffw_azimcorr_ratios}, \ref{fig:ffw_kT_ratio}) suggest strong
suppression due to saturation effects, which is visible both in the
azimuthal decorrelations and the unbalanced $p_{T}$ distributions.

\section*{Acknowledgments}

The authors are grateful N. Armesto, P. Cipriano, S. Jadach, P. van Mechelen, W. S\l ominski, M. Strikman,  and 
M. Trzebi\'nski for discussions. 

This work was supported by the NCBiR grant LIDER/02/35/L-2/10/NCBiR/2011.

\bibliographystyle{abbrv}
\bibliography{fw3j}

\end{document}